\documentclass[review, nopreprintline]{elsarticle}

\usepackage{framed,multirow}
\usepackage{amssymb}
\usepackage{latexsym}
\usepackage{chemformula} 
\usepackage{amsmath}

\DeclareMathOperator*{\vech}{vech}     
\DeclareMathOperator*{\cdf}{cdf}       

\usepackage{url}

\usepackage[T1]{fontenc}
\usepackage[utf8]{inputenc}

\begin{document}

\begin{frontmatter}

\title{Optimisation of the event-based TOF filtered back-projection for online imaging in total-body J-PET}

\author[1]{R.~Y. Shopa\corref{cor1}}
\ead{Roman.Shopa@ncbj.gov.pl}
\cortext[cor1]{Corresponding author at: Department of Complex Systems, National Centre for Nuclear Research, 05-400 Otwock-{\'{S}}wierk, Poland 
  Tel.: +48-22-273-13-06;  
  fax: +48-22-273-16-87;}
\author[1]{K. Klimaszewski}
\author[1]{P. Kopka}
\author[1]{P. Kowalski}
\author[2]{W. Krzemie{\'{n}}}
\author[1]{L. Raczy{\'{n}}ski}
\author[1]{W. Wi{\'{s}}licki}
\author[3,4]{N. Chug}
\author[5]{C. Curceanu}
\author[3,4]{E. Czerwi{\'{n}}ski}
\author[3,4]{M. Dadgar}
\author[3,4]{K. Dulski}
\author[3,4]{A. Gajos}
\author[6]{B.~C. Hiesmayr}
\author[3,4]{K. Kacprzak}
\author[3,4]{{\L}. Kap{\l}on}
\author[3,4]{D. Kisielewska}
\author[3,4]{G. Korcyl}
\author[3,4]{N. Krawczyk}
\author[3,4]{E. Kubicz}
\author[3,4]{Sz. Nied{\'{z}}wiecki}
\author[3,4]{J. Raj}
\author[3,4]{S. Sharma}
\author[3,4]{Shivani}
\author[3,4]{E.~{\L}. St{\c{e}}pie{\'{n}}}
\author[3,4]{F. Tayefi}
\author[3,4]{P. Moskal}\fnref{fn1}
\fntext[fn1]{P. Moskal is a senior author.}

\address[1]{Department of Complex Systems, National Centre for Nuclear Research, 05-400 Otwock-{\'{S}}wierk, Poland}
\address[2]{High Energy Physics Division, National Centre for Nuclear Research, 05-400 Otwock-{\'{S}}wierk, Poland}
\address[3]{Faculty of Physics, Astronomy and Applied Computer Science, Jagiellonian University, prof. Stanis{\l}awa {\L}ojasiewicza 11, 30-348 Cracow, Poland}
\address[4] {Total-Body Jagiellonian-PET Laboratory, Jagiellonian University, Poland}
\address[5]{INFN, Laboratori Nazionali di Frascati, 00044 Frascati, Italy}
\address[6]{Faculty of Physics, University of Vienna, 1090 Vienna, Austria}

\begin{abstract}
We perform a parametric study of the newly developed time-of-flight (TOF) image reconstruction algorithm, proposed for the real-time imaging in total-body Jagiellonian PET (J-PET) scanners. The asymmetric 3D filtering kernel is applied at each most likely position of electron-positron annihilation, estimated from the emissions of back-to-back $\gamma$-photons. The optimisation of its parameters is studied using Monte Carlo simulations of a $1$-mm spherical source, NEMA IEC and XCAT phantoms inside the ideal J-PET scanner. The combination of high-pass filters which included the TOF filtered back-projection (FBP), resulted in spatial resolution, $1.5$ times higher in the axial direction than for the conventional 3D FBP. For realistic $10$-minute scans of NEMA IEC and XCAT, which require a trade-off between the noise and spatial resolution, the need for Gaussian TOF kernel components, coupled with median post-filtering, is demonstrated. The best sets of 3D filter parameters were obtained by the Nelder-Mead minimisation of the mean squared error between the resulting and reference images. The approach allows training the reconstruction algorithm for custom scans, using the IEC phantom, when the temporal resolution is below $50$ ps. The image quality parameters, estimated for the best outcomes, were systematically better than for the non-TOF FBP.
\end{abstract}


\end{frontmatter}


\section{Introduction}
Recent research of new detector materials for positron emission tomography (PET) has opened up avenues for the advanced time-of-flight (TOF) image reconstruction algorithms that provide quality unseen before \citep{Karp2008, Surti2008, Kadrmas2009}, envisioned earlier only by theoretical models \citep{Allemand1980, Mullani1980, Tomitani1981, Yamamoto1982, Budinger1983, Snyder1983, Wong1983}. New scintillating materials available on the market today, such as \ch{LSO:Ce}, \ch{LYSO:Ce}, \ch{LaBr3:Ce}, could potentially achieve coincidence resolving time (CRT) of about $100$~ps \citep{Schaart2010, Gundacker2013, Nemallapudi2015}. In practice, the time of response is limited by a signal readout: even with improved electronics and silicon photomultiplier (PM) matrices (SiPM), clinical TOF-scanners are reaching CRT of $\sim210-400$~ps \citep{Surti2007, Conti2011, Slomka2016, Vandenberghe2016, Grant2016, VanSluis2019, Karp2020}. Further improvement can be achieved using Cherenkov radiation \citep{Miyata2006, Arino-Estrada2019}.

The trend for improving time resolution means that under certain conditions analytical algorithms may achieve quality comparable to the iterative methods \citep{Westerwoudt2014}. Small enough $\text{CRT}\sim10$~ps, predicted in Refs.~\cite{Lecoq2017, Lecoq2020}, potentially allows PET data to be reconstructed by the deposition of TOF events into the image domain, as done in the analytic-DIRECT algorithm \citep{Matej2009, Matej2016}, apply multivariate kernel density estimation (KDE) \citep{Scott1992} or histo-imaging \citep{Vandenberghe2006} over the most likely positions (MLPs) of electron-positron annihilation. The outcome is expected to be less affected by the time of a scan and the geometry of the detector, caused by the distortion during re-projection. However, it is difficult to achieve accuracy below the typical size of a voxel due to the fundamental limits of spatial resolution caused by non-collinearity and positron range \citep{Moses2011}.

The MLP-focused paradigm offers the possibility to process registered events independently. Launched in parallel on powerful computers, it opens perspectives for real-time (online) imaging. It also eliminates the issue of the non-linear noise transition between the projection and image spaces \citep{Walker2011, Nuyts2011, Cloquet2011} -- a problem relevant to iterative methods applied to short scans.

One of the first on-the-fly reconstructions was implemented in INSIDE in-beam PET \citep{Marafini2015, Ferrero2018, Fiorina2018}, for monitoring hadron therapy in clinical conditions. The detector geometry is simple enough to apply iterative algorithms over small datasets ($\sim10^4$ events per time frame). As for the conventional PET scanners, recently reported methods use the pseudo-inverse system response matrix \citep{Lopez-Montes2020} and data-driven direct reconstruction by deep neural networks \citep{Whiteley2021}.

A major trend today is the development of total-body PET scanners with large (up to $2$~m) axial field-of-view (AFOV). This poses major challenges for image reconstruction, concerning processing power, memory size and data acquisition \citep{Badawi2019, Karp2020, Moskal2020, Vandenberghe2020}. A vivid example is the Jagiellonian PET (J-PET) -- a tomograph that utilises the detection of the Compton scattering of back-to-back annihilation $\gamma$-photons inside plastic scintillator strips \citep{Moskal2014, Moskal2015, Moskal2016, Raczynski2014, Raczynski2015, Raczynski2017, Palka2017, Korcyl2018, Moskal2020}. Each scattering hit produces optical signals, registered as two timestamps by PMs attached at opposite ends of a strip, thus making TOF measurements possible. To increase the detection efficiency of plastic scintillators, a multi-layer structure and larger tunnels are used. A total-body $2$-meter J-PET modular scanner, which is now under development \citep{Niedzwiecki2017, Moskal2018, Moskal2019a, Moskal2020, Moskal2020a}, is designed to use a complex readout system: SiPMs with an additional layer of adjacent scintillators -- wavelength shifters (WLS), which should achieve an axial resolution of less than $5$~mm \citep{Smyrski2017}. Moreover, WLS allows partial extraction of depth-of-interaction (DOI) information from the spatial distribution of optical photons inside the strips \citep{MoskalPatent2013}, similarly to the principle, described in Ref. \cite{MarcinkowskiDOI2016}.

The complex physics of detection in J-PET requires big data processing and efficient computation. Therefore, a dedicated field-programmable gate array (FPGA) system-on-chip (SoC) platform has been created. It performs event building, filtering, coincidence search and back-projection \citep{Palka2017, Korcyl2018}. The data, in a list-mode format, can be partitioned using time frames or in an event-by-event way. The principle of operation is compatible with MLP-type reconstructions and, what is the most important, -- with the real-time mode.

A modification of the multivariate KDE, applied to MLP (KDE MLP), using an asymmetric 3D kernel that includes a 2D filtered back-projection (FBP) component, has been introduced in recent works related to the total-body J-PET \citep{Shopa2020, Daria2020}. It integrates separate 3D reconstructions of each emission event and can be launched in parallel. By utilising FPGA, the algorithm could be accelerated to operate on-the-fly during the measurement.

In this work, we shall employ the advanced definition of the 3D kernel for the proposed event-based algorithm, applying regularisation to FBP and adjustable Gaussian or inverse Gaussian filters as TOF components, not strictly defined by CRT as in previous papers. Moreover, a detailed analysis of the kernel properties will be reported, using the simulated data inside the $50$-cm long ideal cylindrical J-PET scanner, for CRT in a $50-500$\,ps range. The main goal is to find the optimal set of parameters for kernel components during quick online scans, aimed to achieve the best trade-off between the spatial resolution, image quality, noise levels and performance speed. 

The structure of the paper is as follows. Section 2 describes the methods and the objects of study. Firstly, image reconstruction algorithms will be presented: KDE MLP and a novel image-domain TOF FBP. Then, the simulation details will be revealed for the J-PET scanners and the phantoms, as well as for the metrics used for the qualitative analysis. Section 3 will focus on the reconstruction results from the viewpoint of kernel properties, optimised for spatial resolution or image quality. Finally, the discussion and the conclusions will be provided. 

\section{Methods}
\subsection{Image reconstruction using time-of-flight}
The measured PET data is expressed as a set of $4$-dimensional projections for conventional scanners and $5$-dimensional -- for TOF scanners. The set of $N$ detected emissions with TOF is expressed as:
\begin{equation}
\lbrace \hat{\textbf{\textit{e}}}_1, \hat{\textbf{\textit{e}}}_2, ..., \hat{\textbf{\textit{e}}}_N\rbrace \subset \mathbb{R}^5, \quad
\hat{\textbf{\textit{e}}}_k = (s,\phi,\zeta,\theta,\Delta t)^\text{T}_k.
\label{eq:projectionTOF}
\end{equation}

\noindent where $s$ and $\phi$ are transaxial coordinates of a particular line-of-response (LOR), $\zeta$ is the axial coordinate of its mid-point, $\theta$ defines the obliqueness angle and $\Delta t$ is the difference between the detection times of annihilation $\gamma$-photons \citep{PETBasicScience2005}.

The granularity of a scanner defines bins for the unique $s$, $\phi$, $\zeta$, $\theta,\Delta t$. TOF information increases their sparseness, in particular for total-body scanners of a large AFOV \citep{Raczynski2020}. Alternatively, the measurements can be processed directly without partitioning into bins (list-mode) or redefined in the spatial domain as MLPs of annihilation points.

\subsubsection{Kernel density estimator applied to most likely position}
Multivariate KDE \citep{Scott1992} can be applied to the set of most likely points $\lbrace \textbf{\textit{x}}_{\text{MLP}}^{(k)}\rbrace$ of $n$ PET emissions ($k=1\ldots n$), estimated for each LOR from its endpoints (scattering hits in J-PET) and TOF difference $\Delta t$ \citep{Vandenberghe2006}. The data is processed as density distributions, avoiding the problem of choosing the anchor points, relevant to histo-imaging \citep{Silverman1986}. It is also justified by the fact that the unknown detector response function, which depends on $\textbf{\textit{x}}_{\text{MLP}}^{(k)}$ and the geometrical orientation of the corresponding LOR, can be perceived as random \citep{Strzelecki2016}.

Kernel density estimator is defined for a 3D dataset $\lbrace \textbf{\textit{x}}_{\text{MLP}}^{(k)}\rbrace$ as
\begin{equation}
\label{eq:KDE}
\hat{f}_{n\mathbf{H}}(\textbf{\textit{x}}) = n^{-1}\sum\limits_{i=1}^{n} \mathcal{K}_\mathbf{H}
\left(\textbf{\textit{x}}-\textbf{\textit{x}}_{\text{MLP}}^{(i)}\right),
\end{equation}  

\noindent where $\textbf{\textit{x}}$ is a vector of voxel coordinates and a symmetric probability density function $\mathcal{K}_\mathbf{H}(x)=|\mathbf{H}|^{-1/2}\mathcal{K}(\mathbf{H}^{-1/2}\textbf{\textit{x}})$ is defined by a 3D Gaussian kernel $\mathcal{K}(\textbf{\textit{x}})$. $\mathbf{H}$ is the bandwidth matrix, symmetric and positive-definite \citep{Chacon2010}, which plays the role of the covariance matrix that controls the amount and orientation of smoothing induced. 

According to Refs. \citep{Duong2003, Duong2007, Chacon2010}, the definition of $\mathcal{K}(\textbf{\textit{x}})$ is not critical for a large $n$, but the choice of $\mathbf{H}$ is crucial. The algorithms dedicated to find an optimal bandwidth matrix are generally based on the minimisation of the mean squared error (MSE) between $\hat{f}_{n\mathbf{H}}(\textbf{\textit{x}})$ and a tractable approximation that replaces the true density $f_n(\textbf{\textit{x}})$. For instance, for a so-called \textit{plug-in} (PI) estimate, a following function is to be minimised:
\begin{equation}
\label{eq:PI}
\text{PI}(\mathbf{H}) = n^{-1}(4\pi)^{-d/2}|\mathbf{H}|^{-1/2}+\frac{1}{4}(\vech\mathbf{H})^\text{T}\hat{\mathbf{\Psi}}_4(\mathbf{G})(\vech{\mathbf{H}}),
\end{equation}  
\begin{equation}
\label{eq:Hpi}
\hat{\mathbf{H}}_\text{PI} = \arg\min\limits_{\mathbf{H}}\text{PI}(\mathbf{H}).
\end{equation}

Here, an estimator $\hat{\mathbf{\Psi}}_4(\mathbf{G})$ substitutes an unknown matrix $\mathbf{\Psi}_4$, composed of integrals with the products of the derivatives of $f_n(\textbf{\textit{x}})$ (see Ref. \cite{Duong2003} for the definition). The elements of $\hat{\mathbf{\Psi}}_4(\mathbf{G})$ are calculated from the "measured" $\textbf{\textit{x}}_{\text{MLP}}^{(i)}$ and a pilot bandwidth matrix, set initially as e.g. $\mathbf{G}=g^2\mathbf{I}$, where $g>0$, $\mathbf{I}$ -- identity matrix \citep{Wand1994}.

A half-vectorisation operator in (\ref{eq:PI}) is defined as follows:
\begin{equation}
\label{eq:vech}
\vech\mathbf{H}=
\vech\begin{bmatrix}
h_{11} & h_{12} & h_{13} \\
h_{12} & h_{22} & h_{23} \\
h_{13} & h_{23} & h_{33} \\
\end{bmatrix} = (h_{11},h_{21},h_{31},h_{22},h_{23},h_{33})^\text{T}.
\end{equation}

We utilise the popular {\small \textsf{R}} package {\small\textsf{'ks'}} \citep{Duong2007}, with two
PI estimators for $\hat{\mathbf{H}}_\text{PI}$: {\small\textsf{samse}}, based on a sum of asymptotic MSE between the elements of $\hat{\mathbf{\Psi}}_4(\mathbf{G})$ and $\mathbf{\Psi}_4$ \citep{Duong2003}, and {\small\textsf{dscalar}} -- a more advanced multistage PI bandwidth selector \citep{Chacon2010}. 

According to Eq.~(\ref{eq:KDE}), \textit{all measured} data in a form of $\textbf{\textit{x}}_{\text{MLP}}^{(i)}$ is required to estimate the intensity of each voxel $\textbf{\textit{x}}$. However, KDE MLP as a reconstruction method could operate in event-by-event mode by changing the order of summation in Eq.~(\ref{eq:KDE}) or partitioning the data into time frames. $\hat{\mathbf{H}}_\text{PI}$ can be predefined or regularly recalculated according to the updated $\hat{\mathbf{\Psi}}_4(\mathbf{G})$ by the newly measured data.

\subsubsection{Event-based time-of-flight FBP in the image domain}
FBP is one of the first reconstruction methods introduced for PET \citep{Helgason1984}, which became a standard required by the National Electrical Manufacturers Association (NEMA) for estimating the spatial resolution \citep{NEMA2012}. In previous works \citep{Shopa2017,Kowalski2018}, we utilised a 3D FBP with re-projection (FBP 3DRP) from the freeware STIR package \citep{STIR2006, STIR2012}. It does not support TOF, multi-layer geometry and the continuous J-PET detectors, and will be used as a reference method in this paper. 

We modify FBP as proposed in Ref. \cite{Conti2005} -- by adding TOF as temporal kernel $h(t)$ that splits 4-dimensional projection functions $p(s,\phi,\zeta,\theta)$ into TOF bins. For the $i$-th bin $h_i(t) \equiv h(t-t_i)$, the filtered function $p^F_i\equiv p^F_i(s,\phi,\zeta,\theta,t)$ reads:
\begin{equation}
p^F_i(s,\phi,\zeta,\theta,t)=\mathcal{F}^{-1}\lbrace W(\omega_s)\mathcal{F} \left[p_i(s,\phi,\zeta,\theta)\right] \rbrace\cdot h(t-t_i),
\label{eq:TOFpF}
\end{equation} 

\noindent where $\mathcal{F}(\cdot)$ and $\mathcal{F}^{-1}(\cdot)$ denote the Fourier and inverse Fourier transform, respectively, the frequency space coordinate $\omega_s$ is related to $s$ which 1-dimensional filter $W(\omega_s)$ is applied to.

We shall treat each emission independently -- in list-mode, instead of accumulating them in bins. The summation is made over LORs $(i=1\ldots N_{\text{LOR}})$, while the time difference $t_i\equiv \Delta t_i =t^{(i)}_2-t^{(i)}_1$ is related to the $i$-th emitted pair of back-to-back $\gamma$-photons. $p_i(s,\phi,\zeta,\theta)$ then becomes a delta-like function (unnormalised unit impulse), and filtering over $s$ will return the filter itself: $p_i(s)\ast w(s)=w(s)$, where $w(s)=\mathcal{F}^{-1}W(\omega_s)$ \citep{Shopa2020}. The eventually reconstructed image $\hat{f}_{N_{\text{LOR}}}(\textbf{\textit{x}})$ is a sum over the $N_\text{LOR}$ back-projections each comprising a single LOR:
\begin{equation}
\label{eq:TOFFBP}
\hat{f}_{N_{\text{LOR}}}(\textbf{\textit{x}}) = \sum_{i=1}^{N_{\text{LOR}}}\hat{f}_i(\textbf{\textit{x}}) = 
	\sum_{i=1}^{N_{\text{LOR}}}\mathcal{B}\lbrace p^F_i(s,\phi,\zeta,\theta,t)\rbrace.
\end{equation}

Each $p^F_i$ can be redefined in the $\mathbb{R}^2$ space of variables $s$ and $t$, only, as the rest remain constant. For conventional PET scanners, this allows to represent $\mathcal{B}\lbrace p^F_i\rbrace$ in the image domain as a 3D kernel constituted by the following components:
\begin{enumerate}
\setlength\itemsep{0.0em}
\item[1)] TOF function $h(t)$, applied along LOR. For convenience, we redefine it as $h_\text{TOF}(l)$, for the distance $l$ from MLP.
\item[2)] FBP filter $w(s)$, aligned perpendicularly to the transverse projection of the LOR.
\end{enumerate}

The unique J-PET design requires an additional function:
\begin{enumerate}
\item[3)] $h_Z(z)$, which represents the axial uncertainty of the two hit positions (Compton scatterings) along scintillator strips.
\end{enumerate}

Despite both $h_\text{TOF}(l)$ and $h_Z(z)$ are defined by the detection times, they are uncorrelated \citep{Niedzwiecki2017}. 

\begin{figure}[!t]
\centering
\includegraphics[width=0.57\textwidth]{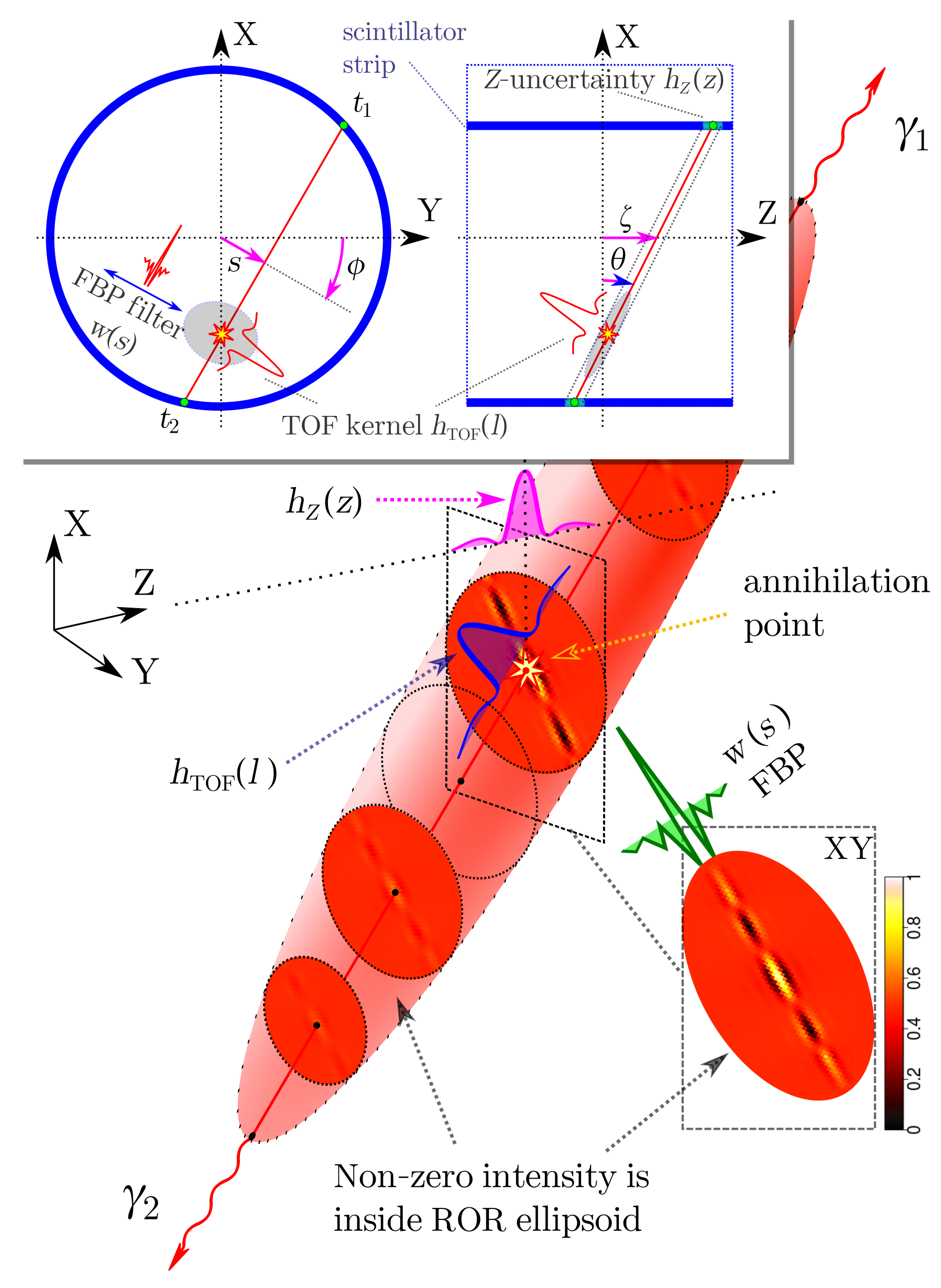}
\caption{Schematic depiction of the three-component kernel -- $w(s)$, $h_\text{TOF}(l)$ and $h_Z(z)$, applied to the MLP of the emission point of back-to-back photons $\gamma_{1,2}$, shown in 3D and as transverse and axial views inside an ideal cylindrical J-PET scanner (inset). $t_1$ and $t_2$ are detection times.}
\label{Fig:LBL_scheme}
\end{figure}

The exemplary representation of a three-component kernel applied over the MLP of an emission point is given in Fig.~\ref{Fig:LBL_scheme}. Projection coordinates and detection times are indicated in the inset on top. Since at some distance from MLP the intensity is effectively zero, it is calculated only inside a small ellipsoid -- region-of-response (ROR), which volume depends on the time resolution and the properties of the filters.

The standard FBP, for a given view $\phi_i$, back-projects $p^F_i$ to the entire (transverse) image plane. To preserve the proper integration over LORs with an additional $h_\text{TOF}(l)$ profile, one should apply weighting normalisation that depends on temporal resolution -- CRT. A TOF model for $w(s)$ function was proposed in Ref. \cite{Zeng2019} as a solution to the following problem:
\begin{equation}
w(s,\phi) {\,\ast\ast\,} h_\text{PSF}(s,\phi) = \delta(s),
\label{eq:PSFconvolve}
\end{equation}

\noindent i.e. the 2D convolution (denoted "$\ast\ast$") of the filter with a point spread function $h_\text{PSF}(s,\phi)$ should return Dirac delta function $\delta(s)$, regardless of the angle $\phi$. A realistic Gaussian $h_\text{PSF}(s,\phi)$ does not depend on $\phi$ either, and the problem is solved in Fourier space, using 1D Hankel transform. Eventually,
\begin{equation}
W(\omega_s) = \frac{\exp{\left[(\pi\tau\omega_s)^2\right]}}{I_0\left[ {(\pi\tau\omega_s)^2}\right]},
\label{eq:FBPfilterTOF}
\end{equation}

\noindent where $\tau$ represents TOF resolution that acts as a regularisation parameter and $I_0(x)=\pi^{-1}\int_0^\pi{\exp{\left[-x\cos\phi\right]}d\phi}$ is the modified Bessel function of the first kind, order 0. Spatial-domain filter $w(s)=\mathcal{F}^{-1} W(\omega_s)$ cannot be expressed in closed form, only calculated numerically.

\begin{figure}[!t]
\centering
\includegraphics[width=0.67\textwidth]{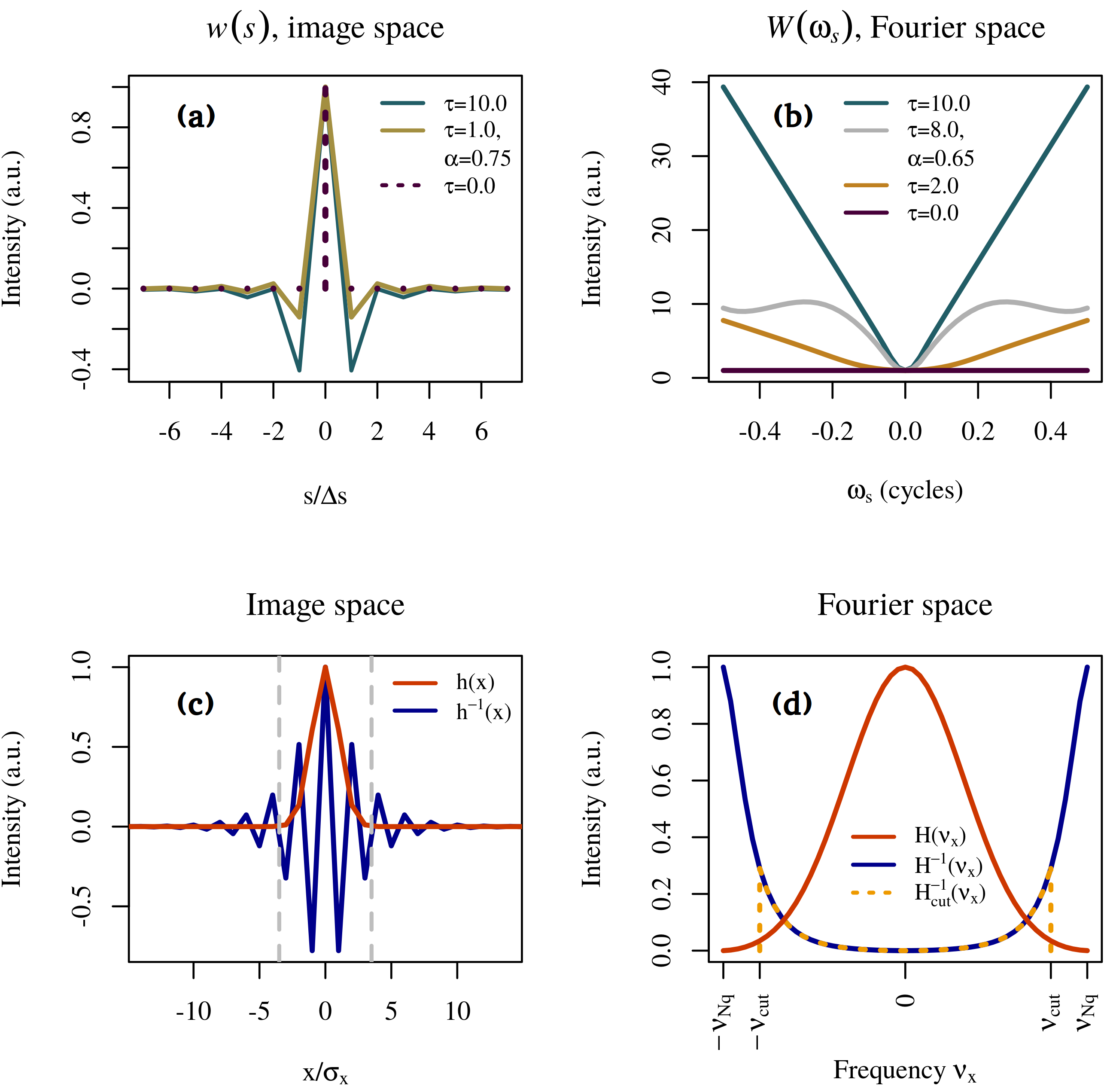}
\caption{Top: FBP filter $w(s)$ (a) and its Fourier representation $W(\omega_s)$ (b), defined by Eqs.~(\ref{eq:FBPfilterTOF})--(\ref{eq:RampApodised}), depending on TOF regularisation $\tau$ and smoothing $\alpha$. For better visualisation, $\tau$ and $\alpha$ are chosen differently. Bottom: Gaussian kernel $h(x)$ and its inverse $h^{-1}(x)$ (c), with the corresponding Fourier transforms $H(\nu_x)$ and $H^{-1}(\nu_x)$, respectively (d). Dashed lines mark $\pm3.5\sigma_x$ truncation range (c) and apodised $H^{-1}_\text{cut}(\nu_x)$ for cut-off frequency $\nu_\text{cut}$ (d).}
\label{Fig:RampAndGaussIndexed}
\end{figure}

Selected examples are shown in Fig.~\ref{Fig:RampAndGaussIndexed}, a-b. The impact of $\tau$ is easier to track in Fourier space. Relatively large $\tau>10.0$ (in cycles$^{-1}$) represent poor timing, as $W(\omega_s)$ approaches $2\pi\tau\vert\omega_s\vert$, behaving as pure ramp (Ram-Lak) filter. Conversely, if $\tau$ is very small, $W(\omega_s)$ becomes a constant $1$ (no filter is required).

The penalisation for the missing data and noise is made by a smoothing window $M(\omega_s)$ \citep{STIR2012}:
\begin{equation}
W(\omega_s) \rightarrow W(\omega_s)\cdot M(\omega_s),
\label{eq:FBPFourier}
\end{equation}
\begin{equation}
M(\omega_s)=
	\begin{cases} 
		\alpha + (1-\alpha)\cdot\cos(\pi\omega_s/\omega_c), &  \vert\omega_s\vert \leq \omega_c \\
		0 & \text{otherwise}
	\end{cases}
\label{eq:RampApodised}
\end{equation} 

\noindent The parameters $\alpha$ and $\omega_c$ (cut-off frequency) can be reduced from the default values that represent ramp filter: $\alpha=1$, $\omega_c=0.5$ (in cycles, corresponds to Nyquist frequency) \citep{FundMedImg2011}. The examples are given in Fig.~\ref{Fig:RampAndGaussIndexed}, a-b.

Unrelated to FBP kernel components $h_\text{TOF}(l)$ and $h_Z(z)$ (Fig.~\ref{Fig:LBL_scheme}) mostly affect the axial direction. We shall utilise two different approaches for their definition: 

\textit{a) Low-pass filtering}, introduced in our work \citep{Shopa2020}, aimed at suppressing noise by using Gaussian cumulative distribution function (CDF) -- $\cdf(x,\mu,\sigma)$. Kernel components can then be expressed as follows:
\begin{equation}
\label{eq:cdfZcdfTOF}
\begin{split}
h_\text{TOF}(l) &= \cdf(l+\Delta l/2,0,\sigma_{\text{TOF}}) - \cdf(l-\Delta l/2,0,\sigma_{\text{TOF}}).\\
h_Z(z)          &= \cdf(z+\Delta z/2,0,\sigma_Z) - \cdf(z-\Delta z/2,0,\sigma_Z),
\end{split}
\end{equation}

\noindent where $\mu=0$ corresponds to MLP, $\sigma_{\text{TOF}}$ and $\sigma_Z$ are standard deviations (SDs), estimated from CRT or freely adjusted, $\Delta l$ and $\Delta z$ are the discretisations of TOF and $Z$-axis, respectively. Eqs.~(\ref{eq:cdfZcdfTOF}) reflect the detection probability in a certain TOF segment or at an axial distance $z$ from LOR. The smoothest integration over CDF profiles implies the following:
\begin{equation}
\sigma_{\text{TOF}}=\Delta l/(2\sqrt{2\log2}),\quad\sigma_Z=\Delta z/(2\sqrt{2\log2}),
\label{eq:SDsAndDeltas}
\end{equation}

\textit{b) High-pass filtering} is focused on the compensation of the smeared hit positions along $Z$ and the MLP along LOR -- a crucial problem for J-PET due to its design. It can be done for an arbitrary profile $h(x)$ by finding its inverse form $h^{-1}(x)$ using the Fourier representation $H(\nu_x)=\mathcal{F}h(x)$ as follows:
\begin{equation}
H^{-1}(\nu_x) = 1/H(\nu_x),\quad h^{-1}(x) = \mathcal{F}^{-1}H^{-1}(\nu_x),
\label{eq:Hinv}
\end{equation}

\noindent where $\nu_x$ is the frequency coordinate for $x$ that could represent $l$ or $z$. For a sufficiently large dataset, $h^{-1}(x)$, applied over a blurred coordinate, will return a delta function. Similarly to $W(\omega_s)$, a smoothing window (\ref{eq:RampApodised}) might be applied to suppress the noise, i.e.  $H^{-1}(\nu_x) \rightarrow H^{-1}(\nu_x)\cdot M(\nu_x)$. 

Fig.~\ref{Fig:RampAndGaussIndexed}, c-d depicts Gaussian $h(x)$ and $H(\nu_x)$ with SD $\sigma_x$, as well as their inverse forms $h^{-1}(x)$ and $H^{-1}(\nu_x)$. An example of the apodised $H^{-1}_\text{cut}(\nu_x)$ by the cut-off $\pm\nu_\text{cut}$ ($\nu_\text{Nq}$ denotes Nyquist frequency) is shown as dashed lines. 

Contrary to the low-pass filtering, the inverse functions $h^{-1}_\text{TOF}(l)$ and $h^{-1}_Z(z)$ must be rigorously calculated from CDF profiles with SDs that reflect the intrinsic resolution of the scanner: 
\begin{equation}
\label{eq:SDsCRT}
\begin{split}
\sigma_{\text{TOF}}	=c_0\cdot\text{CRT}/(4\sqrt{2\log2}), \\
\sigma_Z 			=c_\text{scin}\cdot\text{CRT}/(4\sqrt{2\log{2}}),
\end{split}
\end{equation}

\noindent where $c_0$ is the speed of light in the air, $c_\text{scin}=126\,\text{mm/ns}$ -- the effective propagation velocity of optical signals inside the J-PET scintillator strip \citep{Moskal2016}. 

High-pass filtering increases the minimal size of the allocated ROR volume. The dashed lines in Fig.~\ref{Fig:RampAndGaussIndexed}, c denote the $\pm3.5\sigma$ truncation range, suitable for a Gaussian CDF (red curve) -- see Refs. \citep{Efthimiou2019, Strzelecki2016}. For the inverse $h^{-1}(x)$, however, it ought to be enlarged significantly.

For convenience, we shall use both "event-based TOF FBP" and "TOF FBP" notation in this paper. The script for the algorithm is written in {\small\textsf{R}} using vectorised functions \citep{RCoreTeam2020}. The latest version allows multi-threading, but most of the tests were run in a single-thread mode on a regular node of CI{\'{S}} cluster\footnote[2]{CI{\'{S}} -- Centrum Informatyczne {\'{S}}wierk: \url{https://www.cis.gov.pl/}} (Intel Xeon E5-2680v2, 128~GB RAM).

\subsection{Simulation setup}
\subsubsection{Scanner geometries}
We ran simulations for the selected NEMA sources, phantoms, scanners and interactions of the 511-keV photons with the detectors using the Geant4 Application for Tomographic Emission (GATE) framework \citep{Gate2004, Gate2011}. In order to focus exclusively on the reconstruction methods, an ideal configuration was chosen: a single cylindrical layer of plastic BC-420 scintillator strips placed side-by-side. The datasets simulated for our work \cite{Kowalski2018} were used to study the spatial resolution and validate the earlier results, for a scanner of the radius $R=425.6$~mm, composed of 382 strips with the size 7~mm~$\times$ 20~mm~$\times$ 500~mm. A very similar geometry was defined for the other phantoms, to be consistent with Refs. \cite{Kopka2020, Raczynski2020}: 384 strips, each of 7~mm~$\times$ 19~mm~$\times$ 500~mm size, $R=427.8$~mm. The axial length of $50$~cm is a good middle-point that minimises the parallax effect (see Ref. \cite{Kopka2020}), yet allows generalisation over tomographs with large AFOV. 

Post-simulation smearing of hit positions and times of hits was performed to represent signal readouts with CRT in the $50-500$-ps range. The highest value is close to the results from the early experiments with 3-layer "big barrel" J-PET prototype \citep{Niedzwiecki2017, Pawlik-Niedzwiecka2017, Shopa2020}, the lowest -- reflects a theoretical minimum for BC-420 scintillators with WLS attached, previsioned for the future development of J-PET \citep{Moskal2016}. We shall refer to the latter as "Ideal + WLS". Selected datasets were post-smeared to simulate non-collinearity \citep{Moses2011}, with a rough approximation of positron transport for a number of radionuclides \citep{Moskal2019}, using a two-exponential model (see e.g. Ref. \cite{LeLoirec2007}). The details are given in \ref{A_SR_extended}.

\subsubsection{Phantoms}
\begin{figure}[!t]
\centering
\includegraphics[width=0.65\textwidth]{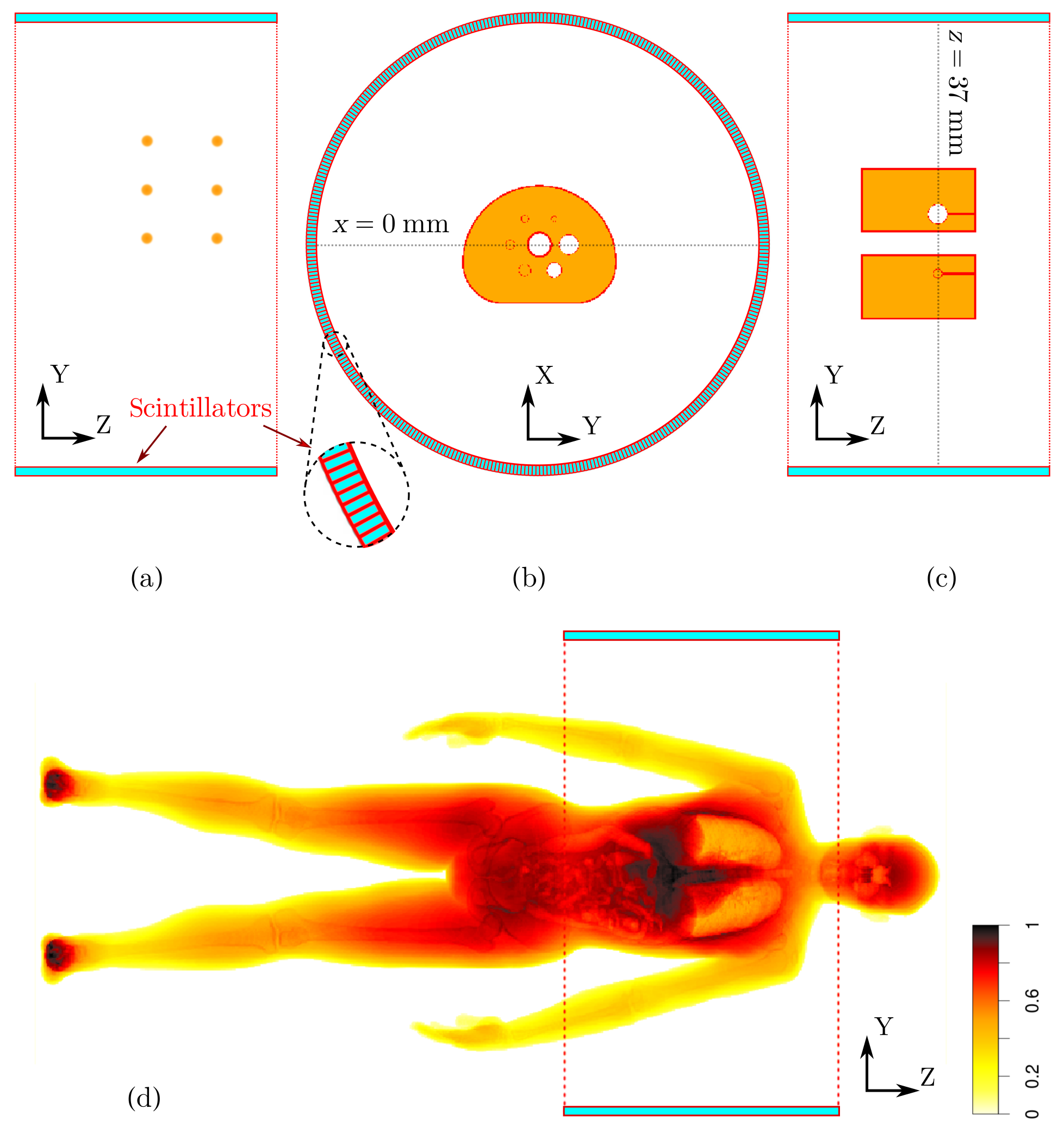}
\caption{Schematic depiction of the simulated objects inside an ideal J-PET scanner: six locations of a $1$-mm spherical source (the diameters are aggrandised) on the axial plane (a, $x=0$~mm), transverse (b, $z=37$~mm) and axial (c, $x=0$~mm) cross-sections of NEMA IEC phantom, coronal view through XCAT female phantom (d). Colours indicate water or plastic for NEMA IEC or a relative attenuation coefficient for XCAT. The magnified $XY$-cut of the scanner shows the composition of the strips.}
\label{Fig:Phantoms}
\end{figure}

For the spatial resolution analysis, a point-like $1$-mm spherical source was defined in GATE inside a $382$-strip ideal J-PET, with an activity of $370$~kBq \citep{Kowalski2018}. A separate simulation was run for the object put at six NEMA locations, varying its transverse ($y=10$~mm, $100$~mm and $200$~mm) and axial position ($z=0$~mm and $187.5$~mm), with fixed $x=0$~mm (Fig.~\ref{Fig:Phantoms}, a). A total of $150,000$ events were acquired per run.

The next simulated object was the NEMA-defined IEC phantom, used for the analysis of the image quality \citep{NEMA2012}. It was aligned axially as shown in Fig.~\ref{Fig:Phantoms}, b-c, with the interior length $220$~mm. Hot lesions, represented by the spheres (with capillaries) of diameters $10$~mm, $13$~mm, $17$~mm and $22$~mm, were filled with radioactive water, which activity was set four times higher than in the background volume. No lung insert was simulated along the central axis, replaced by a cold cylinder. The total injected activity amounted to $53$~MBq for \ch{^{18}F-FDG} dissolved in water. Two datasets were used for the analysis, comprising $\sim40$~mln and $\sim80$~mln coincidences recorded in a $384$-strip $50$-cm long J-PET scanner. It reflects the times of scan of $\sim10$~min or $\sim20$ min, respectively.

Finally, in order to test and evaluate the TOF FBP method for the setup most resembling diagnostic practices, we simulated one bed position of the eXtended CArdiac-Torso (XCAT) static phantom, using the software, developed by \cite{Segars2008, Segars2018}. Separate voxelised images for the activity and attenuation factors were generated from a female template. The phantom was aligned as shown in Fig.~\ref{Fig:Phantoms}, d: the centre of its volume was shifted $-385$~mm along $Z$ from the J-PET central point. The activity concentration was adjusted to match the weight-based $5.3$~MBq/kg level of the simulated NEMA IEC, having reached $350$~MBq for a $66$-kg body mass. We kept the default ratios set for various tissues in the configuration files, which resemble an early distribution of positron emissions during the first $10-20$~min after intravenous administration of \ch{^{18}F}-type radionuclide \citep{Watabe2020}. An additional $12$-mm hot spherical lesion was simulated in the right lung at $(x,y,z)=(-50$~mm, $60$~mm, $60$~mm), four times more active than the neighbouring tissue. The simulated $10$-minute scan produced $119.5$~mln coincidences in total.

\subsubsection{Data correction: sensitivity, scatter, random and attenuation factors}
To compensate for boundary axial effects in the scanners, sensitivity maps were simulated using a hybrid MC/analytical approach from Ref. \cite{Strzelecki2016}. An alternative correction method of re-projection \citep{Kinahan1989} is incorporated in FBP 3DRP, but it is slow and more beneficial for short AFOV. No restriction for maximal obliqueness $\theta$ was imposed.

The data were pre-selected by a $3$-ns time window, to obtain coincident interactions of photons with the detector (refer to Ref. \cite{Kowalski2018} for the discussion on selection criteria). Only true events were taken into account, without scatter and random correction (there are multiple ongoing studies dedicated to the latter, e.g. Ref. \cite{Sharma2020}). The resulting numbers of true coincidences were $\sim10$~mln and $\sim20$~mln for the two datasets of IEC simulation, and $16.5$~mln -- for XCAT.

Attenuation correction for IEC and XCAT was applied on the fly, as a recalculation of a radiological path for each emission by a modified Siddon algorithm \citep{Li2016}, using the predefined attenuation factors (shown in Fig.~\ref{Fig:Phantoms} in different colours). This method could not be applied to KDE MLP, so we employed a simplified solution for IEC (only): true events were resampled according to the statistics for the genuine annihilation points, assigned by GATE. Only the data from the lower half of the phantom ($z\leq0$~mm) were considered, avoiding the regions with spheres. To equalise the attenuated activity levels in the background volume, about one-third of the coincidences had to be dropped, hence a larger subset from the initial $80$-mln statistics was taken to obtain the required 10-mln true events. 

\subsection{Estimation of spatial resolution and image quality}
Spatial resolution, referred to as point spread function (PSF), is estimated from the reconstructed images of a point-like source as the full width at half-maximum measured along the principal axes ($X$, $Y$, $Z$), using 1D profiles built across the voxel with maximum intensity. FBP method is required by NEMA since it preserves the true intrinsic properties of a scanner more fairly than other algorithms.

According to NEMA, image quality measures of contrast recovery coefficient (CRC) and background variability (BV) describe qualitatively the simulated clinical imaging conditions using the IEC phantom \citep{NEMA2012}. Despite some tools for their estimation are available in STIR, a dedicated script was written to ensure consistency in comparing algorithms. 

We assigned the regions-of-interest (ROIs) -- circular areas drawn around the six spheres of the phantom on the transverse plane that intersects their centres (at $z=37$~mm, see Fig.~\ref{Fig:Phantoms}, b-c). For each of them, twelve additional circular ROIs of the shared diameters were concentrically allocated on the background, duplicated on four transverse slices at $\pm10$~mm and $\pm20$~mm from the main plane (60 ROIs in total) \citep{NEMA2012}.

Contrast recovery coefficients $\text{CRC}_{H,d}$ and $\text{CRC}_{C,d}$, for a hot or a cold sphere of a diameter $d$, respectively, are estimated as:
\begin{equation}
\text{CRC}_{H,d} = \frac{\frac{\mu_{H,d}}{\mu_{B,d}}-1}{\alpha-1},\quad 
\text{CRC}_{C,d} = 1-\frac{\mu_{C,d}}{\mu_{B,d}},
\label{eq:CRCs}
\end{equation}

\noindent where $\mu_{H,d}$, $\mu_{C,d}$ and $\mu_{B,d}$ denote the average intensity of the corresponding ROI around a hot or a cold sphere and on the background, respectively, $\alpha = 4$ is the activity ratio between the hot regions and the background.

Background variability for each sphere and signal-to-noise ratio (SNR) for hot spheres only, according to the equation from Ref. \cite{Westerwoudt2014}, are defined as:
\begin{equation}
\text{BV}_{d} = \frac{\sigma_{B,d}}{\mu_{B,d}},\quad \text{SNR}_{H,d} = \frac{\mu_{H,d}-\mu_{B,d}}{\sigma_{B,d}},
\label{eq:BVandSNReq}
\end{equation}

\noindent where SD of the background ROI counts for each sphere is:
\begin{equation}
\sigma_{B,d} = \sqrt{\sum_{i=1}^{N_{\text{ROI}}} (\mu_{B,d,i}-\mu_{B,d})^2/(N_{\text{ROI}}-1)},\, N_{\text{ROI}}=60.
\label{eq:SDROI}
\end{equation}

Eqs.~(\ref{eq:CRCs}) and (\ref{eq:BVandSNReq}) are relevant to a single scan, as in our case. NEMA requires these parameters to be averaged over multiple measurements, hence we used $30$ randomly drawn sets of ROIs with multiple seeds of pseudorandom generators.

\subsection{Kernel optimisation}
The true radiotracer distribution in the phantom, available by the simulation, defines a reference image $f(\textbf{\textit{x}}_i)$ -- ground truth. It was used for the optimisation of 3D kernel parameters of TOF FBP -- by reaching out for minimal MSE between the reconstructed outcome $\hat{f}(\textbf{\textit{x}}_i)$ and $f(\textbf{\textit{x}}_i)$: 
\begin{equation}
\text{MSE}\left(\hat{f}\right) =  \frac{1}{N}\sum_{i=1}^N \left[\hat{f}(\textbf{\textit{x}}_i) - f(\textbf{\textit{x}}_i) \right]^2,
\label{eq:MSE}
\end{equation}

\noindent where $N$ is the total number of voxels $\textbf{\textit{x}}_i$.

We also considered the normalised distribution of the error:
\begin{equation}
\label{eq:ErrFhatF}
\text{Err}(\textbf{\textit{x}}_i)=\frac{\hat{f}(\textbf{\textit{x}}_i)}{\max\hat{f}(\textbf{\textit{x}}_i)} - \frac{f(\textbf{\textit{x}}_i)}{\max f(\textbf{\textit{x}}_i)}.
\end{equation}

The properties of multi-dimensional space formed by kernel parameters are unknown, therefore the minimisation requires non-gradient optimisation methods. We chose the Nelder-Mead (simplex) algorithm, often applied to nonlinear tasks where derivatives cannot be calculated \citep{Nelder1965}. 

In addition to MSE, two alternative image quality metrics were selected: $\text{SNR}_{H,d}$ from Eq. (\ref{eq:BVandSNReq}) and the following macro-parameter of CRC and BV for each sphere of diameter $d$:
\begin{equation}
\label{eq:IQ_Q}
\text{Q}_d=\vert1-\text{CRC}_d\vert + \text{BV}_d.
\end{equation}

Finally, to prevent an over-smoothing during the minimisation, we used a convolutional median post-filter (MPF). It is often applied to eliminate Poisson noise with the advantage of edge preservation (see e.g. Ref. \cite{Fabijanska2011}). Even the simplest cube-shaped $3\times3\times3$ median mask of neighbouring voxels would drastically improve the balance between the resolution, image quality and MSE.

\section{Results}
\subsection{High-pass filtering}
In our previous works \citep{Shopa2020, Daria2020}, the 3D TOF FBP kernel was composed of a ramp filter $W(\omega_s)=\vert\omega_s\vert$ and low-pass Gaussian CDF functions (\ref{eq:cdfZcdfTOF}). To achieve a smoother integration across the axial direction, $\sigma_Z$ was defined by the voxel dimension $\Delta z$ according to Eq.~(\ref{eq:SDsAndDeltas}), while $\sigma_\text{TOF}$ -- by CRT using Eq.~(\ref{eq:SDsCRT}). Such setup, however, appeared to be far from optimal, imposing excessive axial smearing. It was confirmed by the estimated effective SDs from the bandwidth matrix $\hat{\mathbf{H}}_\text{PI}$, optimised for KDE MLP using Scott's rule of thumb \citep{Scott1992}. For the simulated $1$-mm source and $\text{CRT}=235$\,ps (SiPM readout), these SDs were ten times smaller than $\sigma_{\text{TOF}}\approx15$~mm, calculated from CRT. 

Therefore, we considered high-pass filtering: TOF- and $Z$-functions $h^{-1}_\text{TOF}(l)$ and $h^{-1}_Z(z)$, defined in inverse (\ref{eq:Hinv}) form. In this case, $\sigma_Z$ should be calculated by CRT, too, using Eq.~(\ref{eq:SDsCRT}). 

\subsubsection{Spatial resolution} \label{SRsubsection}
\begin{figure}[!t]
\centering
\includegraphics[width=0.67\textwidth]{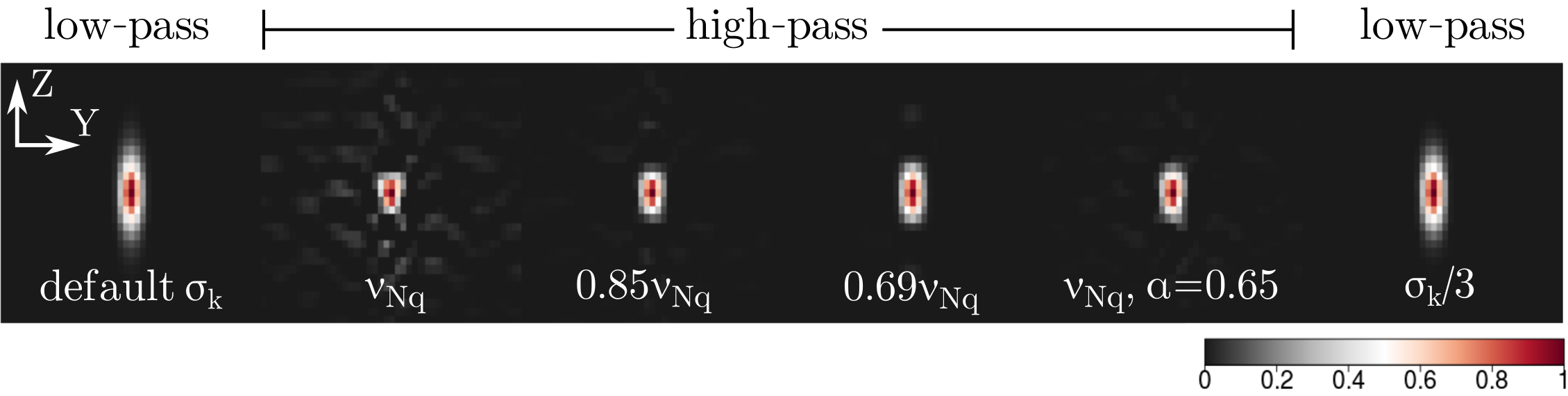}
\caption{Axial cross-sections of the reconstructed by TOF FBP 1-mm source in the ideal J-PET scanner with SiPM readout ($\text{CRT}=235$\,ps), depending on the definition of $h_\text{TOF}(l)$ and $h_Z(z)$ kernel components. Each image is zoomed to $80$\,mm\,$\times$\,$80$\,mm area around the source, $\sigma_k = \sigma_{\text{TOF},Z}$.}
\label{Fig:KernelsResolution}
\end{figure}

Fig.~\ref{Fig:KernelsResolution} shows the axial cross-sections of a $1$-mm spherical source in a 382-strip scanner with SiPM readout, reconstructed using various definitions of TOF- and $Z$-components of the kernel. Voxel size was defined by a virtual $96$-ring, $N$-strip scanner of the radius $R$: $\Delta x=\Delta y=\Delta s/2\approx 1.8$\,mm and $\Delta z\approx2.6$\,mm \cite{Kowalski2018}, where $\Delta s=\pi R/N$ \citep{PETBasicScience2005}. The source was located at ($x_\text{src}$, $y_\text{src}$, $z_\text{src}$) = ($0$\,mm, $100$\,mm, $0$\,mm). TOF FBP regularisation (\ref{eq:FBPfilterTOF}) was not applied as it is redundant for a tiny $1$-mm object.

High-pass filters $h^{-1}_\text{TOF}(l)$ and $h^{-1}_Z(z)$ produce higher axial resolution than the initial low-pass Gaussians (\ref{eq:cdfZcdfTOF}), but also enhance background noise. It could be eliminated by the adjustment of smoothing and apodisation parameters $\alpha$ and $\nu_\text{cut}$ (see Fig.~\ref{Fig:RampAndGaussIndexed}, d). The cut-off $\nu_\text{cut}$ in Fig.~\ref{Fig:KernelsResolution} is represented as a fraction of a Nyquist frequency $\nu_\text{Nq}=1/(2\sigma_k)$. An additional low-pass case is shown on the right: $\sigma_\text{TOF}$ and $\sigma_Z$ are decreased three times from the "initial" values, estimated from CRT and $\Delta z$, respectively. That implies $\sigma_{\text{TOF}}\approx5.0$~mm and $\sigma_Z\approx0.4$~mm for $\text{CRT}=235$\,ps and $\Delta z = 2.6$\,mm. 

\begin{table}[!t]
\renewcommand{\arraystretch}{1.15}
\caption{\label{Tab:SRkernels} PSF values, estimated from the reconstructed $1$-mm source, located at $(x_\text{src},y_\text{src},z_\text{src})$ = ($0$\,mm, $100$\,mm, $0$\,mm) inside the ideal J-PET scanner, depending on the algorithm and kernel choice. Readout -- SiPM.}
\centering
\resizebox{20pc}{!}{ 
\begin{tabular}{lccccc}
\hline 
\hline \\[-2.7ex]
Algorithm \&& \multicolumn{2}{l}{Kernel}  & \multicolumn{3}{c}{PSF (in mm):}\\
filter applied&\multicolumn{2}{l}{parameters} & $X$ & $Y$ & $Z$\\
\hline \\[-2.5ex]
\vspace{0.6mm}
			&$\alpha=$&$\nu_\text{cut}=$		&&&\\
   			& $1.0$&$\nu_\text{Nq}$	& 			$5.5$ & $5.2$ & $9.1$ \\
TOF FBP,		& $1.0$&$0.85\,\nu_\text{Nq}$	& 	$5.8$ & $5.3$ & $10.2$\\
high-pass	& $1.0$&$0.69\,\nu_\text{Nq}$	& 	$5.8$ & $5.4$ & $12.7$\\
			&$0.65$&$\nu_\text{Nq}$				&$5.8$ & $5.3$ & $11.2$\\
			&$0.575$&$\nu_\text{Nq}$			&$5.8$ & $5.4$ & $12.4$\\
\hline \\[-2.5ex]
\vspace{0.6mm}
TOF FBP   & \multicolumn{2}{c}{initial $\sigma_{\text{TOF},Z}$}	&$5.8$ & $5.4$ & $17.0$\\
low-pass (Gaussian)&\multicolumn{2}{c}{$\sigma_{\text{TOF},Z}/3$}& $5.9$ & $5.4$ & $15.8$\\
\hline \\[-2.5ex]
FBP 3DRP 	
& \multirow{2}{*}{--} & \multirow{2}{*}{--} & \multirow{2}{*}{$6.3$} & \multirow{2}{*}{$5.8$} & \multirow{2}{*}{$15.1$} \\
\citep{Kowalski2018} &&&&\\
KDE MLP			&\multicolumn{2}{c}{dscalar}& 				$7.1$ & $6.4$ & $14.8$\\
\hline\\[-5.5ex]
\end{tabular}
} 
\end{table}

\begin{figure*}[!t]
\centering
\includegraphics[width=35pc]{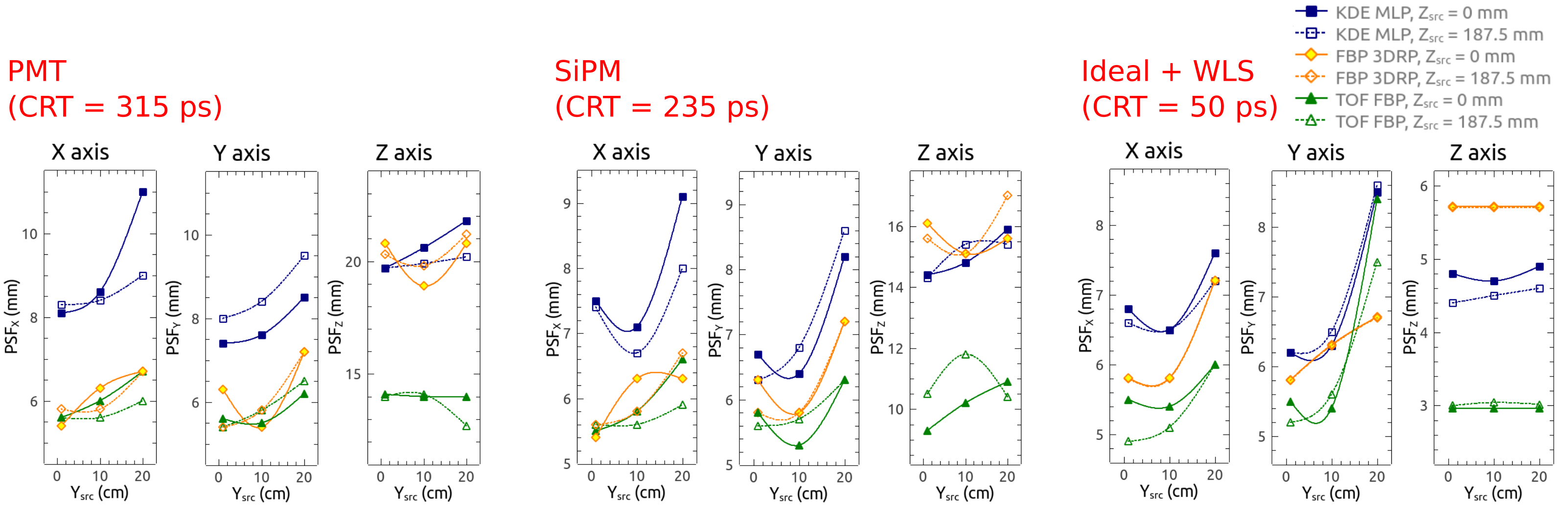}
\caption{PSF values for six locations of 1-mm point-like source in the ideal J-PET scanner, estimated for different algorithms and three readouts.}
\label{Fig:SR_Graphs}
\end{figure*}

Estimated PSF values for the selected kernel definitions are presented in Table~\ref{Tab:SRkernels}. As a reference point, the results obtained by KDE MLP and FBP 3DRP \citep{Kowalski2018} are added. The event-based TOF FBP produces slightly better $\text{PSF}_{X,Y}$, but worse axial resolution for the Gaussian-defined kernel components. The best $\text{PSF}_Z$ is obtained for the ideal ramp and high-pass filters $h^{-1}_\text{TOF}(l)$ and $h^{-1}_Z(z)$, without cut-offs or smoothing, i.e. $\alpha=1$, $\nu_\text{cut}=\nu_\text{Nq}$. However, the imposed noise and distortion are too high (Fig.~\ref{Fig:KernelsResolution}).

As a reasonable trade-off, we have fixed the cut-off at $\nu_\text{cut}=0.85\,\nu_\text{Nq}$ and conducted a detailed study for the source at six NEMA-defined locations (see Fig.~\ref{Fig:Phantoms}, a). The estimated PSF values are shown in Fig.~\ref{Fig:SR_Graphs} for three readouts, as in Ref. \cite{Kowalski2018}: PM tubes (PMT, $\text{CRT}=315$~ps), SiPM ($\text{CRT}=235$~ps) and ideal case with WLS ($\text{CRT}=50$~ps). TOF FBP and FBP 3DRP produce similar results on the transverse plane, outperforming KDE MLP (no FBP filter) even for the best timing. As for axial resolution, high-pass filtering decreases $\text{PSF}_Z$ by about $\times1.5$ times, compared to FBP 3DRP or KDE MLP. More details are revealed in \ref{A_SR_extended}.

\subsubsection{NEMA IEC phantom and image quality} 
As the main object of study, we chose a relatively small $10$-mln statistics of true coincidences (roughly a $10$-min scan), registered from the NEMA IEC phantom in a $384$-strip J-PET scanner. The cut-off $\nu_\text{cut}=0.85\,\nu_\text{Nq}$ used in the previous section is presumably too high, representing the axial resolution achievable for long scans. We thus changed $\nu_\text{cut}$ in a $0.6-0.9\,\nu_\text{Nq}$ range, considered a larger $20$~mln dataset, as well as the reduced smoothing $\alpha<1$. The $235$-ps readout (SiPM) was chosen, relevant to current J-PET prototypes. Unlike previously, a symmetric voxel $\Delta x = \Delta y = \Delta z = 1.8$ mm was used, to avoid distortion caused by post-filtering, and a modified FBP from Eq. (\ref{eq:FBPfilterTOF}). The regularisation $\tau$ was adjusted, with fixed $\omega_c=\omega_\text{Nq}$ and $\alpha=1$.

\begin{figure}[!t]
\centering
\includegraphics[width=0.65\textwidth]{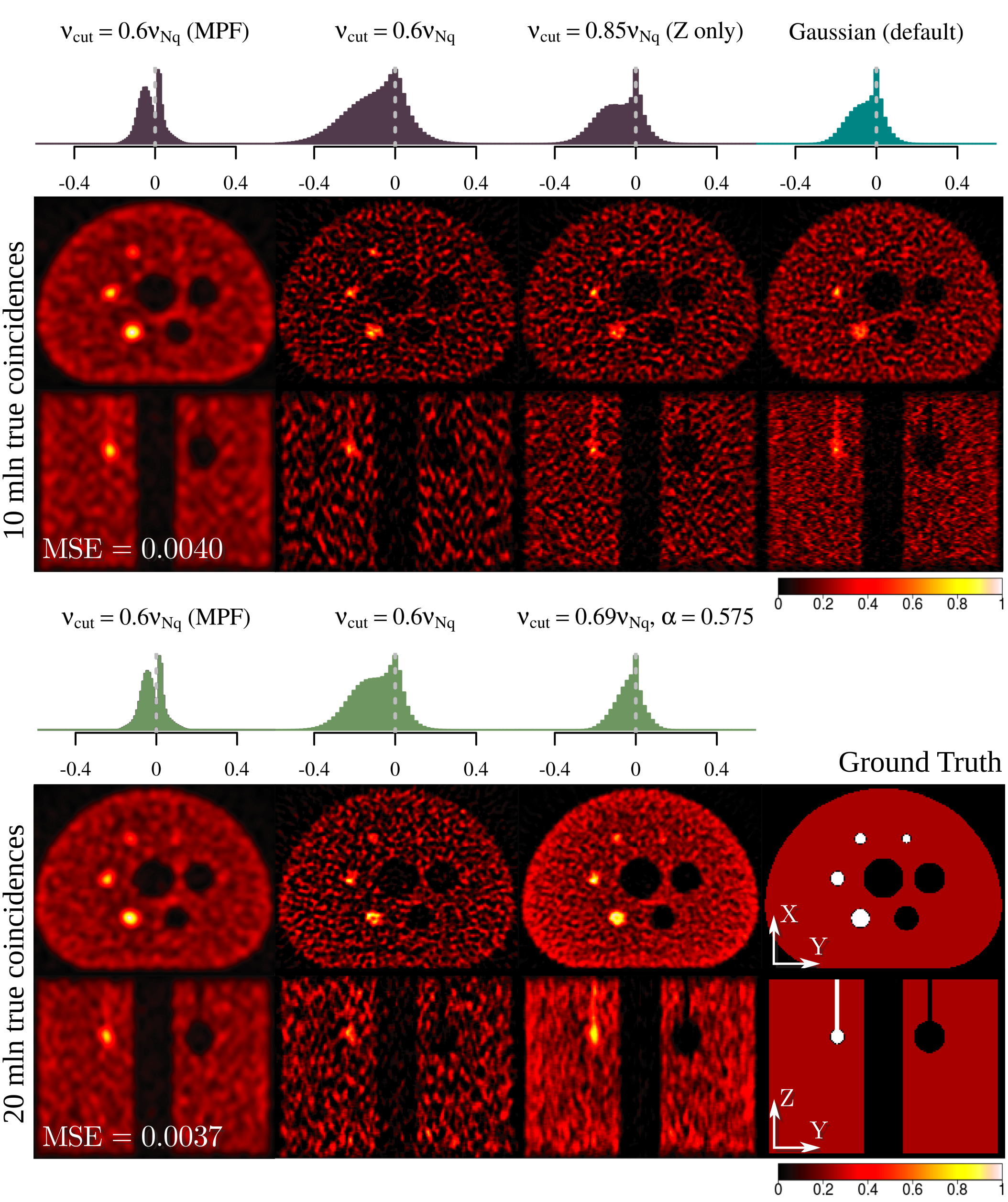}
\caption{Cross-sections (at $z=37$~mm and $x=0.0$~mm) of the NEMA IEC phantom, reconstructed from $10$-mln and $20$-mln datasets, depending on kernel definitions for $h_\text{TOF}(l)$ and $h_Z(z)$. Distributions of the error (\ref{eq:ErrFhatF}) are schematically shown on top of each image, readout -- SiPM ($\text{CRT}=235$\,ps).}
\label{Fig:IEC_dummies_SiPM}
\end{figure}

The results for $\tau=7.5$, explored to be close to the optimal, are presented in Fig.~\ref{Fig:IEC_dummies_SiPM}. The images on the left show the outcomes after MPF has been applied, using a spherical mask of $4$-voxel radius. High-pass filtering produces less random and even quasi-regular intensity levels on the background, compared to the case of low-pass $h_\text{TOF}(l)$ and $h_Z(z)$ (denoted as "default"). 

The histograms of the errors $\text{Err}(\textbf{\textit{x}}_i)$ from Eq. (\ref{eq:ErrFhatF}), shown on top, generally reflect a combination of Gaussian profiles. However, the distributions are more asymmetric than in the "default" case, mainly caused by the negative lobes -- voxels with intensities below zero. Such offset remains even after post-filtering and for a hybrid 3D kernel -- a low-pass $h_\text{TOF}(l)$, combined with $h_Z^{-1}(z)$ (shown for $\nu_\text{cut}=0.85\nu_\text{Nq}$). Improvement can also be achieved by imposing a smoothing window, as depicted for $\alpha=0.575$ and 20-mln dataset, but that would worsen axial resolution to the low-pass levels ($\text{PSF}_Z>15$~mm), which we observed in axial intensity profiles (not shown).

\begin{table}[!t]
\renewcommand{\arraystretch}{1.15}
\caption{\label{Tab:IQ_HP}Image quality parameters, estimated for the three spheres in the NEMA IEC phantom, reconstructed by TOF FBP with various 3D kernel setup. MPF was applied to each image. Readout -- SiPM, standard errors are given in parentheses, $\nu_\text{cut}=0.6\,\nu_\text{Nq}$ for high-pass filters.}
\centering
\resizebox{22.5pc}{!}{ 
\begin{tabular}{lcccccc}
\hline 
\hline \\[-2.7ex]
Dataset \&&\multicolumn{3}{c}{CRC, \%}	&\multicolumn{3}{c}{BV, \%}\\
filters    &13\,mm	&22\,mm	&28\,mm			&13\,mm	&22\,mm	&28\,mm	\\
\hline\\[-2.7ex]
$10$~mln&$43.40$ &$77.42$ &$66.55$ &$10.80$  &$7.57$  &$6.19$\\
\vspace{0.1cm}	
high-pass &\small{$(0.09)$}&\small{$(0.08)$}&\small{$(0.02)$}&\small{$(0.08)$}&\small{$(0.05)$}&\small{$(0.04)$}\\
$20$~mln&$37.24$ &$70.67$ &$71.03$ &$8.99$  &$5.20$  &$4.06$\\
\vspace{0.1cm}	
high-pass &\small{$(0.12)$}&\small{$(0.06)$}&\small{$(0.01)$}&\small{$(0.07)$}&\small{$(0.05)$}&\small{$(0.02)$}\\
$10$~mln&$22.86$ &$46.93$&$74.27$ &$6.89$ &$5.21$ &$4.51$\\
\vspace{0.1cm}	
low-pass&\small{$(0.04)$}&\small{$(0.04)$}&\small{$(0.01)$}&\small{$(0.05)$}&\small{$(0.04)$}&\small{$(0.03)$}\\
\hline\\[-5.5ex]
\end{tabular}
} 
\end{table}

Table~\ref{Tab:IQ_HP} reveals the estimated image quality parameters of NEMA IEC for three spheres and three reconstructions (all with MPF). The first two correspond to $h^{-1}_\text{TOF}(l)$ and $h^{-1}_Z(z)$ ($\nu_\text{cut}=0.6\,\nu_\text{Nq}$) for $10$-mln and $20$-mln subsets of true events, the third -- to the Gaussian CDFs and $10$-mln statistics. High-pass filtering produces higher CRCs, calculated for the hot $13$-mm and $22$-mm spheres, but not for the cold $28$-mm one. BVs are generally better for the reference (low-pass) image, indicating the lesser amount of distortion and noise. Image quality is susceptible to MPF mask, too: CRC for $13$-mm and $22$-mm spheres are lower for the $20$-mln set than for the $10$-mln one.

For the reference TOF FBP reconstruction, made using low-pass filters and MPF, the estimated ${\text{MSE}=3.0\cdot 10^{-3}}$ on a normalised $\left[0,1\right]$ intensity scale was lower than for the images, obtained from $10$-mln statistics using $h^{-1}_\text{TOF}(l)$ and $h^{-1}_Z(z)$ (e.g. $\text{MSE}=\,4.0\cdot10^{-3}$ for $\nu_\text{cut}=0.6\,\nu_\text{Nq}$ shown in Fig.~\ref{Fig:IEC_dummies_SiPM} on the top right). Lower $\text{MSE}=2.5\cdot10^{-3}$ was achieved only for a $20$-mln dataset with a substantial smoothing $\alpha=0.575$.

Worse MSE, spatial resolution, deteriorated by the apodisation and the smoothing, noise issues with asymmetric $\text{Err}(\textbf{\textit{x}}_i)$ distributions and image quality, rather not superior to the alternative kernel definition, do not favour the usage of high-pass filtering for online scans. Moreover, the inverse Gaussians $h^{-1}_\text{TOF}(l)$ and $h^{-1}_Z(z)$ require a larger ROR span (see Fig.~\ref{Fig:RampAndGaussIndexed}, c), which slows the computing performance. On the other hand, such filters would be beneficial for tomographs of higher sensitivity and longer diagnostic times. We shall focus on a low-pass representation of $h_\text{TOF}(l)$ and $h_Z(z)$ in the further analysis.

\subsection{Low-pass filtering: minimisation of MSE for NEMA IEC}
The TOF- and $Z$-components of a 3D TOF FBP kernel, defined as Gaussian CDFs (\ref{eq:cdfZcdfTOF}), produce a smoother background and more predictable noise than high-pass filters. In order to optimise the algorithm, a Nelder-Mead minimisation of MSE between the reconstructed NEMA IEC phantom and the ground truth was conducted, using a $10$-mln simulated subset of true events. Two J-PET readouts were considered: "ideal" with WLS ($50$-ps timing) and SiPM ($235$\,ps).

The role of axial function $h_Z(z)$ appeared to be substantially smaller than $h_\text{TOF}(l)$ or FBP filter $w(s)$, therefore we set $\sigma_Z = 0.76$\,mm, estimated by Eq. (\ref{eq:SDsAndDeltas}) from the voxel size $\Delta x = \Delta y = \Delta z = 1.8$~mm. Five free parameters have been chosen for the minimisation, with the following constraints: 
\begin{gather*}
0.5 \leq \alpha \leq 1.0,\quad 0 < \omega_c \leq 0.5,\\
0 < \tau \leq 15.0, \quad\sigma_\text{TOF} > 0,\quad \Delta l > 0.
\end{gather*}

As already revealed, we considered two alternative metrics -- $\text{SNR}_{H,d}$ and $\text{Q}_d$ (Eqs. (\ref{eq:BVandSNReq}) and (\ref{eq:IQ_Q}), respectively). The average values $\overline{\text{Q}}$ (for all spheres) and $\overline{\text{SNR}}$ (hot spheres only) were also tracked, as points of reference during the optimisation.

Technical details of the Nelder-Mead method are given in \ref{A_NMprocedure}. The post-filtering was essential for the algorithm to converge, because of reduced available data in the $10$-min scan. Without MPF, the contribution of noise was such that the outcomes were highly blurred even for $\text{CRT}=50$~ps, with typical values $\omega_c<0.25$ and $\sigma_\text{TOF}>50.0$\,mm for the ultimate iteration -- evidently inconsistent with the spatial resolution. 

\begin{figure*}[!t]
\centering
\includegraphics[width=0.82\textwidth]{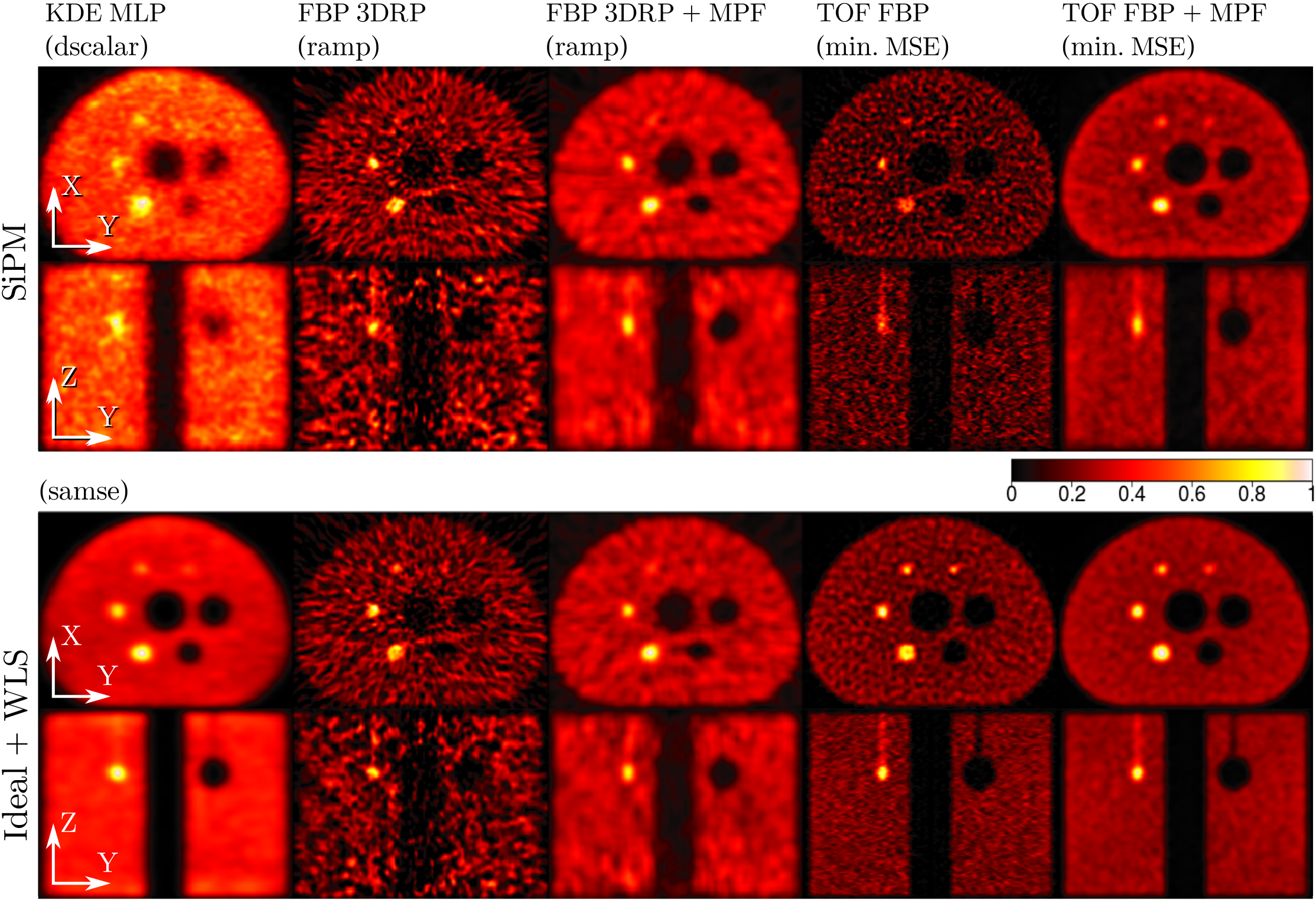}
\caption{Transverse ($z=37$~mm) and axial ($x=0$~mm) cross-sections of the reconstructed NEMA IEC phantom, for two readouts and various algorithms. The MPF masks were ball-shaped for TOF FBP (a $3$-voxel radius for ideal readout, $4$-voxel -- for SiPM) and cube-shaped ($7\times7\times7$) -- for FPB 3DRP.}
\label{Fig:IEC_MutiAlgorithms_SiPM_WLS50}
\end{figure*}

Cross-sections of the images, reconstructed by various algorithms and parametrisation, are depicted in Fig.~\ref{Fig:IEC_MutiAlgorithms_SiPM_WLS50}. The results for TOF FBP, taken from the ultimate Nelder-Mead iteration, look much better: even for $\text{CRT}=235$\,ps (SiPM), the 10-mm hot sphere is distinguishable, the intensity on the background is relatively smooth, the loss of axial resolution -- acceptable. 

The values for the optimised free parameters and the metrics, estimated after having applied MPF, are aggregated in Table~\ref{Tab:ParametrisationParamsIEC}. Three cases are shown: for the ultimate Nelder-Mead iteration ($\text{MSE}_\text{min}$), for the best $\overline{\text{Q}}_\text{min}$ and the "initial" settings: an ideal ramp filter, $\sigma_\text{TOF}$ calculated from CRT and $\Delta l$ defined by Eq (\ref{eq:SDsAndDeltas}). A notably higher $\tau$, but lesser apodisation and smoothing are required for SiPM, compared to the ideal/WLS readout. The regularisation is crucial for $\text{CRT}=50$\,ps, as the initial choice $\tau=15.0$ obscures $W(\omega_s)$ in Eq. (\ref{eq:FBPfilterTOF}), resulting in high MSE. Conversely, TOF profile $h_\text{TOF}(l)$ is apparently unrelated to CRT, narrower for SiPM and larger for the $50$-ps case. 

\begin{table}[!t]
\renewcommand{\arraystretch}{1.15}
\caption{\label{Tab:ParametrisationParamsIEC}Parameters of the 3D TOF FBP kernel and the estimated metrics for the reconstructed NEMA IEC phantom. MPF was applied in all cases.}
\centering
\resizebox{22.5pc}{!}{ 
\begin{tabular}{lcccccc}
\hline 
\hline \\[-2.7ex]
\multirow{2}{*}{Parameters} & \multicolumn{3}{c}{SiPM} & \multicolumn{3}{c}{Ideal\,+\,WLS} \\
& Ini & $\text{MSE}_\text{min}$ & $\overline{\text{Q}}_\text{min}$ & Ini & $\text{MSE}_\text{min}$ &$\overline{\text{Q}}_\text{min}$\\
\hline\\[-2.7ex]
$\alpha$ (a.u.) 			&$1.00$&$0.95$&$0.89$&$1.00$&$0.85$&$0.95$\\
$\omega_c$ (cycles)			&$0.50$&$0.44$&$0.36$&$0.50$&$0.37$&$0.45$\\
$\tau$ (cycles$^{-1}$) 		&$15.0$&$8.5$&$15.0$&$15.0$&$3.6$&$12.5$\\
$\sigma_\text{TOF}$ (mm) 	&$15.0$&$8.8$&$8.8$&$3.2$&$5.9$&$5.0$\\
$\Delta l$ (mm) 			&$17.6$&$12.2$&$10.3$&$3.7$&$10.9$&$15.0$\\
\hline\\[-2.7ex]
MSE$\times10^3$ (a.u.) 			&$3.5$&$2.4$&$8.4$&$16.7$&$1.4$&$11.3$\\
$\overline{\text{Q}}$ (a.u.) 	&$0.56$&$0.55$&$0.45$&$0.35$&$0.44$&$0.32$\\
\vspace{0.1cm}
$\overline{\text{SNR}}$ (a.u.) &$14.6$&$14.7$&$12.1$&$34.4$&$28.8$&$35.2$\\
\hline\\[-5.5ex]
\end{tabular}
} 
\end{table}

The reference results were inferior to TOF FBP. For FBP 3DRP, MSE ranged from $4.8\cdot10^{-3}$ (ideal timing) to $8.3\cdot10^{-3}$ (SiPM), with $\overline{\text{Q}}\sim0.7$ and $\overline{\text{SNR}}\sim9.0$ regardless of the readout. KDE MLP, though relatively good for $\text{CRT}=50$\,ps, is still worse than TOF FBP ($\text{MSE}\sim2.5\cdot10^{-3}$, $\overline{\text{Q}}\sim0.45$ and $\overline{\text{SNR}}\sim30.0$). As for the SiPM case, it is evidently too blurred, as seen from both Fig.~\ref{Fig:IEC_MutiAlgorithms_SiPM_WLS50} and the estimated $\text{MSE}>25.0\cdot10^{-3}$.

\begin{table}[!t]
\renewcommand{\arraystretch}{1.15}
\caption{\label{Tab:IQTOFBP}Estimated image quality parameters for the three spheres in the NEMA IEC phantom, reconstructed by the event-based TOF FBP + MPF. Standard errors for 30 various choices of ROIs are given in parentheses.}
\centering
\resizebox{22.5pc}{!}{ 
\begin{tabular}{lcccccc}
\hline 
\hline \\[-2.7ex]
Metric&\multicolumn{3}{c}{CRC, \%}	&\multicolumn{3}{c}{BV, \%}\\
chosen&13\,mm	&22\,mm	&28\,mm			&13\,mm	&22\,mm	&28\,mm	\\
\hline\\[-2.7ex]
\multicolumn{7}{c}{SiPM}\\
\multirow{2}{*}{$\text{MSE}_\text{min}$}
&$24.47$ &$52.14$ &$87.80$ &$7.88$  &$5.83$  &$5.03$\\
&\small{$(0.06)$}&\small{$(0.05)$}&\small{$(0.01)$}&\small{$(0.06)$}&\small{$(0.06)$}&\small{$(0.04)$}\\
\multirow{2}{*}{$\overline{\text{Q}}_\text{min}$}
&$49.52$ &$111.58$&$99.62$ &$18.32$ &$12.53$ &$10.24$\\
\vspace{0.1cm}	
&\small{$(0.29)$}&\small{$(0.41)$}&\small{$(0.01)$}&\small{$(0.26)$}&\small{$(0.25)$}&\small{$(0.20)$}\\
\multicolumn{7}{c}{Ideal + WLS}\\
\multirow{2}{*}{$\text{MSE}_\text{min}$}
&$37.84$&$62.79$&$89.34$&$5.11$&$4.41$&$4.08$\\
&\small{$(0.04)$}&\small{$(0.02)$}&\small{$(0.01)$}&\small{$(0.04)$}&\small{$(0.02)$}&\small{$(0.01)$}\\
\multirow{2}{*}{$\overline{\text{Q}}_\text{min}$}
&$67.68$&$137.88$&$99.86$&$9.62$&$7.26$&$6.31$\\
&\small{$(0.23)$}&\small{$(0.30)$}&\small{$(0.01)$}&\small{$(0.14)$}&\small{$(0.13)$}&\small{$(0.09)$}\\
\hline\\[-5.5ex]
\end{tabular}
} 
\end{table}

Finally, the NEMA-defined image quality parameters were analysed. Table~\ref{Tab:IQTOFBP} reveals the estimated CRC and BV for three spheres of the phantom ($13$~mm, $22$~mm and $28$~mm), reconstructed by the event-based TOF FBP with MPF applied afterwards. The 3D kernel was defined accordingly to $\text{MSE}_\text{min}$ and $\overline{\text{Q}}_\text{min}$ in Table~\ref{Tab:ParametrisationParamsIEC}. The $\text{CRC}\left(\text{BV}\right)$ combinations for $\text{MSE}_\text{min}$ are better than for the "low-pass" reconstruction in Table~\ref{Tab:IQ_HP} and for the reference images, obtained by FBP 3DRP and KDE MLP (see \ref{A_IQ_curves}). For $\overline{\text{Q}}_\text{min}$, CRC exceeds $100\%$ for the $22$-mm sphere, while BV values are much higher than for $\text{MSE}_\text{min}$.

\subsection{A fast scan of the XCAT phantom}
The most complex case, but also decisive for diagnostics, was to reconstruct the torso part of a motionless female XCAT phantom (Fig.~\ref{Fig:Phantoms}, d) and thus validate the event-based TOF FBP. A simulated $10$-min scan produced $16.5$~mln true events. Similarly to NEMA IEC, we used symmetric voxels $\Delta x = \Delta y = \Delta z = 1.8$~mm and considered two readouts -- idealistic\,+ WLS and SiPM. The volume chosen for the analysis was limited to a range $x = \pm150$~mm, $y = \pm300$~mm, $z = \pm200$~mm.

A Nelder-Mead optimisation was run in the same way as for NEMA IEC: FBP filter (\ref{eq:FBPfilterTOF}) with TOF regularisation, low-pass Gaussian (\ref{eq:cdfZcdfTOF}) used for TOF- and $Z$-components, free parameters $\alpha$, $\omega_c$, $\tau$, $\sigma_\text{TOF}$ and $\Delta l$ (with constraints), a fixed $\sigma_Z = 0.76$\,mm (see \ref{A_NMprocedure}). For consistency, the initial $4$-vertex simplex included the set $(\alpha,\omega_c,\tau,\sigma_\text{TOF},\Delta l)^\text{T}$ taken from Table~\ref{Tab:ParametrisationParamsIEC}, optimised for the IEC phantom. The post-filtering was applied three times per image, using a ball-shaped MPF mask of the decreasing radii: $2$~voxels $\rightarrow 1$~voxel $\rightarrow 1$~voxel. 

\begin{table}[!t]
\renewcommand{\arraystretch}{1.15}
\caption{\label{Tab:ParametrisationParamsXCAT}Kernel parameters of the event-based TOF FBP used for the reconstruction of the XCAT phantom. Estimated MSE and effective PSF values are shown below. MPF was applied in all cases.}
\centering
\resizebox{25pc}{!}{ 
\begin{tabular}{lcccc}
\hline 
\hline \\[-2.7ex]
\vspace{0.1cm}
\multirow{2}{*}{Parameters} & \multicolumn{2}{c}{SiPM} & \multicolumn{2}{c}{Ideal\,+\,WLS} \\
\vspace{0.1cm}
& $\text{MSE}_\text{min}^{(XCAT)}$ & $\text{MSE}_\text{min}^{(IEC)}$ & $\text{MSE}_\text{min}^{(XCAT)}$ & $\text{MSE}_\text{min}^{(IEC)}$\\
\hline\\[-2.7ex]
$\alpha$ (a.u.) 			&$0.63$&$0.95$&$0.90$&$0.85$\\
$\omega_c$ (cycles)			&$0.27$&$0.44$&$0.36$&$0.37$\\
$\tau$ (cycles$^{-1}$) 		&$9.0$&$8.5$&$3.1$&$3.6$\\
$\sigma_\text{TOF}$ (mm) 	&$20.3$&$8.8$&$5.1$&$5.9$\\
$\Delta l$ (mm) 			&$22.6$&$12.2$&$10.0$&$10.9$\\
\hline\\[-2.7ex]
\vspace{0.05cm}
MSE$\times10^3$ (a.u.) 			&$5.0$&$5.5$&$3.8$&$3.9$\\
$\text{PSF}_X$ (mm) 			&$11.8-12.9$&	$6.5-7.4$&					$7.5-8.0$&	$7.6-8.2$\\
$\text{PSF}_Y$ (mm) 			&$11.3-11.9$&	$6.3-7.4$&	$7.4-8.2$&	$7.5-8.3$\\
\vspace{0.1cm}
$\text{PSF}_Z$ (mm) 			&$15.1-17.5$&	$15.0-17.2$&					$4.4-5.6$&			$4.5-5.7$\\
\hline\\[-5.5ex]
\end{tabular}
} 
\end{table}

Final results from the Nelder-Mead optimisation are given in Table~\ref{Tab:ParametrisationParamsXCAT}, compared with the best set of $\alpha$, $\omega_c$, $\tau$, $\sigma_\text{TOF}$ and $\Delta l$, taken from Table~\ref{Tab:ParametrisationParamsIEC}. Two sets are denoted according to the metrics used -- $\text{MSE}_\text{min}^{(XCAT)}$ and $\text{MSE}_\text{min}^{(IEC)}$, respectively. The difference is minor for the ideal readout, but rather significant for SiPM. Gaussian CDF profiles $h_\text{TOF}(l)$ with larger SDs are required, as well as more smoothing and apodisation -- reduced $\alpha$ and $\omega_c$ -- for FBP filter. On the contrary, if CRT is low, TOF FBP is less sensitive to kernel components (except $\tau$), allowing to preserve spatial resolution and image quality.

We have also estimated the effective PSF values for the considered 3D TOF FBP kernels, optimised for two phantoms, using the simulated $1$-mm source. The ranges between the results, calculated for six NEMA locations, are given at the bottom of Table~\ref{Tab:ParametrisationParamsXCAT}. The transverse $\text{PSF}_{X,Y}$ is more sensitive to FBP filter $w(s)$, compared to the impact of $h_\text{TOF}(l)$: even for $\sigma_\text{TOF}>20$~mm (SiPM), the calculated $\text{PSF}_Z$ is consistent with the "default" value ${\sim17}$~mm from Table~\ref{Tab:SRkernels}. The $50$-ps timing preserve a relatively good axial resolution of ${\sim5}$~mm, therefore we did not advocate for the practical usage of high-pass TOF- and $Z$-filters for big phantoms in this case.

\begin{figure*}[!t]
\centering
\includegraphics[width=35pc]{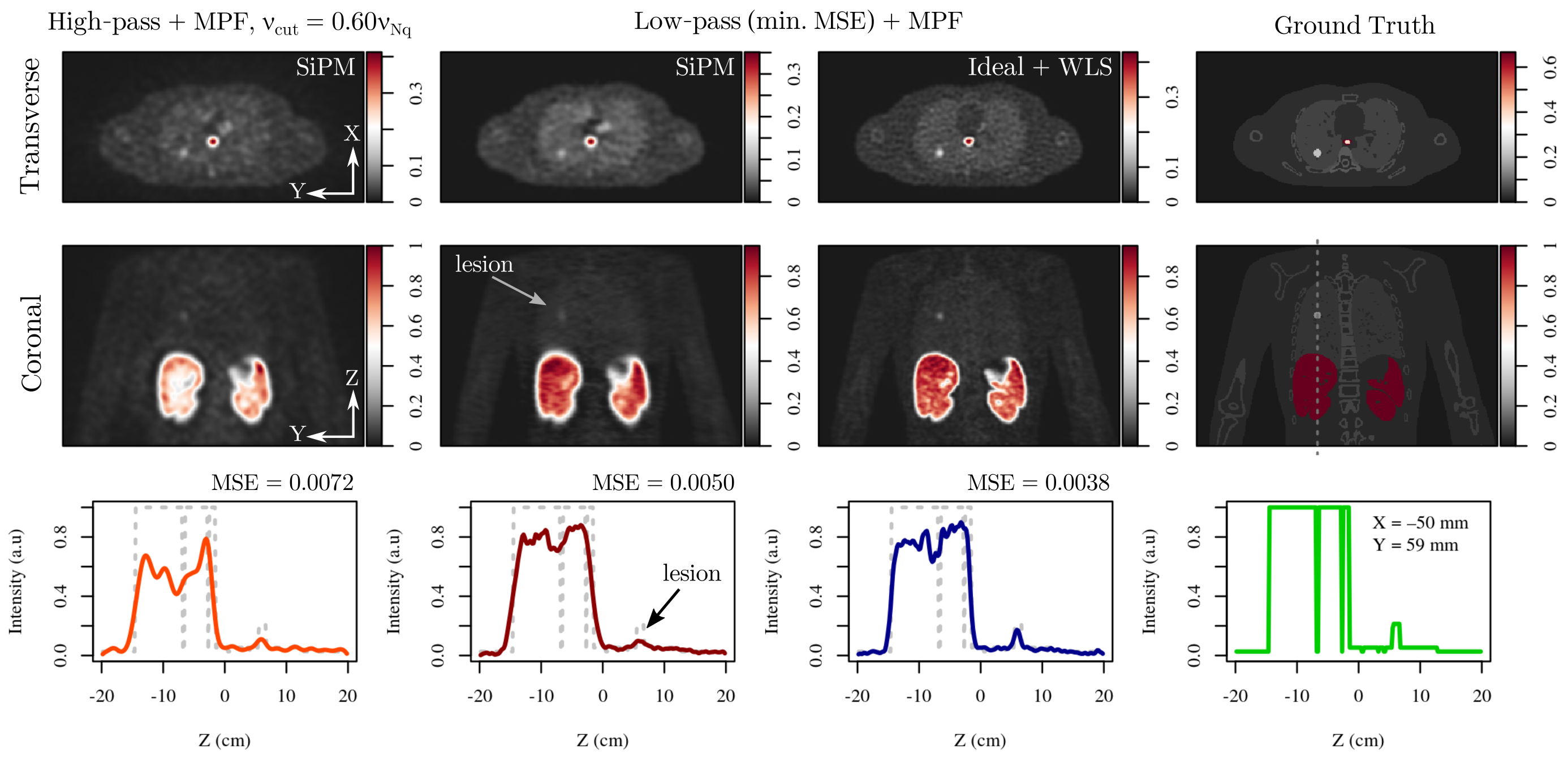}
\caption{Transverse (top row, $z=61$~mm) and coronal (middle row, $x=-50$~mm, $z\in\pm200$~mm) cross-sections of the XCAT phantom (torso part), reconstructed by TOF FBP with various configurations of the 3D kernel. Axial intensity profiles along the line indicated in the coronal cross-section of the ground truth are shown on the bottom row. The reference curve is replicated in each graph as a grey dashed line.}
\label{Fig:XCAT_reconstructions}
\end{figure*}

Selected cross-sections of the reconstructed XCAT phantom across a $12$-mm simulated lesion inside the right lung are presented in Fig.~\ref{Fig:XCAT_reconstructions}, along with the reference (ground truth). MPF has been applied to all images as described above. Two middle columns represent the results from the final iteration with the lowest MSE. On the left, the alternate reconstruction for SiPM readout is shown -- using high-pass $h^{-1}_\text{TOF}(l)$ and $h^{-1}_Z(z)$ filters with the cut-off $\nu_\text{cut}=0.6\nu_\text{Nq}$ and FBP kernel, adjusted according to $\text{MSE}_\text{min}^{(XCAT)}$ in Table~\ref{Tab:ParametrisationParamsXCAT}. Unfortunately, this configuration produced even more distortion than for the simulated IEC phantom, with blurred edges and the worst MSE.

The lesion is visible both in cross-sections and axial intensity profiles, shown on the bottom row. The reason for the difference between the optimised kernel parameters in Table~\ref{Tab:ParametrisationParamsXCAT} for SiPM is evident: most of the tissues in XCAT phantom had relatively small activity ($\sim 1.3 \text{ kBq/cm}^3$), more than $4$ times lower than in the background volume of NEMA IEC, despite the averages were almost the same. For early stages after the injection of a radioisotope, as implemented in the simulation, the emissions are disproportionally distributed, the majority of which occurred in distinct "hot" ($48 \text{ kBq/cm}^3$) regions -- liver, kidneys, cardiac muscle etc. Nevertheless, we still managed to observe the anomaly even for the worse $\text{CRT}=235$\,ps.

\section{Discussion}
The novelty of the proposed TOF FBP is that each emission is processed separately by a hybrid 3D filter that includes $h_Z(z)$ -- the uncertainty of hits inside scintillator strips. It is possible to compensate for the loss of axial resolution in J-PET scanners or to achieve a trade-off between the bias and the variance, similarly to the bandwidth selection in KDE. The method is event-centric and focused on the practical side of online imaging. The script optimisation, the assessment of target performance and benchmark analysis remain the issues for future works.

\subsection{Relation to other methods}
There is a similar image domain analytic-DIRECT algorithm \citep{Matej2009}. It calculates a unique 3D filtering kernel for each bin, which, if convoluted with unfiltered back-projection, results in a 3D Dirac delta function. However, such a model cannot operate on-the-fly: all the measured data is required to be partitioned beforehand into view bins. And despite the analytic-DIRECT reconstruction allows to estimate MLP more accurately than TOF FBP from a mathematical viewpoint, it is impractical for total-body J-PET scanners with elongated detector strips, multi-layer composition and DOI information, potentially accessible with the help of WLS.

An important benefit of the even-based TOF FBP is that it can be employed for an arbitrary geometry (large AFOV, multiple layers) and is capable to account for DOI. Alternative methods, based on the estimation of the system response matrix, e.g. recent on-the-fly algorithm, proposed by \cite{Lopez-Montes2020}, use models that ignore or oversimplify these factors. 

If we consider realistic CRT of J-PET tomographs \citep{Kowalski2018, Moskal2016}, the iterative algorithms outperform the methods based on closed-form formulas, such as TOF FBP, so the main motivation to employ the latter is real-time imaging. The reconstructions could also serve as an initial guess for iterative or other methods, e.g. a data-driven FastPET that utilises deep neural networks \citep{Whiteley2021}. 

\subsection{TOF FBP optimisation in clinical applications}
As seen from the results of our study, the best set of the 3D TOF FBP kernel parameters depends on the phantom's shape, activity, the time elapsed after the administration of a radionuclide, duration of the scan etc. This means that a custom configuration ought to be trained on known objects such as XCAT, either simulated or purposely built, according to the chosen diagnostic scenario. In real medical practice, however, it could be difficult to match the setup, obtained for a generic phantom, to a particular patient undergoing a PET scan. Nevertheless, kernel parameters can be refined afterwards by the Nelder-Mead method, where an unknown ground truth is replaced by a high-quality reconstruction made using e.g. iterative MLEM or total variation regularisation \citep{Raczynski2020}. In the long run, this would create a database with diverse TOF FPB setups.

On the other hand, the role of the object of scan decreases with the improvement of CRT, which could make the choice of parameters simpler. As seen from Table~\ref{Tab:ParametrisationParamsXCAT} for $\text{CRT}=50$~ps, the optimised sets of $(\alpha,\omega_c,\sigma_\text{TOF},\Delta l)^\text{T}$ are similar for two phantoms. Current J-PET prototypes are yet to achieve the timing ${<200}$~ps, but the envisioned temporal resolution may allow to optimise TOF FBP even using a simulated NEMA IEC.

\subsection{The role of CRT and the type of filters}
The intrinsic properties of J-PET tomographs are highly sensitive to the temporal characteristics of the scintillators and the readout, as seen from the analysis of spatial resolution given in subsection \ref{SRsubsection}. The explored usage of high-pass filters $h^{-1}_\text{TOF}(l)$ and $h^{-1}_Z(z)$, aimed at the reduction of axial smearing, raises the issues related to noise, distortion, performance and a minimal number of detected emissions. 

First of all, an enlarged ROR is required with a much higher number of voxels to update per event, if compared to low-pass filters. We assessed the minimal increase factor as $5-6$: the longest axis of the ellipsoid (along LOR -- see Fig.~\ref{Fig:LBL_scheme}) will be incremented from $3.5\sigma_\text{TOF}$ to about $13\sigma_\text{TOF}$. For $235$-ps timing, this corresponds to the increase from $\sim50$~mm up to $320$~mm, respectively, or $9$ times slower performance speed.

The minimal size of ROR also depends on FBP parameters $\alpha$, $\omega_c$ and filter discretisation $\Delta s$, but of a lesser scale. For the default $\alpha=1$ and $\omega_c$ at Nyquist frequency, the truncation region in the direction, perpendicular to LOR on $XY$-plane (see Fig.~\ref{Fig:LBL_scheme}), will be about $\pm9\Delta s$ ($\pm32$~mm), enlarged to $\pm19\Delta s$ ($\pm68$~mm) if apodised and/or smoothed.

Quasi-regular noise is another consequence of using the functions $h_\text{TOF}^{-1}(l)$ and $h_Z^{-1}(z)$. Even for twice as large $20$-mln statistics of true emissions, detected from the NEMA IEC phantom, one has to utilise smoothing and/or apodisation windows (see Fig.~\ref{Fig:IEC_dummies_SiPM}). It nevertheless might not reduce noise and distortion to acceptable levels, producing images worse than for low-pass filters and with asymmetric distributions of $\text{Err}(\textbf{\textit{x}}_i)$, as explored for XCAT (Fig.~\ref{Fig:XCAT_reconstructions}).

\subsection{The choice of TOF regularisation $\tau$}
\begin{figure}[!t]
\centering
\includegraphics[width=0.68\textwidth]{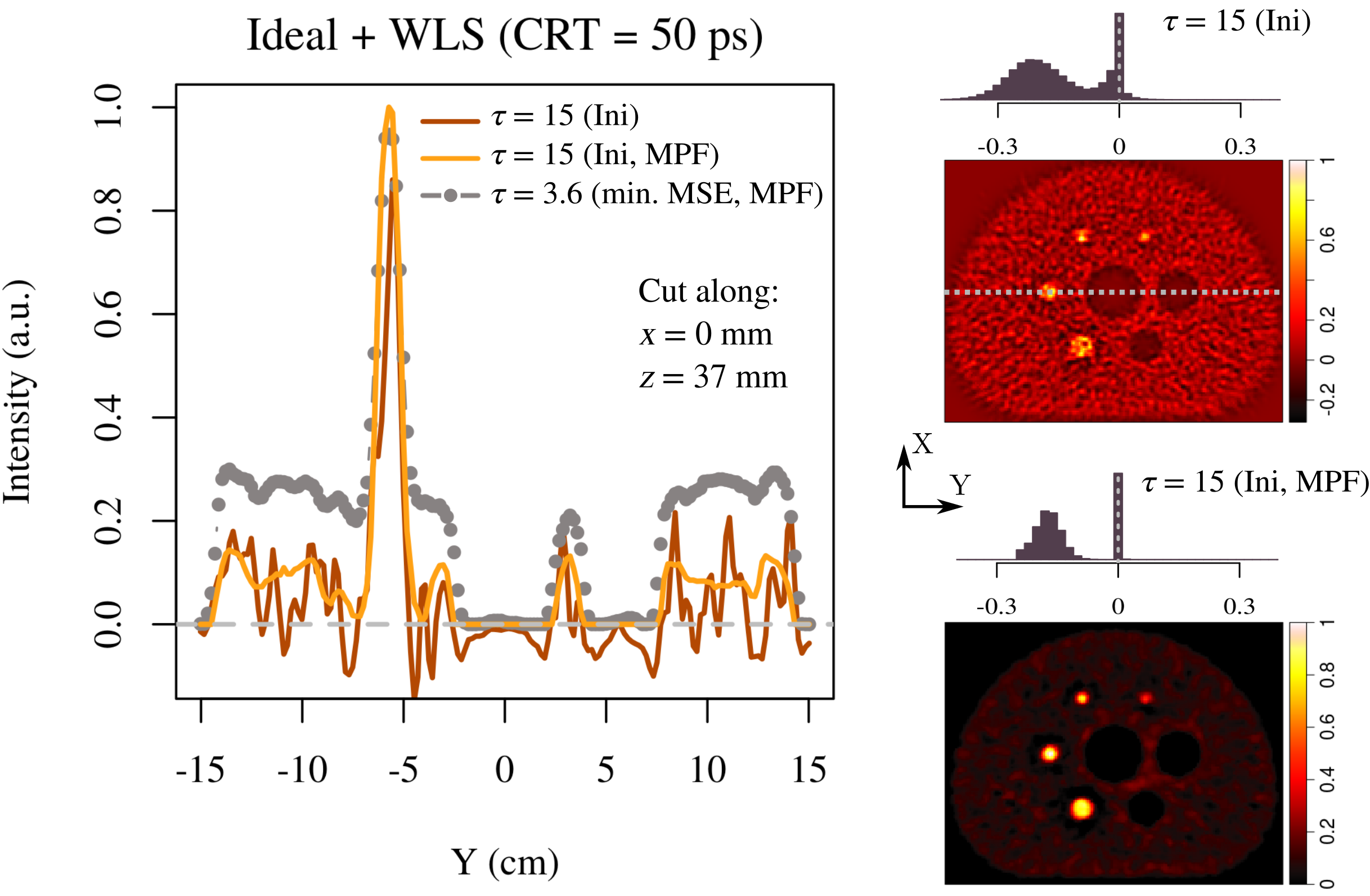}
\caption{Intensity profiles built along $Y$ (left) and transverse cross-sections (right, $z=37$~mm) of the reconstructed NEMA IEC phantom, depending on kernel configuration. Distributions of $\text{Err}(\textbf{\textit{x}}_i)$ are shown on top of each image on the right, dotted line denotes $x=0$~mm. Readout -- ideal + WLS.}
\label{Fig:YcsectionsMulti}
\end{figure}

Lowering of the FBP regularisation factor $\tau$ is essential for $\text{CRT}=50$\,ps, which is demonstrated in Fig.~\ref{Fig:YcsectionsMulti}. For a ramp-like filter ($\tau=15.0$, referenced as "Ini" in Table~\ref{Tab:ParametrisationParamsIEC}), the intensity inside the volume of the IEC phantom will be suppressed deeply below its expected $0.25$-level (on a normalised scale),  with many voxels at negative levels. The consequence of these voxels being coerced to zero afterwards, as done after FBP, is seen in Fig.~\ref{Fig:YcsectionsMulti}, right. The transverse cross-sections are corrupted, leading to asymmetric distributions of $\text{Err}(\textbf{\textit{x}}_i)$, in particular the Gaussian component of the background intensity. Note the difference in intensity colour palettes. Conversely, the optimal $\tau=3.6$ obtained for minimal MSE (see Table~\ref{Tab:ParametrisationParamsIEC}), preserves the proper level of activity inside the volume of the phantom.

Higher $\tau=8.5$, used for SiPM, means that the FBP filter contributes more to boundary effects at the edges of the NEMA IEC, especially after MPF is applied (Fig.~\ref{Fig:IEC_MutiAlgorithms_SiPM_WLS50}, top). High sensitivity, envisioned in the future total-body J-PET prototypes, would rectify the effect by acquiring more data during short scans.

Image quality is less sensitive to $\tau$ and boundary effects than MSE, due to the rigorous NEMA requirements to where ROIs can be drawn \citep{NEMA2012}. Similarly, circular intensity profiles of NEMA IEC are less representative than those built along axes, as in Fig.~\ref{Fig:YcsectionsMulti}, left. As a consequence, CRC, BV and SNR do not account for the irregularities at the edges of the phantom. That explains why the minimal $\overline{\text{Q}}$ for both readouts was obtained for $\tau>12.0$. Such a locality issue creates major obstacles in using these parameters as minimisation criteria.

\section{Conclusions}
We performed a comprehensive study to find the optimal configuration for the event-based TOF-FBP reconstruction algorithm, based on the most likely position (MLP) and an asymmetric 3D kernel approach. The benefit of direct operation in the image domain lays the foundation for real-time reconstruction using the FPGA SoC platform developed for J-PET scanners. We have shown that the optimisation of the TOF FBP setup strongly depends on the duration of a scan and the metric defined for the minimisation. The role of the object being scanned, while being significant at temporal resolution $>200$~ps, is rather low at $50$~ps, allowing to train and configure the PET setup using simple phantoms, such as NEMA IEC.

High-pass filters, chosen for the non-transverse kernel components of TOF FBP, although improving axial resolution, are not suitable for short-time clinical scans due to excessive noise and distortion imposed, indicated by the analysis of the reconstructed torso part of a simulated XCAT phantom. On the contrary, low-pass filters produce substantially lower noise and require a much smaller number of voxels to update per event, ultimately boosting the performance. The negative impact of axial smearing, caused by them, could be compensated by a small CRT and WLS in the prospective J-PET tomographs.

The best parameters of a hybrid 3D filter, found via the Nelder-Mead method for NEMA IEC and XCAT phantoms, are uncorrelated with intrinsic scanner properties, such as CRT or PSF, and resulted in reasonably good reconstructions, superior to non-TOF FBP 3DRP from the STIR framework. We demonstrated the crucial role of post-filtering during the iteration process for the datasets, registered during $10$-min scans. At the same time, we showed that CRC, BV and SNR cannot be utilised as reliable optimisation criteria because of their locality.  

FBP filtering may no longer be needed for CRT below 50~ps and can be replaced by KDE MLP with symmetric positive kernels. This would narrow the optimisation down to a rigorous problem of bandwidth selection, which could be solved analytically. A proper implementation for KDE, however, would require its event-based redefinition and correction factors embedded within, which stand as challenges for future works.

\section*{Declaration of Competing Interest}
The authors declare that they have no known competing financial interests or personal relationships that could have appeared to influence the work reported in this paper.

\section*{Acknowledgements}
This work was supported by the Foundation for Polish Science through MPD and TEAM POIR.04.04.00-00-4204/17 programmes, the National Science Centre through grants No. 2016/21/B/ST2/01222, 2017/25/N/NZ1/00861 and 2019/35/B/ST2/03562, the Ministry for Science and Higher Education through grants No. 6673/IA/SP/2016, 7150/E-338/SPUB/2017/1 and 7150/E-338/M/2017, the EU and MSHE through grant No. POIG.02.03.00-161 00-013/09 and the Austrian Science Fund (FWF-P26783). The research was also funded by the Priority Research Area SciMat under the program Excellence Initiative -- Research University at the Jagiellonian University in Krak{\'{o}}w.

\appendix
\section{The impact of CRT, positron range and non-collinearity on spatial resolution} \label{A_SR_extended}
\begin{figure}[!t]
\centering
\includegraphics[width=0.65\textwidth]{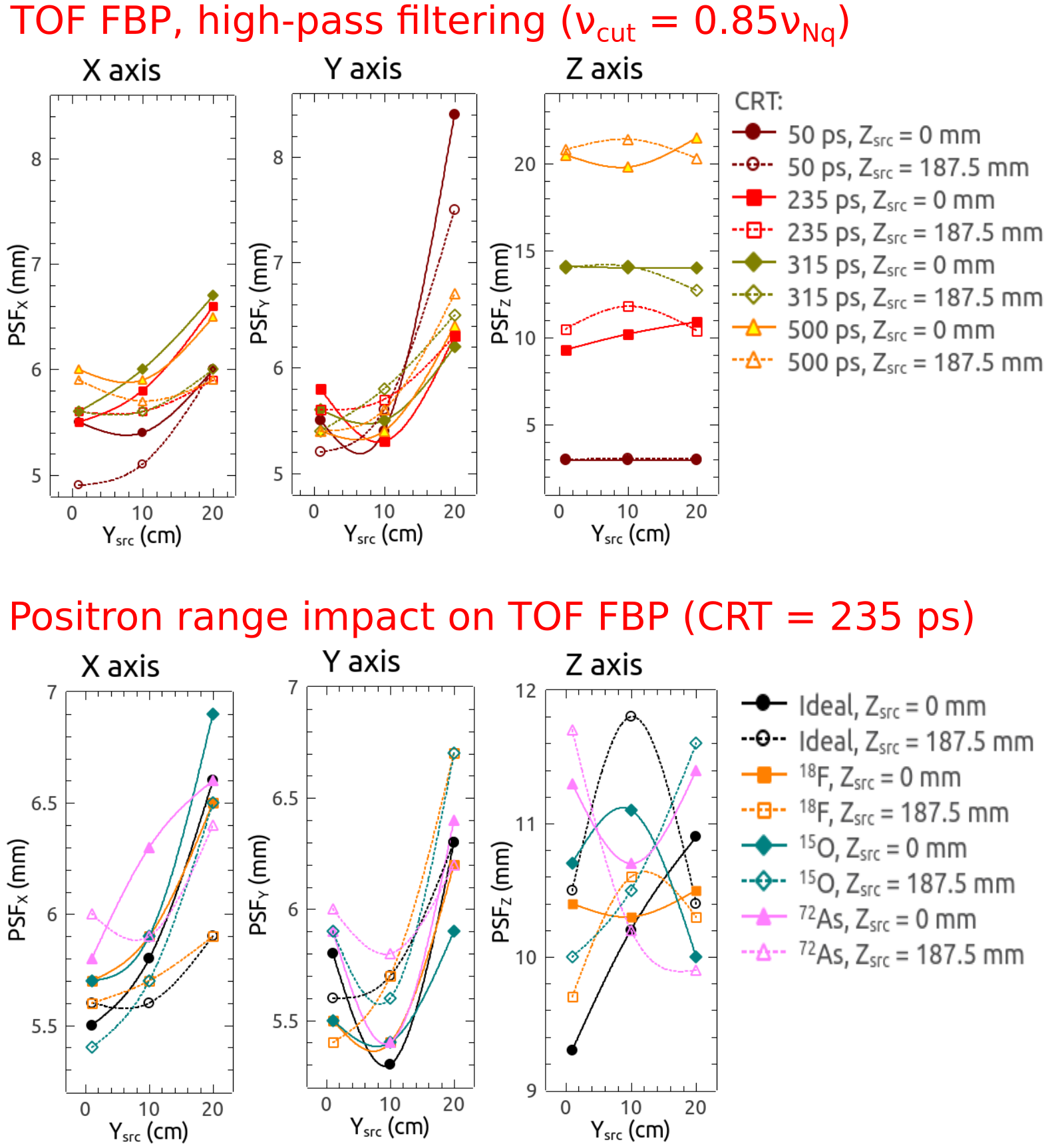}
\caption{Estimated PSFs for the reconstructed by TOF FBP $1$-mm source at six locations in the ideal J-PET scanner, depending on CRT (top) and positron range/non-collinearity for various radionuclides (bottom).
}
\label{Fig:A_SR_Extended}
\end{figure}

Fig.\ref{Fig:A_SR_Extended}, top shows the estimated PSF values for the reconstructed point-like source in the $382$-strip ideal J-PET scanner with temporal resolution up to $500$~ps. The 3D TOF FBP kernel comprised the high-pass filters $h^{-1}_\text{TOF}(l)$ and $h^{-1}_Z(z)$ with cut-off $\nu_\text{cut}=0.85\,\nu_\text{Nq}$. The benefit of using such filters is questionable for $\text{CRT}=50$~ps: despite the excellent $\text{PSF}_Z\approx3$~mm, more noise and distortion were observed. For $y_\text{src}=200$~mm, this has lead to $\text{PSF}_Y>7.5$~mm which can be improved only by further smoothing or apodisation at a cost of worse axial resolution. 

To consider a more realistic case, a 3D two-exponential probability distribution model was utilised to smear the true annihilation points (extracted from GATE) due to positron range \citep{LeLoirec2007, Carter2020}. The positions and times of hits were adjusted accordingly. Next, to emulate non-collinearity, the emission angles were blurred by a spherical Gaussian model of the magnitude (FWHM) $\Delta\varphi = 0.0044$\,rad \citep{Moses2011}. As a final step, the axial and temporal post-smearing were applied to the updated positions and times of hits. The high-pass version of the 3D TOF FBP kernel ($\nu_\text{cut}=0.85\,\nu_\text{Nq}$) remained unchanged for the reconstruction. 

Overall, the spatial resolution was minimally affected, as seen in Fig.\ref{Fig:A_SR_Extended}, bottom ($\text{CRT}=235$~ps). In comparison, noise and distortion were higher for the beta-plus radioactive isotopes ${}^{15}\ch{O}$ and ${}^{72}\ch{As}$, more pronounced for $50$-ps timing. These tendencies are consistent with the early experiments using the ${}^{22}\ch{Na}$ radionuclide \citep{Shopa2020}, where more distortion was observed, compared to simulations, yet transverse PSFs remained similar. In this work, positron range and non-collinearity have been neglected in the studies of NEMA IEC and XCAT phantoms, since we deal with a typical $\text{PSF}\approx5$\,mm across all axes and assume ${}^{18}\ch{F-FDG}$ used for our simulations. Other isotopes require further investigation, using more accurate models.

\section{Nelder-Mead minimisation algorithm} \label{A_NMprocedure}
An initial $4$-vertex simplex $\lbrace \lambda_i^{(0)}\rbrace$, used for the corresponding event-based TOF FBP reconstructions $\lbrace\hat{f}_i^{(0)}(\textbf{\textit{x}})\rbrace$, comprised four randomly selected sets of free parameters: 
\begin{equation}
\lambda_i^{(0)} = (\alpha^{(0)},\omega^{(0)}_c,\tau^{(0)},\sigma^{(0)}_\text{TOF},\Delta l^{(0)})^\text{T}_i,\quad i=1\ldots 4.
\label{eq:NMvertices}
\end{equation}

During the $k$-th iteration, MPF was applied to each $\hat{f}_i^{(k)}(\textbf{\textit{x}})$ up to four times using ball-shaped masks of decreasing radii (e.g. $4$~voxels $\rightarrow 2$~voxels $\rightarrow 1$~voxel). The resulting post-filtered image $\hat{f}_{i,\text{MPF}}^{(k)}(\textbf{\textit{x}})$ reflected the combination of such masks with the lowest MSE between $\hat{f}_{i,\text{MPF}}^{(k)}(\textbf{\textit{x}})$ and the ground truth $f(\textbf{\textit{x}})$. 

According to the Nelder-Mead method \citep{Nelder1965}, the worst vertex $\lbrace \lambda_i^{(k)}\rbrace$ for the $\hat{f}_{i,\text{MPF}}^{(k)}(\textbf{\textit{x}})$ with the highest MSE is replaced by a new one using \textit{reflection}, \textit{expansion} or \textit{contraction} through the centroid of the remaining $3$ vertices. If none of the new vertices produces a better image, the simplex is \textit{shrunk} towards the best $\lbrace \lambda_i^{(k)}\rbrace$. 

When tolerance for MSE is reached between $\lbrace \lambda_i^{(k)}\rbrace$, the best vertex is assigned as the final choice. We set it as $5.0\times10^{-5}$ (for a normalised $\left[0,1\right]$ intensity scale). 

\begin{figure}[!t]
\centering
\includegraphics[width=0.67\textwidth]{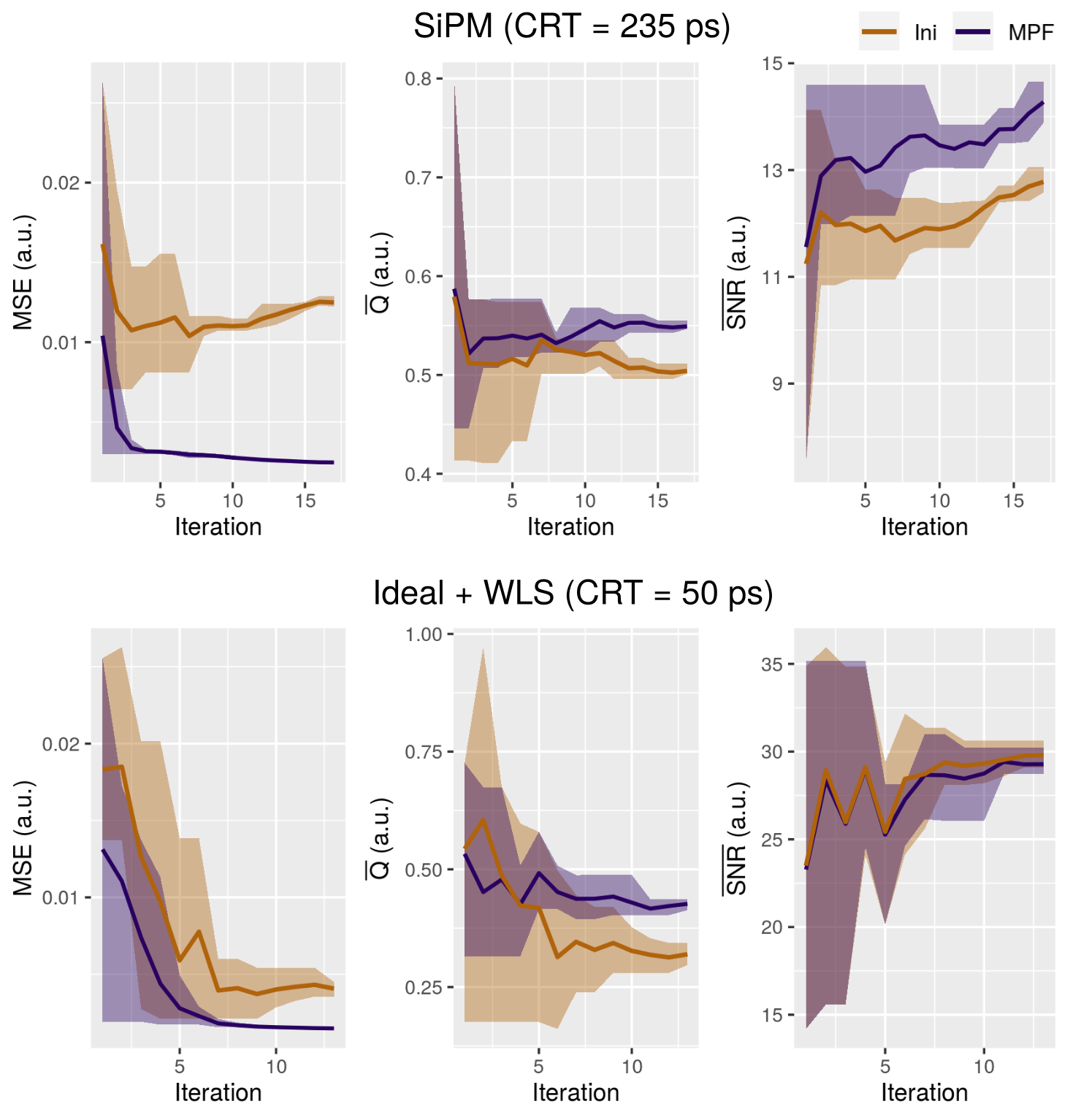}
\caption{Nelder-Mead minimisation progress for the metrics, calculated from the initial reconstructions $\hat{f}_i^{(k)}(\textbf{\textit{x}})$ (Ini) of the NEMA IEC phantom and after having applied MPF -- $\hat{f}_{i,\text{MPF}}^{(k)}(\textbf{\textit{x}})$. The lines denote average values over simplex vertices, the ribbons -- the min-max ranges for each iteration.}
\label{Fig:A_NelderMeadCurves}
\end{figure}

Fig.~\ref{Fig:A_NelderMeadCurves} shows the iteration progress of the Nelder-Mead algorithm, executed for the simulated NEMA IEC phantom. The average values of MSE, $\overline{\text{Q}}$ and $\overline{\text{SNR}}$, calculated for both $\hat{f}_i^{(k)}(\textbf{\textit{x}})$ and $\hat{f}_{i,\text{MPF}}^{(k)}(\textbf{\textit{x}})$ from a simplex $\lbrace \lambda_i^{(k)}\rbrace$, are shown as lines, the ranges between the vertices -- as ribbons. The role of the median filter could be seen as MSEs for $\hat{f}_i^{(k)}(\textbf{\textit{x}})$ and $\hat{f}_{i,\text{MPF}}^{(k)}(\textbf{\textit{x}})$ exhibit different trends, almost opposite for SiPM readout.

As seen from the figure, using image quality metrics as an alternative for minimisation is questionable. The averaged $\overline{\text{Q}}$ changes slowly over iterations and is mainly affected by the hot spheres of the phantom. It is sensitive to post-filtering, too: $\overline{\text{Q}}$-curves for the images without MPF are systematically lower. The lowest (the best) estimate for SiPM has been obtained far from the optimal vertex, rendering $\overline{\text{Q}}$ a bad minimisation metric. 

Another criterium -- $\overline{\text{SNR}}$ -- strongly depends on the background -- see Eq. (\ref{eq:BVandSNReq}). Despite the general trend to increase over minimisation, the best (the highest) $\overline{\text{SNR}}$ values have been obtained during earlier iterations, similarly to $\overline{\text{Q}}$. It is probably a consequence of a high regularisation $\tau$: although more distortion and the enhanced negative lobes are observed, $\overline{\text{SNR}}$ would raise due to lower background intensity.

\section{The evolution of image quality over the minimisation procedure} \label{A_IQ_curves}
\begin{figure}[!t]
\centering
\includegraphics[width=0.65\textwidth]{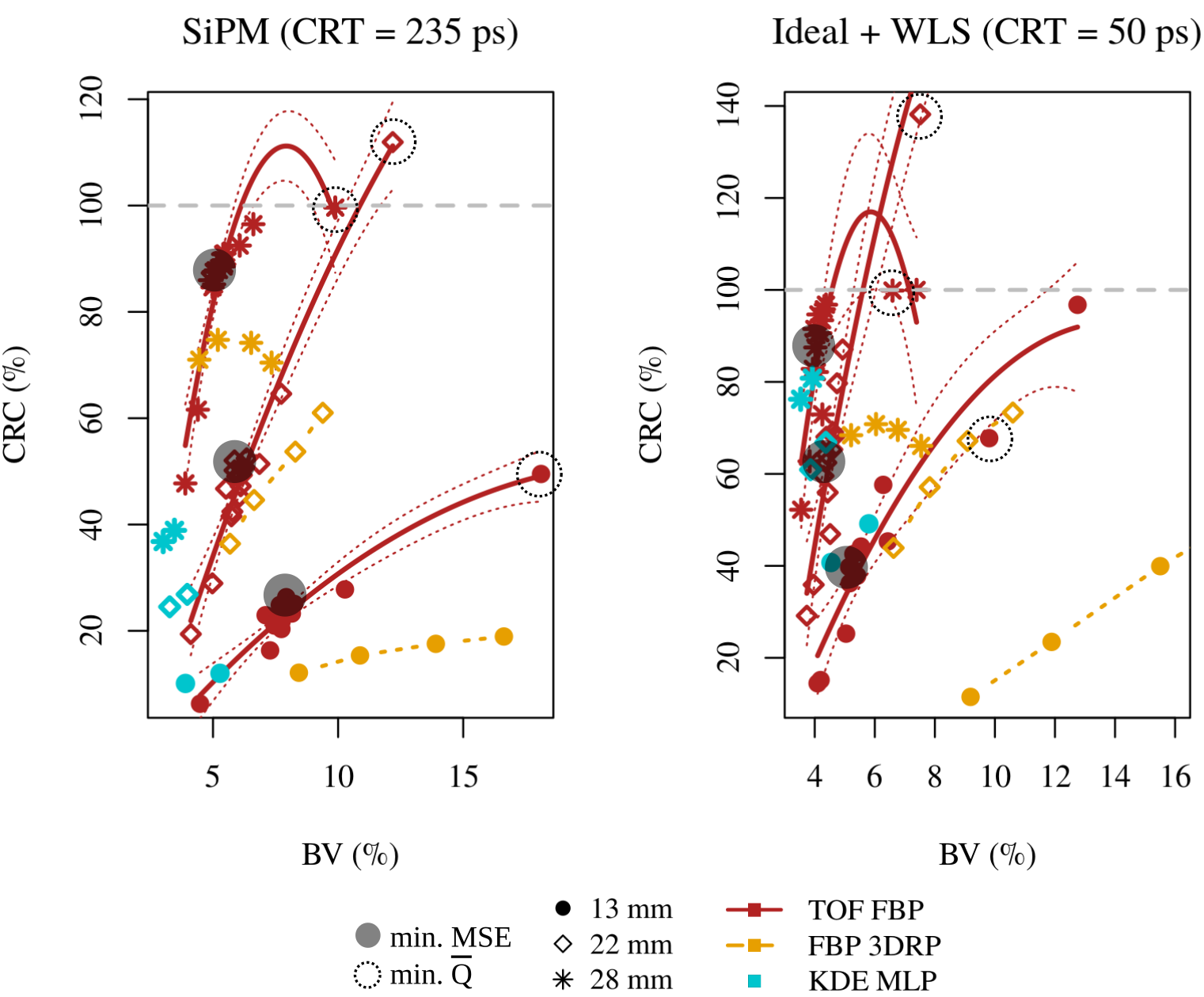}
\caption{CRC over BV for various algorithms, estimated for three spheres of NEMA IEC phantom and two readouts of J-PET scanner. Curves denote 2-nd order polynomial fits ($\text{CI}=95\%$) over the vertices taken from Nelder-Mead simplex. Ideal CRC = 100\% is marked by a dashed line.}
\label{Fig:A_IQcurvesMinimumsMedian}
\end{figure}

$\text{CRC}\left(\text{BV}\right)$ dependencies, estimated for two hot ($13$\,mm and $22$\,mm) and one cold ($28$\,mm) spheres in the NEMA IEC phantom, have been analysed using a set of the worst simplex vertices taken from each iteration. Fig.~\ref{Fig:A_IQcurvesMinimumsMedian} shows the results for TOF FBP + MPF, compared with other algorithms: KDE MLP using two bandwidth selection methods and FBP 3DRP with Hamming window, i.e. a product of the signal and a cosine profile \citep{FundMedImg2011}, with $\alpha=0.54$, $\omega_c=0.5$. Some points are dropped for better visualisation.

The best outcomes in terms of MSE and $\overline{\text{Q}}$ are emphasised by grey and open circles, respectively. Line+scatter plots represent various sizes of the MPF mask defined in STIR -- up to a cube of $7\times7\times7$ voxels. It affects image quality for FBP 3DRP more than $\alpha$ and $\omega_c$: the reconstructions shown in Fig.~\ref{Fig:IEC_MutiAlgorithms_SiPM_WLS50}, obtained for the ideal ramp filter look similar and produce almost identical curves to those for the Hamming window.

The event-based TOF FBP, optimised for the minimal MSE between the reconstruction and the ground truth (grey circles), generally produces superior outcomes, closer to the ideal $\text{CRC}=100\%$ and $\text{BV}=0\%$. Instead, the points for the best average $\overline{\text{Q}}$ (open circles) are rather far -- a consequence of a high $\tau=12.5-15.0$ (see Table~\ref{Tab:ParametrisationParamsIEC}), which leads to CRC above 100\%.

For $50$-ps readout, KDE MLP produces better image quality than FBP 3DRP, regardless of bandwidth selection of $\hat{\mathbf{H}}_\text{PI}$ matrix (Fig.~\ref{Fig:A_IQcurvesMinimumsMedian}, right). We remind, though, that KDE MLP required a larger dataset for the simplified attenuation correction, with about a third of events being dropped. 


\begin{thebibliography}{92}
\providecommand{\url}[1]{\texttt{#1}}
\providecommand{\href}[2]{#2}
\providecommand{\path}[1]{#1}
\providecommand{\DOIprefix}{doi:}
\providecommand{\ArXivprefix}{arXiv:}
\providecommand{\URLprefix}{URL: }
\providecommand{\Pubmedprefix}{pmid:}
\providecommand{\doi}[1]{\href{http://dx.doi.org/#1}{\path{#1}}}
\providecommand{\Pubmed}[1]{\href{pmid:#1}{\path{#1}}}
\providecommand{\bibinfo}[2]{#2}
\ifx\xfnm\undefined \def\xfnm[#1]{\unskip,\space#1}\fi
\bibitem[{Allemand et~al.(1980)Allemand, Gresset and Vacher}]{Allemand1980}
\bibinfo{author}{Allemand\xfnm[ R.]}, \bibinfo{author}{Gresset\xfnm[ C.]},
  \bibinfo{author}{Vacher\xfnm[ J.]}.
\newblock \bibinfo{title}{{Potential advantages of a cesium fluoride
  scintillator for a time-of-flight positron camera}}.
\newblock \bibinfo{journal}{J Nucl Med}
  \bibinfo{year}{1980};\bibinfo{volume}{21}(\bibinfo{number}{2}):\bibinfo{pages}{153--155}.
\newblock \URLprefix \url{http://www.ncbi.nlm.nih.gov/pubmed/6965404}.
\bibitem[{Ari{\~{n}}o-Estrada et~al.(2019)Ari{\~{n}}o-Estrada, Mitchell, Kim,
  Du, Kwon, Cirignano, Shah and Cherry}]{Arino-Estrada2019}
\bibinfo{author}{Ari{\~{n}}o-Estrada\xfnm[ G.]},
  \bibinfo{author}{Mitchell\xfnm[ G.S.]}, \bibinfo{author}{Kim\xfnm[ H.]},
  \bibinfo{author}{Du\xfnm[ J.]}, \bibinfo{author}{Kwon\xfnm[ S.I.]},
  \bibinfo{author}{Cirignano\xfnm[ L.J.]}, \bibinfo{author}{Shah\xfnm[ K.S.]},
  \bibinfo{author}{Cherry\xfnm[ S.R.]}.
\newblock \bibinfo{title}{{First Cerenkov charge-induction (CCI) TlBr detector
  for TOF-PET and proton range verification}}.
\newblock \bibinfo{journal}{Physics in Medicine {\&} Biology}
  \bibinfo{year}{2019};\bibinfo{volume}{64}(\bibinfo{number}{17}):\bibinfo{pages}{175001}.
\newblock \URLprefix
  \url{https://iopscience.iop.org/article/10.1088/1361-6560/ab35c4}.
  \DOIprefix\doi{10.1088/1361-6560/ab35c4}.
\bibitem[{Badawi et~al.(2019)Badawi, Shi, Hu, Chen, Xu, Price, Ding, Spencer,
  Nardo, Liu, Bao, Jones, Li and Cherry}]{Badawi2019}
\bibinfo{author}{Badawi\xfnm[ R.D.]}, \bibinfo{author}{Shi\xfnm[ H.]},
  \bibinfo{author}{Hu\xfnm[ P.]}, \bibinfo{author}{Chen\xfnm[ S.]},
  \bibinfo{author}{Xu\xfnm[ T.]}, \bibinfo{author}{Price\xfnm[ P.M.]},
  \bibinfo{author}{Ding\xfnm[ Y.]}, \bibinfo{author}{Spencer\xfnm[ B.A.]},
  \bibinfo{author}{Nardo\xfnm[ L.]}, \bibinfo{author}{Liu\xfnm[ W.]},
  \bibinfo{author}{Bao\xfnm[ J.]}, \bibinfo{author}{Jones\xfnm[ T.]},
  \bibinfo{author}{Li\xfnm[ H.]}, \bibinfo{author}{Cherry\xfnm[ S.R.]}.
\newblock \bibinfo{title}{{First Human Imaging Studies with the EXPLORER
  Total-Body PET Scanner}}.
\newblock \bibinfo{journal}{J Nucl Med}
  \bibinfo{year}{2019};\bibinfo{volume}{60}(\bibinfo{number}{3}):\bibinfo{pages}{299--303}.
\newblock \URLprefix
  \url{http://jnm.snmjournals.org/lookup/doi/10.2967/jnumed.119.226498}.
  \DOIprefix\doi{10.2967/jnumed.119.226498}.
\bibitem[{Bailey et~al.(2005)Bailey, Townsend, Valk and
  Maisey}]{PETBasicScience2005}
\bibinfo{editor}{Bailey\xfnm[ D.L.]}, \bibinfo{editor}{Townsend\xfnm[ D.W.]},
  \bibinfo{editor}{Valk\xfnm[ P.E.]}, \bibinfo{editor}{Maisey\xfnm[ M.N.]},
  editors.
\newblock \bibinfo{title}{{Positron Emission Tomography}}.
\newblock \bibinfo{address}{London}: \bibinfo{publisher}{Springer-Verlag},
  \bibinfo{year}{2005}.
\newblock \URLprefix \url{http://link.springer.com/10.1007/b136169}.
  \DOIprefix\doi{10.1007/b136169}.
\bibitem[{Budinger(1983)}]{Budinger1983}
\bibinfo{author}{Budinger\xfnm[ T.F.]}.
\newblock \bibinfo{title}{{Time-of-flight positron emission tomography: status
  relative to conventional PET.}}
\newblock \bibinfo{journal}{J Nucl Med}
  \bibinfo{year}{1983};\bibinfo{volume}{24}(\bibinfo{number}{1}):\bibinfo{pages}{73--78}.
\newblock \URLprefix \url{http://www.ncbi.nlm.nih.gov/pubmed/6336778}.
\bibitem[{Carter et~al.(2020)Carter, Kesner, Pratt, Sanders, Massicano, Cutler,
  Lapi and Lewis}]{Carter2020}
\bibinfo{author}{Carter\xfnm[ L.M.]}, \bibinfo{author}{Kesner\xfnm[ A.L.]},
  \bibinfo{author}{Pratt\xfnm[ E.C.]}, \bibinfo{author}{Sanders\xfnm[ V.A.]},
  \bibinfo{author}{Massicano\xfnm[ A.V.F.]}, \bibinfo{author}{Cutler\xfnm[
  C.S.]}, \bibinfo{author}{Lapi\xfnm[ S.E.]}, \bibinfo{author}{Lewis\xfnm[
  J.S.]}.
\newblock \bibinfo{title}{{The Impact of Positron Range on PET Resolution,
  Evaluated with Phantoms and PHITS Monte Carlo Simulations for Conventional
  and Non-conventional Radionuclides}}.
\newblock \bibinfo{journal}{Molecular Imaging and Biology}
  \bibinfo{year}{2020};\bibinfo{volume}{22}(\bibinfo{number}{1}):\bibinfo{pages}{73--84}.
\newblock \URLprefix \url{http://link.springer.com/10.1007/s11307-019-01337-2}.
  \DOIprefix\doi{10.1007/s11307-019-01337-2}.
\bibitem[{Chac{\'{o}}n and Duong(2010)}]{Chacon2010}
\bibinfo{author}{Chac{\'{o}}n\xfnm[ J.E.]}, \bibinfo{author}{Duong\xfnm[ T.]}.
\newblock \bibinfo{title}{{Multivariate plug-in bandwidth selection with
  unconstrained pilot bandwidth matrices}}.
\newblock \bibinfo{journal}{TEST}
  \bibinfo{year}{2010};\bibinfo{volume}{19}(\bibinfo{number}{2}):\bibinfo{pages}{375--398}.
\newblock \URLprefix \url{http://link.springer.com/10.1007/s11749-009-0168-4}.
  \DOIprefix\doi{10.1007/s11749-009-0168-4}.
\bibitem[{Cloquet and Defrise(2011)}]{Cloquet2011}
\bibinfo{author}{Cloquet\xfnm[ C.]}, \bibinfo{author}{Defrise\xfnm[ M.]}.
\newblock \bibinfo{title}{{Does OSEM achieve the lowest variance?}}
\newblock In: \bibinfo{booktitle}{2011 IEEE Nucl. Sci. Symp. Conf. Rec.}
  \bibinfo{publisher}{IEEE}; \bibinfo{year}{2011}. p.
  \bibinfo{pages}{2360--2365}.
\newblock \URLprefix \url{http://ieeexplore.ieee.org/document/6153880/}.
  \DOIprefix\doi{10.1109/NSSMIC.2011.6153880}.
\bibitem[{Conti(2011)}]{Conti2011}
\bibinfo{author}{Conti\xfnm[ M.]}.
\newblock \bibinfo{title}{{Focus on time-of-flight PET: the benefits of
  improved time resolution.}}
\newblock \bibinfo{journal}{EJNMMI}
  \bibinfo{year}{2011};\bibinfo{volume}{38}(\bibinfo{number}{6}):\bibinfo{pages}{1147--1157}.
\newblock \URLprefix \url{http://www.ncbi.nlm.nih.gov/pubmed/21229244}.
  \DOIprefix\doi{10.1007/s00259-010-1711-y}.
\bibitem[{Conti et~al.(2005)Conti, Bendriem, Casey, Chen, Kehren, Michel and
  Panin}]{Conti2005}
\bibinfo{author}{Conti\xfnm[ M.]}, \bibinfo{author}{Bendriem\xfnm[ B.]},
  \bibinfo{author}{Casey\xfnm[ M.]}, \bibinfo{author}{Chen\xfnm[ M.]},
  \bibinfo{author}{Kehren\xfnm[ F.]}, \bibinfo{author}{Michel\xfnm[ C.]},
  \bibinfo{author}{Panin\xfnm[ V.]}.
\newblock \bibinfo{title}{{First experimental results of time-of-flight
  reconstruction on an LSO PET scanner}}.
\newblock \bibinfo{journal}{Phys Med Biol}
  \bibinfo{year}{2005};\bibinfo{volume}{50}(\bibinfo{number}{19}):\bibinfo{pages}{4507--4526}.
\newblock \URLprefix
  \url{https://iopscience.iop.org/article/10.1088/0031-9155/50/19/006}.
  \DOIprefix\doi{10.1088/0031-9155/50/19/006}.
\bibitem[{Duong(2007)}]{Duong2007}
\bibinfo{author}{Duong\xfnm[ T.]}.
\newblock \bibinfo{title}{{ks: Kernel Density Estimation and Kernel
  Discriminant Analysis for Multivariate Data in R}}.
\newblock \bibinfo{journal}{J Stat Softw}
  \bibinfo{year}{2007};\bibinfo{volume}{21}(\bibinfo{number}{7}):\bibinfo{pages}{1--16}.
\newblock \URLprefix \url{http://www.jstatsoft.org/v21/i07/}.
  \DOIprefix\doi{10.18637/jss.v021.i07}.
\bibitem[{Duong and Hazelton(2003)}]{Duong2003}
\bibinfo{author}{Duong\xfnm[ T.]}, \bibinfo{author}{Hazelton\xfnm[ M.]}.
\newblock \bibinfo{title}{{Plug-in bandwidth matrices for bivariate kernel
  density estimation}}.
\newblock \bibinfo{journal}{J Nonparametr Stat}
  \bibinfo{year}{2003};\bibinfo{volume}{15}(\bibinfo{number}{1}):\bibinfo{pages}{17--30}.
\newblock \URLprefix
  \url{http://www.tandfonline.com/doi/abs/10.1080/10485250306039}.
  \DOIprefix\doi{10.1080/10485250306039}.
\bibitem[{Efthimiou et~al.(2019)Efthimiou, Emond, Wadhwa, Cawthorne, Tsoumpas
  and Thielemans}]{Efthimiou2019}
\bibinfo{author}{Efthimiou\xfnm[ N.]}, \bibinfo{author}{Emond\xfnm[ E.]},
  \bibinfo{author}{Wadhwa\xfnm[ P.]}, \bibinfo{author}{Cawthorne\xfnm[ C.]},
  \bibinfo{author}{Tsoumpas\xfnm[ C.]}, \bibinfo{author}{Thielemans\xfnm[ K.]}.
\newblock \bibinfo{title}{{Implementation and validation of time-of-flight PET
  image reconstruction module for listmode and sinogram projection data in the
  STIR library}}.
\newblock \bibinfo{journal}{Phys Med Biol}
  \bibinfo{year}{2019};\bibinfo{volume}{64}(\bibinfo{number}{3}):\bibinfo{pages}{035004}.
\newblock \URLprefix
  \url{https://iopscience.iop.org/article/10.1088/1361-6560/aaf9b9}.
  \DOIprefix\doi{10.1088/1361-6560/aaf9b9}.
\bibitem[{Fabija{\'{n}}ska and Sankowski(2011)}]{Fabijanska2011}
\bibinfo{author}{Fabija{\'{n}}ska\xfnm[ A.]}, \bibinfo{author}{Sankowski\xfnm[
  D.]}.
\newblock \bibinfo{title}{{Noise adaptive switching median-based filter for
  impulse noise removal from extremely corrupted images}}.
\newblock \bibinfo{journal}{IET Image Proc}
  \bibinfo{year}{2011};\bibinfo{volume}{5}(\bibinfo{number}{5}):\bibinfo{pages}{472--480}.
\newblock \URLprefix
  \url{https://digital-library.theiet.org/content/journals/10.1049/iet-ipr.2009.0178}.
  \DOIprefix\doi{10.1049/iet-ipr.2009.0178}.
\bibitem[{Ferrara and Mansi(2011)}]{FundMedImg2011}
\bibinfo{author}{Ferrara\xfnm[ R.]}, \bibinfo{author}{Mansi\xfnm[ L.]}.
\newblock \bibinfo{title}{{Paul Suetens (ed): Fundamentals of Medical Imaging
  (2nd edition)}}.
\newblock \bibinfo{journal}{EJNMMI}
  \bibinfo{year}{2011};\bibinfo{volume}{38}(\bibinfo{number}{2}):\bibinfo{pages}{409--409}.
\newblock \URLprefix \url{http://link.springer.com/10.1007/s00259-010-1694-8}.
  \DOIprefix\doi{10.1007/s00259-010-1694-8}.
\bibitem[{Ferrero et~al.(2018)Ferrero, Fiorina, Morrocchi, Pennazio, Baroni,
  Battistoni, Belcari, Camarlinghi, Ciocca, {Del Guerra}, Donetti, Giordanengo,
  Giraudo, Patera, Peroni, Rivetti, Rolo, Rossi, Rosso, Sportelli, Tampellini,
  Valvo, Wheadon, Cerello and Bisogni}]{Ferrero2018}
\bibinfo{author}{Ferrero\xfnm[ V.]}, \bibinfo{author}{Fiorina\xfnm[ E.]},
  \bibinfo{author}{Morrocchi\xfnm[ M.]}, \bibinfo{author}{Pennazio\xfnm[ F.]},
  \bibinfo{author}{Baroni\xfnm[ G.]}, \bibinfo{author}{Battistoni\xfnm[ G.]},
  \bibinfo{author}{Belcari\xfnm[ N.]}, \bibinfo{author}{Camarlinghi\xfnm[ N.]},
  \bibinfo{author}{Ciocca\xfnm[ M.]}, \bibinfo{author}{{Del Guerra}\xfnm[ A.]},
  \bibinfo{author}{Donetti\xfnm[ M.]}, \bibinfo{author}{Giordanengo\xfnm[ S.]},
  \bibinfo{author}{Giraudo\xfnm[ G.]}, \bibinfo{author}{Patera\xfnm[ V.]},
  \bibinfo{author}{Peroni\xfnm[ C.]}, \bibinfo{author}{Rivetti\xfnm[ A.]},
  \bibinfo{author}{Rolo\xfnm[ M.D.d.R.]}, \bibinfo{author}{Rossi\xfnm[ S.]},
  \bibinfo{author}{Rosso\xfnm[ V.]}, \bibinfo{author}{Sportelli\xfnm[ G.]},
  \bibinfo{author}{Tampellini\xfnm[ S.]}, \bibinfo{author}{Valvo\xfnm[ F.]},
  \bibinfo{author}{Wheadon\xfnm[ R.]}, \bibinfo{author}{Cerello\xfnm[ P.]},
  \bibinfo{author}{Bisogni\xfnm[ M.G.]}.
\newblock \bibinfo{title}{{Online proton therapy monitoring: clinical test of a
  Silicon-photodetector-based in-beam PET}}.
\newblock \bibinfo{journal}{Sci Rep}
  \bibinfo{year}{2018};\bibinfo{volume}{8}(\bibinfo{number}{1}):\bibinfo{pages}{4100}.
\newblock \URLprefix \url{http://www.nature.com/articles/s41598-018-22325-6}.
  \DOIprefix\doi{10.1038/s41598-018-22325-6}.
\bibitem[{Fiorina et~al.(2018)Fiorina, Ferrero, Pennazio, Baroni, Battistoni,
  Belcari, Cerello, Camarlinghi, Ciocca, {Del Guerra}, Donetti, Ferrari,
  Giordanengo, Giraudo, Mairani, Morrocchi, Peroni, Rivetti, {Da Rocha Rolo},
  Rossi, Rosso, Sala, Sportelli, Tampellini, Valvo, Wheadon and
  Bisogni}]{Fiorina2018}
\bibinfo{author}{Fiorina\xfnm[ E.]}, \bibinfo{author}{Ferrero\xfnm[ V.]},
  \bibinfo{author}{Pennazio\xfnm[ F.]}, \bibinfo{author}{Baroni\xfnm[ G.]},
  \bibinfo{author}{Battistoni\xfnm[ G.]}, \bibinfo{author}{Belcari\xfnm[ N.]},
  \bibinfo{author}{Cerello\xfnm[ P.]}, \bibinfo{author}{Camarlinghi\xfnm[ N.]},
  \bibinfo{author}{Ciocca\xfnm[ M.]}, \bibinfo{author}{{Del Guerra}\xfnm[ A.]},
  \bibinfo{author}{Donetti\xfnm[ M.]}, \bibinfo{author}{Ferrari\xfnm[ A.]},
  \bibinfo{author}{Giordanengo\xfnm[ S.]}, \bibinfo{author}{Giraudo\xfnm[ G.]},
  \bibinfo{author}{Mairani\xfnm[ A.]}, \bibinfo{author}{Morrocchi\xfnm[ M.]},
  \bibinfo{author}{Peroni\xfnm[ C.]}, \bibinfo{author}{Rivetti\xfnm[ A.]},
  \bibinfo{author}{{Da Rocha Rolo}\xfnm[ M.]}, \bibinfo{author}{Rossi\xfnm[
  S.]}, \bibinfo{author}{Rosso\xfnm[ V.]}, \bibinfo{author}{Sala\xfnm[ P.]},
  \bibinfo{author}{Sportelli\xfnm[ G.]}, \bibinfo{author}{Tampellini\xfnm[
  S.]}, \bibinfo{author}{Valvo\xfnm[ F.]}, \bibinfo{author}{Wheadon\xfnm[ R.]},
  \bibinfo{author}{Bisogni\xfnm[ M.]}.
\newblock \bibinfo{title}{{Monte Carlo simulation tool for online treatment
  monitoring in hadrontherapy with in-beam PET: A patient study}}.
\newblock \bibinfo{journal}{Phys Med}
  \bibinfo{year}{2018};\bibinfo{volume}{51}:\bibinfo{pages}{71--80}.
\newblock \URLprefix
  \url{https://linkinghub.elsevier.com/retrieve/pii/S1120179718304587}.
  \DOIprefix\doi{10.1016/j.ejmp.2018.05.002}.
\bibitem[{Grant et~al.(2016)Grant, Deller, Khalighi, Maramraju, Delso and
  Levin}]{Grant2016}
\bibinfo{author}{Grant\xfnm[ A.M.]}, \bibinfo{author}{Deller\xfnm[ T.W.]},
  \bibinfo{author}{Khalighi\xfnm[ M.M.]}, \bibinfo{author}{Maramraju\xfnm[
  S.H.]}, \bibinfo{author}{Delso\xfnm[ G.]}, \bibinfo{author}{Levin\xfnm[
  C.S.]}.
\newblock \bibinfo{title}{{NEMA NU 2-2012 performance studies for the
  SiPM-based ToF-PET component of the GE SIGNA PET/MR system.}}
\newblock \bibinfo{journal}{Med Phys}
  \bibinfo{year}{2016};\bibinfo{volume}{43}(\bibinfo{number}{5}):\bibinfo{pages}{2334}.
\newblock \URLprefix \url{http://www.ncbi.nlm.nih.gov/pubmed/27147345}.
  \DOIprefix\doi{10.1118/1.4945416}.
\bibitem[{Gundacker et~al.(2013)Gundacker, Auffray, Frisch, Jarron, Knapitsch,
  Meyer, Pizzichemi and Lecoq}]{Gundacker2013}
\bibinfo{author}{Gundacker\xfnm[ S.]}, \bibinfo{author}{Auffray\xfnm[ E.]},
  \bibinfo{author}{Frisch\xfnm[ B.]}, \bibinfo{author}{Jarron\xfnm[ P.]},
  \bibinfo{author}{Knapitsch\xfnm[ A.]}, \bibinfo{author}{Meyer\xfnm[ T.]},
  \bibinfo{author}{Pizzichemi\xfnm[ M.]}, \bibinfo{author}{Lecoq\xfnm[ P.]}.
\newblock \bibinfo{title}{{Time of flight positron emission tomography towards
  100ps resolution with L(Y)SO: an experimental and theoretical analysis}}.
\newblock \bibinfo{journal}{JINST}
  \bibinfo{year}{2013};\bibinfo{volume}{8}(\bibinfo{number}{07}):\bibinfo{pages}{P07014--P07014}.
\newblock \URLprefix
  \url{https://iopscience.iop.org/article/10.1088/1748-0221/8/07/P07014}.
  \DOIprefix\doi{10.1088/1748-0221/8/07/P07014}.
\bibitem[{Helgason(1984)}]{Helgason1984}
\bibinfo{author}{Helgason\xfnm[ S.]}.
\newblock \bibinfo{title}{{Groups and geometric analysis: integral geometry,
  invariant differential operators, and spherical functions}}.
\newblock \bibinfo{address}{Orlando, FL}: \bibinfo{publisher}{Academic Press},
  \bibinfo{year}{1984}.
\bibitem[{Jan et~al.(2011)Jan, Benoit, Becheva, Carlier, Cassol, Descourt,
  Frisson, Grevillot, Guigues, Maigne, Morel, Perrot, Rehfeld, Sarrut, Schaart,
  Stute, Pietrzyk, Visvikis, Zahra and Buvat}]{Gate2011}
\bibinfo{author}{Jan\xfnm[ S.]}, \bibinfo{author}{Benoit\xfnm[ D.]},
  \bibinfo{author}{Becheva\xfnm[ E.]}, \bibinfo{author}{Carlier\xfnm[ T.]},
  \bibinfo{author}{Cassol\xfnm[ F.]}, \bibinfo{author}{Descourt\xfnm[ P.]},
  \bibinfo{author}{Frisson\xfnm[ T.]}, \bibinfo{author}{Grevillot\xfnm[ L.]},
  \bibinfo{author}{Guigues\xfnm[ L.]}, \bibinfo{author}{Maigne\xfnm[ L.]},
  \bibinfo{author}{Morel\xfnm[ C.]}, \bibinfo{author}{Perrot\xfnm[ Y.]},
  \bibinfo{author}{Rehfeld\xfnm[ N.]}, \bibinfo{author}{Sarrut\xfnm[ D.]},
  \bibinfo{author}{Schaart\xfnm[ D.R.]}, \bibinfo{author}{Stute\xfnm[ S.]},
  \bibinfo{author}{Pietrzyk\xfnm[ U.]}, \bibinfo{author}{Visvikis\xfnm[ D.]},
  \bibinfo{author}{Zahra\xfnm[ N.]}, \bibinfo{author}{Buvat\xfnm[ I.]}.
\newblock \bibinfo{title}{{GATE V6: a major enhancement of the GATE simulation
  platform enabling modelling of CT and radiotherapy}}.
\newblock \bibinfo{journal}{Phys Med Biol}
  \bibinfo{year}{2011};\bibinfo{volume}{56}(\bibinfo{number}{4}):\bibinfo{pages}{881--901}.
\newblock \URLprefix \url{http://www.ncbi.nlm.nih.gov/pubmed/21248393}.
  \DOIprefix\doi{10.1088/0031-9155/56/4/001}.
\bibitem[{Jan et~al.(2004)Jan, Santin, Strul, Staelens, Assi{\'{e}}, Autret,
  Avner, Barbier, Bardi{\`{e}}s, Bloomfield, Brasse, Breton, Bruyndonckx,
  Buvat, Chatziioannou, Choi, Chung, Comtat, Donnarieix, Ferrer, Glick,
  Groiselle, Guez, Honore, Kerhoas-Cavata, Kirov, Kohli, Koole, Krieguer,
  van~der Laan, Lamare, Largeron, Lartizien, Lazaro, Maas, Maigne, Mayet,
  Melot, Merheb, Pennacchio, Perez, Pietrzyk, Rannou, Rey, Schaart,
  Schmidtlein, Simon, Song, Vieira, Visvikis, de~Walle, Wie{\"{e}}rs and
  Morel}]{Gate2004}
\bibinfo{author}{Jan\xfnm[ S.]}, \bibinfo{author}{Santin\xfnm[ G.]},
  \bibinfo{author}{Strul\xfnm[ D.]}, \bibinfo{author}{Staelens\xfnm[ S.]},
  \bibinfo{author}{Assi{\'{e}}\xfnm[ K.]}, \bibinfo{author}{Autret\xfnm[ D.]},
  \bibinfo{author}{Avner\xfnm[ S.]}, \bibinfo{author}{Barbier\xfnm[ R.]},
  \bibinfo{author}{Bardi{\`{e}}s\xfnm[ M.]}, \bibinfo{author}{Bloomfield\xfnm[
  P.M.]}, \bibinfo{author}{Brasse\xfnm[ D.]}, \bibinfo{author}{Breton\xfnm[
  V.]}, \bibinfo{author}{Bruyndonckx\xfnm[ P.]}, \bibinfo{author}{Buvat\xfnm[
  I.]}, \bibinfo{author}{Chatziioannou\xfnm[ A.F.]},
  \bibinfo{author}{Choi\xfnm[ Y.]}, \bibinfo{author}{Chung\xfnm[ Y.H.]},
  \bibinfo{author}{Comtat\xfnm[ C.]}, \bibinfo{author}{Donnarieix\xfnm[ D.]},
  \bibinfo{author}{Ferrer\xfnm[ L.]}, \bibinfo{author}{Glick\xfnm[ S.J.]},
  \bibinfo{author}{Groiselle\xfnm[ C.J.]}, \bibinfo{author}{Guez\xfnm[ D.]},
  \bibinfo{author}{Honore\xfnm[ P.F.]}, \bibinfo{author}{Kerhoas-Cavata\xfnm[
  S.]}, \bibinfo{author}{Kirov\xfnm[ A.S.]}, \bibinfo{author}{Kohli\xfnm[ V.]},
  \bibinfo{author}{Koole\xfnm[ M.]}, \bibinfo{author}{Krieguer\xfnm[ M.]},
  \bibinfo{author}{van~der Laan\xfnm[ D.J.]}, \bibinfo{author}{Lamare\xfnm[
  F.]}, \bibinfo{author}{Largeron\xfnm[ G.]}, \bibinfo{author}{Lartizien\xfnm[
  C.]}, \bibinfo{author}{Lazaro\xfnm[ D.]}, \bibinfo{author}{Maas\xfnm[ M.C.]},
  \bibinfo{author}{Maigne\xfnm[ L.]}, \bibinfo{author}{Mayet\xfnm[ F.]},
  \bibinfo{author}{Melot\xfnm[ F.]}, \bibinfo{author}{Merheb\xfnm[ C.]},
  \bibinfo{author}{Pennacchio\xfnm[ E.]}, \bibinfo{author}{Perez\xfnm[ J.]},
  \bibinfo{author}{Pietrzyk\xfnm[ U.]}, \bibinfo{author}{Rannou\xfnm[ F.R.]},
  \bibinfo{author}{Rey\xfnm[ M.]}, \bibinfo{author}{Schaart\xfnm[ D.R.]},
  \bibinfo{author}{Schmidtlein\xfnm[ C.R.]}, \bibinfo{author}{Simon\xfnm[ L.]},
  \bibinfo{author}{Song\xfnm[ T.Y.]}, \bibinfo{author}{Vieira\xfnm[ J.M.]},
  \bibinfo{author}{Visvikis\xfnm[ D.]}, \bibinfo{author}{de~Walle\xfnm[ R.V.]},
  \bibinfo{author}{Wie{\"{e}}rs\xfnm[ E.]}, \bibinfo{author}{Morel\xfnm[ C.]}.
\newblock \bibinfo{title}{{GATE: a simulation toolkit for PET and SPECT}}.
\newblock \bibinfo{journal}{Phys Med Biol}
  \bibinfo{year}{2004};\bibinfo{volume}{49}(\bibinfo{number}{19}):\bibinfo{pages}{4543--4561}.
\newblock \URLprefix
  \url{https://iopscience.iop.org/article/10.1088/0031-9155/49/19/007}.
  \DOIprefix\doi{10.1088/0031-9155/49/19/007}.
\bibitem[{Kadrmas et~al.(2009)Kadrmas, Casey, Conti, Jakoby, Lois and
  Townsend}]{Kadrmas2009}
\bibinfo{author}{Kadrmas\xfnm[ D.J.]}, \bibinfo{author}{Casey\xfnm[ M.E.]},
  \bibinfo{author}{Conti\xfnm[ M.]}, \bibinfo{author}{Jakoby\xfnm[ B.W.]},
  \bibinfo{author}{Lois\xfnm[ C.]}, \bibinfo{author}{Townsend\xfnm[ D.W.]}.
\newblock \bibinfo{title}{{Impact of time-of-flight on PET tumor detection}}.
\newblock \bibinfo{journal}{J Nucl Med}
  \bibinfo{year}{2009};\bibinfo{volume}{50}(\bibinfo{number}{8}):\bibinfo{pages}{1315--1323}.
\newblock \URLprefix \url{http://www.ncbi.nlm.nih.gov/pubmed/19617317}.
  \DOIprefix\doi{10.2967/jnumed.109.063016}.
\bibitem[{Karp et~al.(2008)Karp, Surti, Daube-Witherspoon and
  Muehllehner}]{Karp2008}
\bibinfo{author}{Karp\xfnm[ J.S.]}, \bibinfo{author}{Surti\xfnm[ S.]},
  \bibinfo{author}{Daube-Witherspoon\xfnm[ M.E.]},
  \bibinfo{author}{Muehllehner\xfnm[ G.]}.
\newblock \bibinfo{title}{{Benefit of Time-of-Flight in PET: Experimental and
  Clinical Results}}.
\newblock \bibinfo{journal}{J Nucl Med}
  \bibinfo{year}{2008};\bibinfo{volume}{49}(\bibinfo{number}{3}):\bibinfo{pages}{462--470}.
\newblock \URLprefix
  \url{http://jnm.snmjournals.org/cgi/doi/10.2967/jnumed.107.044834}.
  \DOIprefix\doi{10.2967/jnumed.107.044834}.
\bibitem[{Karp et~al.(2020)Karp, Viswanath, Geagan, Muehllehner, Pantel, Parma,
  Perkins, Schmall, Werner and Daube-Witherspoon}]{Karp2020}
\bibinfo{author}{Karp\xfnm[ J.S.]}, \bibinfo{author}{Viswanath\xfnm[ V.]},
  \bibinfo{author}{Geagan\xfnm[ M.J.]}, \bibinfo{author}{Muehllehner\xfnm[
  G.]}, \bibinfo{author}{Pantel\xfnm[ A.R.]}, \bibinfo{author}{Parma\xfnm[
  M.J.]}, \bibinfo{author}{Perkins\xfnm[ A.E.]}, \bibinfo{author}{Schmall\xfnm[
  J.P.]}, \bibinfo{author}{Werner\xfnm[ M.E.]},
  \bibinfo{author}{Daube-Witherspoon\xfnm[ M.E.]}.
\newblock \bibinfo{title}{{PennPET Explorer: Design and Preliminary Performance
  of a Whole-Body Imager}}.
\newblock \bibinfo{journal}{J Nucl Med}
  \bibinfo{year}{2020};\bibinfo{volume}{61}(\bibinfo{number}{1}):\bibinfo{pages}{136--143}.
\newblock \URLprefix
  \url{http://jnm.snmjournals.org/lookup/doi/10.2967/jnumed.119.229997}.
  \DOIprefix\doi{10.2967/jnumed.119.229997}.
\bibitem[{Kinahan and Rogers(1989)}]{Kinahan1989}
\bibinfo{author}{Kinahan\xfnm[ P.E.]}, \bibinfo{author}{Rogers\xfnm[ J.G.]}.
\newblock \bibinfo{title}{{Analytic 3D image reconstruction using all detected
  events}}.
\newblock \bibinfo{journal}{IEEE Trans Nucl Sci}
  \bibinfo{year}{1989};\bibinfo{volume}{36}(\bibinfo{number}{1}):\bibinfo{pages}{964--968}.
\newblock \URLprefix \url{http://ieeexplore.ieee.org/document/34585/}.
  \DOIprefix\doi{10.1109/23.34585}.
\bibitem[{Kopka and Klimaszewski(2020)}]{Kopka2020}
\bibinfo{author}{Kopka\xfnm[ P.]}, \bibinfo{author}{Klimaszewski\xfnm[ K.]}.
\newblock \bibinfo{title}{{Reconstruction of the NEMA IEC Body Phantom from
  J-PET Total-body Scanner Simulation Using STIR}}.
\newblock \bibinfo{journal}{APPB}
  \bibinfo{year}{2020};\bibinfo{volume}{51}(\bibinfo{number}{1}):\bibinfo{pages}{357}.
\newblock \URLprefix
  \url{http://www.actaphys.uj.edu.pl/findarticle?series=Reg\&vol=51\&page=357}.
  \DOIprefix\doi{10.5506/APhysPolB.51.357}.
\bibitem[{Korcyl et~al.(2018)Korcyl, Hiesmayr, Jasinska, Kacprzak,
  Kajetanowicz, Kisielewska, Kowalski, Kozik, Krawczyk, Krzemien, Kubicz,
  Bialas, Mohammed, Niedzwiecki, Pawlik-Niedzwiecka, Palka, Raczynski, Rajda,
  Rudy, Salabura, Sharma, Sharma, Curceanu, Shopa, Skurzok, Silarski,
  Strzempek, Wieczorek, Wislicki, Zaleski, Zgardzinska, Zielinski, Moskal,
  Czerwinski, Dulski, Flak, Gajos, Glowacz and Gorgol}]{Korcyl2018}
\bibinfo{author}{Korcyl\xfnm[ G.]}, \bibinfo{author}{Hiesmayr\xfnm[ B.C.]},
  \bibinfo{author}{Jasinska\xfnm[ B.]}, \bibinfo{author}{Kacprzak\xfnm[ K.]},
  \bibinfo{author}{Kajetanowicz\xfnm[ M.]}, \bibinfo{author}{Kisielewska\xfnm[
  D.]}, \bibinfo{author}{Kowalski\xfnm[ P.]}, \bibinfo{author}{Kozik\xfnm[
  T.]}, \bibinfo{author}{Krawczyk\xfnm[ N.]}, \bibinfo{author}{Krzemien\xfnm[
  W.]}, \bibinfo{author}{Kubicz\xfnm[ E.]}, \bibinfo{author}{Bialas\xfnm[ P.]},
  \bibinfo{author}{Mohammed\xfnm[ M.]}, \bibinfo{author}{Niedzwiecki\xfnm[
  S.]}, \bibinfo{author}{Pawlik-Niedzwiecka\xfnm[ M.]},
  \bibinfo{author}{Palka\xfnm[ M.]}, \bibinfo{author}{Raczynski\xfnm[ L.]},
  \bibinfo{author}{Rajda\xfnm[ P.]}, \bibinfo{author}{Rudy\xfnm[ Z.]},
  \bibinfo{author}{Salabura\xfnm[ P.]}, \bibinfo{author}{Sharma\xfnm[ N.G.]},
  \bibinfo{author}{Sharma\xfnm[ S.]}, \bibinfo{author}{Curceanu\xfnm[ C.]},
  \bibinfo{author}{Shopa\xfnm[ R.Y.]}, \bibinfo{author}{Skurzok\xfnm[ M.]},
  \bibinfo{author}{Silarski\xfnm[ M.]}, \bibinfo{author}{Strzempek\xfnm[ P.]},
  \bibinfo{author}{Wieczorek\xfnm[ A.]}, \bibinfo{author}{Wislicki\xfnm[ W.]},
  \bibinfo{author}{Zaleski\xfnm[ R.]}, \bibinfo{author}{Zgardzinska\xfnm[ B.]},
  \bibinfo{author}{Zielinski\xfnm[ M.]}, \bibinfo{author}{Moskal\xfnm[ P.]},
  \bibinfo{author}{Czerwinski\xfnm[ E.]}, \bibinfo{author}{Dulski\xfnm[ K.]},
  \bibinfo{author}{Flak\xfnm[ B.]}, \bibinfo{author}{Gajos\xfnm[ A.]},
  \bibinfo{author}{Glowacz\xfnm[ B.]}, \bibinfo{author}{Gorgol\xfnm[ M.]}.
\newblock \bibinfo{title}{{Evaluation of Single-Chip, Real-Time Tomographic
  Data Processing on FPGA SoC Devices}}.
\newblock \bibinfo{journal}{IEEE Trans Med Imagin}
  \bibinfo{year}{2018};\bibinfo{volume}{37}(\bibinfo{number}{11}):\bibinfo{pages}{2526--2535}.
\newblock \URLprefix \url{https://ieeexplore.ieee.org/document/8360475/}.
  \DOIprefix\doi{10.1109/TMI.2018.2837741}.
\bibitem[{Kowalski et~al.(2018)Kowalski, Wi{\'{s}}licki, Shopa,
  Raczy{\'{n}}ski, Klimaszewski, Curcenau, Czerwi{\'{n}}ski, Dulski, Gajos,
  Gorgol, Gupta-Sharma, Hiesmayr, Jasi{\'{n}}ska, Kap{\l}on,
  Kisielewska-Kami{\'{n}}ska, Korcyl, Kozik, Krzemie{\'{n}}, Kubicz, Mohammed,
  Nied{\'{z}}wiecki, Pa{\l}ka, Pawlik-Nied{\'{z}}wiecka, Raj, Rakoczy, Rudy,
  Sharma, Shivani, Silarski, Skurzok, Zgardzi{\'{n}}ska, Zieli{\'{n}}ski and
  Moskal}]{Kowalski2018}
\bibinfo{author}{Kowalski\xfnm[ P.]}, \bibinfo{author}{Wi{\'{s}}licki\xfnm[
  W.]}, \bibinfo{author}{Shopa\xfnm[ R.Y.]},
  \bibinfo{author}{Raczy{\'{n}}ski\xfnm[ L.]},
  \bibinfo{author}{Klimaszewski\xfnm[ K.]}, \bibinfo{author}{Curcenau\xfnm[
  C.]}, \bibinfo{author}{Czerwi{\'{n}}ski\xfnm[ E.]},
  \bibinfo{author}{Dulski\xfnm[ K.]}, \bibinfo{author}{Gajos\xfnm[ A.]},
  \bibinfo{author}{Gorgol\xfnm[ M.]}, \bibinfo{author}{Gupta-Sharma\xfnm[ N.]},
  \bibinfo{author}{Hiesmayr\xfnm[ B.]}, \bibinfo{author}{Jasi{\'{n}}ska\xfnm[
  B.]}, \bibinfo{author}{Kap{\l}on\xfnm[ {\L}.]},
  \bibinfo{author}{Kisielewska-Kami{\'{n}}ska\xfnm[ D.]},
  \bibinfo{author}{Korcyl\xfnm[ G.]}, \bibinfo{author}{Kozik\xfnm[ T.]},
  \bibinfo{author}{Krzemie{\'{n}}\xfnm[ W.]}, \bibinfo{author}{Kubicz\xfnm[
  E.]}, \bibinfo{author}{Mohammed\xfnm[ M.]},
  \bibinfo{author}{Nied{\'{z}}wiecki\xfnm[ S.]},
  \bibinfo{author}{Pa{\l}ka\xfnm[ M.]},
  \bibinfo{author}{Pawlik-Nied{\'{z}}wiecka\xfnm[ M.]},
  \bibinfo{author}{Raj\xfnm[ J.]}, \bibinfo{author}{Rakoczy\xfnm[ K.]},
  \bibinfo{author}{Rudy\xfnm[ Z.]}, \bibinfo{author}{Sharma\xfnm[ S.]},
  \bibinfo{author}{Shivani\xfnm[ S.]}, \bibinfo{author}{Silarski\xfnm[ M.]},
  \bibinfo{author}{Skurzok\xfnm[ M.]}, \bibinfo{author}{Zgardzi{\'{n}}ska\xfnm[
  B.]}, \bibinfo{author}{Zieli{\'{n}}ski\xfnm[ M.]},
  \bibinfo{author}{Moskal\xfnm[ P.]}.
\newblock \bibinfo{title}{{Estimating the NEMA characteristics of the J-PET
  tomograph using the GATE package}}.
\newblock \bibinfo{journal}{Phys Med Biol}
  \bibinfo{year}{2018};\bibinfo{volume}{63}(\bibinfo{number}{16}):\bibinfo{pages}{165008}.
\newblock \URLprefix
  \url{https://iopscience.iop.org/article/10.1088/1361-6560/aad29b}.
  \DOIprefix\doi{10.1088/1361-6560/aad29b}.
\bibitem[{{Le Loirec} and Champion(2007)}]{LeLoirec2007}
\bibinfo{author}{{Le Loirec}\xfnm[ C.]}, \bibinfo{author}{Champion\xfnm[ C.]}.
\newblock \bibinfo{title}{{Track structure simulation for positron emitters of
  medical interest. Part I: The case of the allowed decay isotopes}}.
\newblock \bibinfo{journal}{Nuclear Instruments and Methods in Physics Research
  Section A: Accelerators, Spectrometers, Detectors and Associated Equipment}
  \bibinfo{year}{2007};\bibinfo{volume}{582}(\bibinfo{number}{2}):\bibinfo{pages}{644--653}.
\newblock \URLprefix
  \url{https://linkinghub.elsevier.com/retrieve/pii/S0168900207018578}.
  \DOIprefix\doi{10.1016/j.nima.2007.08.159}.
\bibitem[{Lecoq(2017)}]{Lecoq2017}
\bibinfo{author}{Lecoq\xfnm[ P.]}.
\newblock \bibinfo{title}{{Pushing the Limits in Time-of-Flight PET Imaging}}.
\newblock \bibinfo{journal}{IEEE Trans Radiat Plasma Med Sci}
  \bibinfo{year}{2017};\bibinfo{volume}{1}(\bibinfo{number}{6}):\bibinfo{pages}{473--485}.
\newblock \URLprefix \url{https://ieeexplore.ieee.org/document/8049484/}.
  \DOIprefix\doi{10.1109/TRPMS.2017.2756674}.
\bibitem[{Lecoq et~al.(2020)Lecoq, Morel, Prior, Visvikis, Gundacker, Auffray,
  Kri{\v{z}}an, Turtos, Thers, Charbon, Varela, {de La Taille}, Rivetti,
  Breton, Pratte, Nuyts, Surti, Vandenberghe, Marsden, Parodi, Benlloch and
  Benoit}]{Lecoq2020}
\bibinfo{author}{Lecoq\xfnm[ P.]}, \bibinfo{author}{Morel\xfnm[ C.]},
  \bibinfo{author}{Prior\xfnm[ J.O.]}, \bibinfo{author}{Visvikis\xfnm[ D.]},
  \bibinfo{author}{Gundacker\xfnm[ S.]}, \bibinfo{author}{Auffray\xfnm[ E.]},
  \bibinfo{author}{Kri{\v{z}}an\xfnm[ P.]}, \bibinfo{author}{Turtos\xfnm[
  R.M.]}, \bibinfo{author}{Thers\xfnm[ D.]}, \bibinfo{author}{Charbon\xfnm[
  E.]}, \bibinfo{author}{Varela\xfnm[ J.]}, \bibinfo{author}{{de La
  Taille}\xfnm[ C.]}, \bibinfo{author}{Rivetti\xfnm[ A.]},
  \bibinfo{author}{Breton\xfnm[ D.]}, \bibinfo{author}{Pratte\xfnm[ J.F.]},
  \bibinfo{author}{Nuyts\xfnm[ J.]}, \bibinfo{author}{Surti\xfnm[ S.]},
  \bibinfo{author}{Vandenberghe\xfnm[ S.]}, \bibinfo{author}{Marsden\xfnm[
  P.]}, \bibinfo{author}{Parodi\xfnm[ K.]}, \bibinfo{author}{Benlloch\xfnm[
  J.M.]}, \bibinfo{author}{Benoit\xfnm[ M.]}.
\newblock \bibinfo{title}{{Roadmap toward the 10 ps time-of-flight PET
  challenge}}.
\newblock \bibinfo{journal}{Physics in Medicine {\&} Biology}
  \bibinfo{year}{2020};\bibinfo{volume}{65}(\bibinfo{number}{21}):\bibinfo{pages}{21RM01}.
\newblock \URLprefix
  \url{https://iopscience.iop.org/article/10.1088/1361-6560/ab9500}.
  \DOIprefix\doi{10.1088/1361-6560/ab9500}.
\bibitem[{Li et~al.(2016)Li, Wang, Hou, Yang and Wang}]{Li2016}
\bibinfo{author}{Li\xfnm[ S.]}, \bibinfo{author}{Wang\xfnm[ M.]},
  \bibinfo{author}{Hou\xfnm[ H.]}, \bibinfo{author}{Yang\xfnm[ J.]},
  \bibinfo{author}{Wang\xfnm[ X.]}.
\newblock \bibinfo{title}{{Fast algorithm for calculating the radiological path
  in fan-beam CT image reconstruction}}.
\newblock \bibinfo{journal}{Optik}
  \bibinfo{year}{2016};\bibinfo{volume}{127}(\bibinfo{number}{5}):\bibinfo{pages}{2973--2977}.
\newblock \URLprefix
  \url{https://linkinghub.elsevier.com/retrieve/pii/S0030402615019439}.
  \DOIprefix\doi{10.1016/j.ijleo.2015.12.034}.
\bibitem[{L{\'{o}}pez-Montes et~al.(2020)L{\'{o}}pez-Montes, Galve, Udias,
  Cal-Gonz{\'{a}}lez, Vaquero, Desco and Herraiz}]{Lopez-Montes2020}
\bibinfo{author}{L{\'{o}}pez-Montes\xfnm[ A.]}, \bibinfo{author}{Galve\xfnm[
  P.]}, \bibinfo{author}{Udias\xfnm[ J.M.]},
  \bibinfo{author}{Cal-Gonz{\'{a}}lez\xfnm[ J.]},
  \bibinfo{author}{Vaquero\xfnm[ J.J.]}, \bibinfo{author}{Desco\xfnm[ M.]},
  \bibinfo{author}{Herraiz\xfnm[ J.L.]}.
\newblock \bibinfo{title}{{Real-Time 3D PET Image with Pseudoinverse
  Reconstruction}}.
\newblock \bibinfo{journal}{Applied Sciences}
  \bibinfo{year}{2020};\bibinfo{volume}{10}(\bibinfo{number}{8}):\bibinfo{pages}{2829}.
\newblock \URLprefix \url{https://www.mdpi.com/2076-3417/10/8/2829}.
  \DOIprefix\doi{10.3390/app10082829}.
\bibitem[{Marafini et~al.(2015)Marafini, Attili, Battistoni, Belcari, Bisogni,
  Camarlinghi, Cappucci, Cecchetti, Cerello, Ciciriello, Cirrone, Coli, Corsi,
  Cuttone, {De Lucia}, Ferretti, Faccini, Fiorina, Frallicciardi, Giraudo,
  Kostara, Kraan, Licciulli, Liu, Marino, Marzocca, Matarrese, Morone,
  Morrocchi, Muraro, Patera, Pennazio, Peroni, Piersanti, Piliero, Pirrone,
  Rivetti, Romano, Rosso, Sala, Sarti, Sciubba, Sportelli, Voena, Wheadon and
  {Del Guerra}}]{Marafini2015}
\bibinfo{author}{Marafini\xfnm[ M.]}, \bibinfo{author}{Attili\xfnm[ A.]},
  \bibinfo{author}{Battistoni\xfnm[ G.]}, \bibinfo{author}{Belcari\xfnm[ N.]},
  \bibinfo{author}{Bisogni\xfnm[ M.]}, \bibinfo{author}{Camarlinghi\xfnm[ N.]},
  \bibinfo{author}{Cappucci\xfnm[ F.]}, \bibinfo{author}{Cecchetti\xfnm[ M.]},
  \bibinfo{author}{Cerello\xfnm[ P.]}, \bibinfo{author}{Ciciriello\xfnm[ F.]},
  \bibinfo{author}{Cirrone\xfnm[ G.]}, \bibinfo{author}{Coli\xfnm[ S.]},
  \bibinfo{author}{Corsi\xfnm[ F.]}, \bibinfo{author}{Cuttone\xfnm[ G.]},
  \bibinfo{author}{{De Lucia}\xfnm[ E.]}, \bibinfo{author}{Ferretti\xfnm[ S.]},
  \bibinfo{author}{Faccini\xfnm[ R.]}, \bibinfo{author}{Fiorina\xfnm[ E.]},
  \bibinfo{author}{Frallicciardi\xfnm[ P.]}, \bibinfo{author}{Giraudo\xfnm[
  G.]}, \bibinfo{author}{Kostara\xfnm[ E.]}, \bibinfo{author}{Kraan\xfnm[ A.]},
  \bibinfo{author}{Licciulli\xfnm[ F.]}, \bibinfo{author}{Liu\xfnm[ B.]},
  \bibinfo{author}{Marino\xfnm[ N.]}, \bibinfo{author}{Marzocca\xfnm[ C.]},
  \bibinfo{author}{Matarrese\xfnm[ G.]}, \bibinfo{author}{Morone\xfnm[ C.]},
  \bibinfo{author}{Morrocchi\xfnm[ M.]}, \bibinfo{author}{Muraro\xfnm[ S.]},
  \bibinfo{author}{Patera\xfnm[ V.]}, \bibinfo{author}{Pennazio\xfnm[ F.]},
  \bibinfo{author}{Peroni\xfnm[ C.]}, \bibinfo{author}{Piersanti\xfnm[ L.]},
  \bibinfo{author}{Piliero\xfnm[ M.]}, \bibinfo{author}{Pirrone\xfnm[ G.]},
  \bibinfo{author}{Rivetti\xfnm[ A.]}, \bibinfo{author}{Romano\xfnm[ F.]},
  \bibinfo{author}{Rosso\xfnm[ V.]}, \bibinfo{author}{Sala\xfnm[ P.]},
  \bibinfo{author}{Sarti\xfnm[ A.]}, \bibinfo{author}{Sciubba\xfnm[ A.]},
  \bibinfo{author}{Sportelli\xfnm[ G.]}, \bibinfo{author}{Voena\xfnm[ C.]},
  \bibinfo{author}{Wheadon\xfnm[ R.]}, \bibinfo{author}{{Del Guerra}\xfnm[
  A.]}.
\newblock \bibinfo{title}{{The INSIDE Project: Innovative Solutions for In-Beam
  Dosimetry in Hadrontherapy}}.
\newblock \bibinfo{journal}{APPA}
  \bibinfo{year}{2015};\bibinfo{volume}{127}(\bibinfo{number}{5}):\bibinfo{pages}{1465--1467}.
\newblock \URLprefix
  \url{http://przyrbwn.icm.edu.pl/APP/PDF/127/a127z5p06.pdf}.
  \DOIprefix\doi{10.12693/APhysPolA.127.1465}.
\bibitem[{Marcinkowski et~al.(2016)Marcinkowski, Mollet, {Van Holen} and
  Vandenberghe}]{MarcinkowskiDOI2016}
\bibinfo{author}{Marcinkowski\xfnm[ R.]}, \bibinfo{author}{Mollet\xfnm[ P.]},
  \bibinfo{author}{{Van Holen}\xfnm[ R.]}, \bibinfo{author}{Vandenberghe\xfnm[
  S.]}.
\newblock \bibinfo{title}{{Sub-millimetre DOI detector based on monolithic LYSO
  and digital SiPM for a dedicated small-animal PET system}}.
\newblock \bibinfo{journal}{Phys Med Biol}
  \bibinfo{year}{2016};\bibinfo{volume}{61}(\bibinfo{number}{5}):\bibinfo{pages}{2196--2212}.
\newblock \URLprefix
  \url{https://iopscience.iop.org/article/10.1088/0031-9155/61/5/2196}.
  \DOIprefix\doi{10.1088/0031-9155/61/5/2196}.
\bibitem[{Matej et~al.(2016)Matej, Daube-Witherspoon and Karp}]{Matej2016}
\bibinfo{author}{Matej\xfnm[ S.]}, \bibinfo{author}{Daube-Witherspoon\xfnm[
  M.E.]}, \bibinfo{author}{Karp\xfnm[ J.S.]}.
\newblock \bibinfo{title}{{Analytic TOF PET reconstruction algorithm within
  DIRECT data partitioning framework}}.
\newblock \bibinfo{journal}{Phys Med Biol}
  \bibinfo{year}{2016};\bibinfo{volume}{61}(\bibinfo{number}{9}):\bibinfo{pages}{3365--3386}.
\newblock \URLprefix
  \url{https://iopscience.iop.org/article/10.1088/0031-9155/61/9/3365}.
  \DOIprefix\doi{10.1088/0031-9155/61/9/3365}.
\bibitem[{Matej et~al.(2009)Matej, Surti, Jayanthi, Daube-Witherspoon, Lewitt
  and Karp}]{Matej2009}
\bibinfo{author}{Matej\xfnm[ S.]}, \bibinfo{author}{Surti\xfnm[ S.]},
  \bibinfo{author}{Jayanthi\xfnm[ S.]},
  \bibinfo{author}{Daube-Witherspoon\xfnm[ M.]}, \bibinfo{author}{Lewitt\xfnm[
  R.]}, \bibinfo{author}{Karp\xfnm[ J.]}.
\newblock \bibinfo{title}{{Efficient 3-D TOF PET Reconstruction Using
  View-Grouped Histo-Images: DIRECT -- Direct Image Reconstruction for TOF}}.
\newblock \bibinfo{journal}{IEEE Trans Med Imaging}
  \bibinfo{year}{2009};\bibinfo{volume}{28}(\bibinfo{number}{5}):\bibinfo{pages}{739--751}.
\newblock \URLprefix \url{http://ieeexplore.ieee.org/document/4749317/}.
  \DOIprefix\doi{10.1109/TMI.2008.2012034}.
\bibitem[{Miyata et~al.(2006)Miyata, Tomita, Watanabe, Kawarabayashi and
  Iguchi}]{Miyata2006}
\bibinfo{author}{Miyata\xfnm[ M.]}, \bibinfo{author}{Tomita\xfnm[ H.]},
  \bibinfo{author}{Watanabe\xfnm[ K.]}, \bibinfo{author}{Kawarabayashi\xfnm[
  J.]}, \bibinfo{author}{Iguchi\xfnm[ T.]}.
\newblock \bibinfo{title}{{Development of TOF-PET using Cherenkov Radiation}}.
\newblock \bibinfo{journal}{Journal of Nuclear Science and Technology}
  \bibinfo{year}{2006};\bibinfo{volume}{43}(\bibinfo{number}{4}):\bibinfo{pages}{339--343}.
\newblock \URLprefix
  \url{http://www.tandfonline.com/doi/abs/10.1080/18811248.2006.9711101}.
  \DOIprefix\doi{10.1080/18811248.2006.9711101}.
\bibitem[{Moses(2011)}]{Moses2011}
\bibinfo{author}{Moses\xfnm[ W.W.]}.
\newblock \bibinfo{title}{{Fundamental limits of spatial resolution in PET}}.
\newblock \bibinfo{journal}{Nuclear Instruments and Methods in Physics Research
  Section A: Accelerators, Spectrometers, Detectors and Associated Equipment}
  \bibinfo{year}{2011};\bibinfo{volume}{648}:\bibinfo{pages}{S236--S240}.
\newblock \URLprefix
  \url{https://linkinghub.elsevier.com/retrieve/pii/S0168900210026276}.
  \DOIprefix\doi{10.1016/j.nima.2010.11.092}.
\bibitem[{Moskal(2018)}]{Moskal2018}
\bibinfo{author}{Moskal\xfnm[ P.]}.
\newblock \bibinfo{title}{{J-PET: Towards total-body modular PET from plastic
  scintillators}}.
\newblock In: \bibinfo{booktitle}{Abstracts of the Total Body PET Conference
  2018, EJNMMI Physics}. volume~\bibinfo{volume}{5}; \bibinfo{year}{2018}.
  p.~\bibinfo{pages}{S2}.
\newblock \URLprefix
  \url{https://ejnmmiphys.springeropen.com/articles/10.1186/s40658-018-0218-7}.
  \DOIprefix\doi{10.1186/s40658-018-0218-7}.
\bibitem[{{Moskal} et~al.(2020){Moskal}, {Bednarski}, {Nied{\'{z}}wiecki},
  {Silarski}, {Czerwi{\'{n}}ski}, {Kozik}, {Chhokar}, {Ba{\l}a}, {Curceanu},
  {Del Grande}, {Dadgar}, {Dulski}, {Gajos}, {Gorgol}, {Gupta-Sharma},
  {Hiesmayr}, {Jasi{\'{n}}ska}, {Kacprzak}, {Kap{\l}on}, {Karimi},
  {Kisielewska}, {Klimaszewski}, {Korcyl}, {Kowalski}, {Krawczyk},
  {Krzemie{\'{n}}}, {Kubicz}, {Mohammed}, {Pa{\l}ka},
  {Pawlik-Nied{\'{z}}wiecka}, {Raczy{\'{n}}ski}, {Raj}, {Sharma}, {Shivani},
  {Shopa}, {Skurzok}, {St{\c{e}}pie{\'{n}}}, {Wi{\'{s}}licki} and
  {Zgardzi{\'{n}}ska}}]{Moskal2020a}
\bibinfo{author}{{Moskal}\xfnm[ P.]}, \bibinfo{author}{{Bednarski}\xfnm[ T.]},
  \bibinfo{author}{{Nied{\'{z}}wiecki}\xfnm[ S.]},
  \bibinfo{author}{{Silarski}\xfnm[ M.]},
  \bibinfo{author}{{Czerwi{\'{n}}ski}\xfnm[ E.]},
  \bibinfo{author}{{Kozik}\xfnm[ T.]}, \bibinfo{author}{{Chhokar}\xfnm[ J.]},
  \bibinfo{author}{{Ba{\l}a}\xfnm[ M.]}, \bibinfo{author}{{Curceanu}\xfnm[
  C.]}, \bibinfo{author}{{Del Grande}\xfnm[ R.]},
  \bibinfo{author}{{Dadgar}\xfnm[ M.]}, \bibinfo{author}{{Dulski}\xfnm[ K.]},
  \bibinfo{author}{{Gajos}\xfnm[ A.]}, \bibinfo{author}{{Gorgol}\xfnm[ M.]},
  \bibinfo{author}{{Gupta-Sharma}\xfnm[ N.]}, \bibinfo{author}{{Hiesmayr}\xfnm[
  B.C.]}, \bibinfo{author}{{Jasi{\'{n}}ska}\xfnm[ B.]},
  \bibinfo{author}{{Kacprzak}\xfnm[ K.]}, \bibinfo{author}{{Kap{\l}on}\xfnm[
  {\L}.]}, \bibinfo{author}{{Karimi}\xfnm[ H.]},
  \bibinfo{author}{{Kisielewska}\xfnm[ D.]},
  \bibinfo{author}{{Klimaszewski}\xfnm[ K.]}, \bibinfo{author}{{Korcyl}\xfnm[
  G.]}, \bibinfo{author}{{Kowalski}\xfnm[ P.]},
  \bibinfo{author}{{Krawczyk}\xfnm[ N.]},
  \bibinfo{author}{{Krzemie{\'{n}}}\xfnm[ W.]}, \bibinfo{author}{{Kubicz}\xfnm[
  E.]}, \bibinfo{author}{{Mohammed}\xfnm[ M.]},
  \bibinfo{author}{{Pa{\l}ka}\xfnm[ M.]},
  \bibinfo{author}{{Pawlik-Nied{\'{z}}wiecka}\xfnm[ M.]},
  \bibinfo{author}{{Raczy{\'{n}}ski}\xfnm[ L.]}, \bibinfo{author}{{Raj}\xfnm[
  J.]}, \bibinfo{author}{{Sharma}\xfnm[ S.]}, \bibinfo{author}{{Shivani}\xfnm[
  S.]}, \bibinfo{author}{{Shopa}\xfnm[ R.Y.]}, \bibinfo{author}{{Skurzok}\xfnm[
  M.]}, \bibinfo{author}{{St{\c{e}}pie{\'{n}}}\xfnm[ E.]},
  \bibinfo{author}{{Wi{\'{s}}licki}\xfnm[ W.]},
  \bibinfo{author}{{Zgardzi{\'{n}}ska}\xfnm[ B.]}.
\newblock \bibinfo{title}{{Synchronisation and calibration of the 24-modules
  J-PET prototype with 300 mm axial field of view}}.
\newblock \bibinfo{journal}{IEEE Transactions on Instrumentation and
  Measurement} \bibinfo{year}{2020};:\bibinfo{pages}{1--1}\URLprefix
  \url{https://ieeexplore.ieee.org/document/9173708/}.
  \DOIprefix\doi{10.1109/TIM.2020.3018515}.
\bibitem[{Moskal et~al.(2019{\natexlab{a}})Moskal, Jasi{\'{n}}ska,
  St{\c{e}}pie{\'{n}} and Bass}]{Moskal2019a}
\bibinfo{author}{Moskal\xfnm[ P.]}, \bibinfo{author}{Jasi{\'{n}}ska\xfnm[ B.]},
  \bibinfo{author}{St{\c{e}}pie{\'{n}}\xfnm[ E.{\L}.]},
  \bibinfo{author}{Bass\xfnm[ S.D.]}.
\newblock \bibinfo{title}{{Positronium in medicine and biology}}.
\newblock \bibinfo{journal}{Nature Reviews Physics}
  \bibinfo{year}{2019}{\natexlab{a}};\bibinfo{volume}{1}(\bibinfo{number}{9}):\bibinfo{pages}{527--529}.
\newblock \URLprefix \url{http://www.nature.com/articles/s42254-019-0078-7}.
  \DOIprefix\doi{10.1038/s42254-019-0078-7}.
\bibitem[{Moskal et~al.(2019{\natexlab{b}})Moskal, Kisielewska, Curceanu,
  Czerwi{\'{n}}ski, Dulski, Gajos, Gorgol, Hiesmayr, Jasi{\'{n}}ska, Kacprzak,
  Kap{\l}on, Korcyl, Kowalski, Krzemie{\'{n}}, Kozik, Kubicz, Mohammed,
  Nied{\'{z}}wiecki, Pa{\l}ka, Pawlik-Nied{\'{z}}wiecka, Raczy{\'{n}}ski, Raj,
  Sharma, Shivani, Shopa, Silarski, Skurzok, St{\c{e}}pie{\'{n}},
  Wi{\'{s}}licki and Zgardzi{\'{n}}ska}]{Moskal2019}
\bibinfo{author}{Moskal\xfnm[ P.]}, \bibinfo{author}{Kisielewska\xfnm[ D.]},
  \bibinfo{author}{Curceanu\xfnm[ C.]}, \bibinfo{author}{Czerwi{\'{n}}ski\xfnm[
  E.]}, \bibinfo{author}{Dulski\xfnm[ K.]}, \bibinfo{author}{Gajos\xfnm[ A.]},
  \bibinfo{author}{Gorgol\xfnm[ M.]}, \bibinfo{author}{Hiesmayr\xfnm[ B.]},
  \bibinfo{author}{Jasi{\'{n}}ska\xfnm[ B.]}, \bibinfo{author}{Kacprzak\xfnm[
  K.]}, \bibinfo{author}{Kap{\l}on\xfnm[ {\L}.]}, \bibinfo{author}{Korcyl\xfnm[
  G.]}, \bibinfo{author}{Kowalski\xfnm[ P.]},
  \bibinfo{author}{Krzemie{\'{n}}\xfnm[ W.]}, \bibinfo{author}{Kozik\xfnm[
  T.]}, \bibinfo{author}{Kubicz\xfnm[ E.]}, \bibinfo{author}{Mohammed\xfnm[
  M.]}, \bibinfo{author}{Nied{\'{z}}wiecki\xfnm[ S.]},
  \bibinfo{author}{Pa{\l}ka\xfnm[ M.]},
  \bibinfo{author}{Pawlik-Nied{\'{z}}wiecka\xfnm[ M.]},
  \bibinfo{author}{Raczy{\'{n}}ski\xfnm[ L.]}, \bibinfo{author}{Raj\xfnm[ J.]},
  \bibinfo{author}{Sharma\xfnm[ S.]}, \bibinfo{author}{Shivani\xfnm[]},
  \bibinfo{author}{Shopa\xfnm[ R.Y.]}, \bibinfo{author}{Silarski\xfnm[ M.]},
  \bibinfo{author}{Skurzok\xfnm[ M.]},
  \bibinfo{author}{St{\c{e}}pie{\'{n}}\xfnm[ E.]},
  \bibinfo{author}{Wi{\'{s}}licki\xfnm[ W.]},
  \bibinfo{author}{Zgardzi{\'{n}}ska\xfnm[ B.]}.
\newblock \bibinfo{title}{{Feasibility study of the positronium imaging with
  the J-PET tomograph}}.
\newblock \bibinfo{journal}{Physics in Medicine {\&} Biology}
  \bibinfo{year}{2019}{\natexlab{b}};\bibinfo{volume}{64}(\bibinfo{number}{5}):\bibinfo{pages}{055017}.
\newblock \URLprefix
  \url{https://iopscience.iop.org/article/10.1088/1361-6560/aafe20}.
  \DOIprefix\doi{10.1088/1361-6560/aafe20}.
\bibitem[{Moskal et~al.(2020)Moskal, Kisielewska, Shopa, Bura, Chhokar,
  Curceanu, Czerwi{\'{n}}ski, Dadgar, Dulski, Gajewski, Gajos, Gorgol, {Del
  Grande}, Hiesmayr, Jasi{\'{n}}ska, Kacprzak, Kami{\'{n}}ska, Kap{\l}on,
  Karimi, Korcyl, Kowalski, Krawczyk, Krzemie{\'{n}}, Kozik, Kubicz,
  Ma{\l}czak, Mohammed, Nied{\'{z}}wiecki, Pa{\l}ka, Pawlik-Nied{\'{z}}wiecka,
  P{\c{e}}dziwiatr, Raczy{\'{n}}ski, Raj, Ruci{\'{n}}ski, Sharma, Shivani,
  Silarski, Skurzok, St{\c{e}}pie{\'{n}}, Vandenberghe, Wi{\'{s}}licki and
  Zgardzi{\'{n}}ska}]{Daria2020}
\bibinfo{author}{Moskal\xfnm[ P.]}, \bibinfo{author}{Kisielewska\xfnm[ D.]},
  \bibinfo{author}{Shopa\xfnm[ R.Y.]}, \bibinfo{author}{Bura\xfnm[ Z.]},
  \bibinfo{author}{Chhokar\xfnm[ J.]}, \bibinfo{author}{Curceanu\xfnm[ C.]},
  \bibinfo{author}{Czerwi{\'{n}}ski\xfnm[ E.]}, \bibinfo{author}{Dadgar\xfnm[
  M.]}, \bibinfo{author}{Dulski\xfnm[ K.]}, \bibinfo{author}{Gajewski\xfnm[
  J.]}, \bibinfo{author}{Gajos\xfnm[ A.]}, \bibinfo{author}{Gorgol\xfnm[ M.]},
  \bibinfo{author}{{Del Grande}\xfnm[ R.]}, \bibinfo{author}{Hiesmayr\xfnm[
  B.C.]}, \bibinfo{author}{Jasi{\'{n}}ska\xfnm[ B.]},
  \bibinfo{author}{Kacprzak\xfnm[ K.]}, \bibinfo{author}{Kami{\'{n}}ska\xfnm[
  A.]}, \bibinfo{author}{Kap{\l}on\xfnm[ {\L}.]}, \bibinfo{author}{Karimi\xfnm[
  H.]}, \bibinfo{author}{Korcyl\xfnm[ G.]}, \bibinfo{author}{Kowalski\xfnm[
  P.]}, \bibinfo{author}{Krawczyk\xfnm[ N.]},
  \bibinfo{author}{Krzemie{\'{n}}\xfnm[ W.]}, \bibinfo{author}{Kozik\xfnm[
  T.]}, \bibinfo{author}{Kubicz\xfnm[ E.]}, \bibinfo{author}{Ma{\l}czak\xfnm[
  P.]}, \bibinfo{author}{Mohammed\xfnm[ M.]},
  \bibinfo{author}{Nied{\'{z}}wiecki\xfnm[ S.]},
  \bibinfo{author}{Pa{\l}ka\xfnm[ M.]},
  \bibinfo{author}{Pawlik-Nied{\'{z}}wiecka\xfnm[ M.]},
  \bibinfo{author}{P{\c{e}}dziwiatr\xfnm[ M.]},
  \bibinfo{author}{Raczy{\'{n}}ski\xfnm[ L.]}, \bibinfo{author}{Raj\xfnm[ J.]},
  \bibinfo{author}{Ruci{\'{n}}ski\xfnm[ A.]}, \bibinfo{author}{Sharma\xfnm[
  S.]}, \bibinfo{author}{Shivani\xfnm[ S.]}, \bibinfo{author}{Silarski\xfnm[
  M.]}, \bibinfo{author}{Skurzok\xfnm[ M.]},
  \bibinfo{author}{St{\c{e}}pie{\'{n}}\xfnm[ E.{\L}.]},
  \bibinfo{author}{Vandenberghe\xfnm[ S.]},
  \bibinfo{author}{Wi{\'{s}}licki\xfnm[ W.]},
  \bibinfo{author}{Zgardzi{\'{n}}ska\xfnm[ B.]}.
\newblock \bibinfo{title}{{Performance assessment of the 2 $\gamma$positronium
  imaging with the total-body PET scanners}}.
\newblock \bibinfo{journal}{EJNMMI Phys}
  \bibinfo{year}{2020};\bibinfo{volume}{7}(\bibinfo{number}{44}):\bibinfo{pages}{1}.
\newblock \URLprefix \url{http://arxiv.org/abs/1911.06841
  https://ejnmmiphys.springeropen.com/articles/10.1186/s40658-020-00307-w}.
  \DOIprefix\doi{10.1186/s40658-020-00307-w}.
\bibitem[{Moskal et~al.(2014)Moskal, Nied{\'{z}}wiecki, Bednarski,
  Czerwi{\'{n}}ski, Kap{\l}on, Kubicz, Moskal, Pawlik-Nied{\'{z}}wiecka,
  Sharma, Silarski, Zieli{\'{n}}ski, Zo{\'{n}}, Bia{\l}as, Gajos, Kochanowski,
  Korcyl, Kowal, Kowalski, Kozik, Krzemie{\'{n}}, Molenda, Pa{\l}ka,
  Raczy{\'{n}}ski, Rudy, Salabura, S{\l}omski, Smyrski, Strzelecki, Wieczorek
  and Wi{\'{s}}licki}]{Moskal2014}
\bibinfo{author}{Moskal\xfnm[ P.]}, \bibinfo{author}{Nied{\'{z}}wiecki\xfnm[
  S.]}, \bibinfo{author}{Bednarski\xfnm[ T.]},
  \bibinfo{author}{Czerwi{\'{n}}ski\xfnm[ E.]},
  \bibinfo{author}{Kap{\l}on\xfnm[ {\L}.]}, \bibinfo{author}{Kubicz\xfnm[ E.]},
  \bibinfo{author}{Moskal\xfnm[ I.]},
  \bibinfo{author}{Pawlik-Nied{\'{z}}wiecka\xfnm[ M.]},
  \bibinfo{author}{Sharma\xfnm[ N.G.]}, \bibinfo{author}{Silarski\xfnm[ M.]},
  \bibinfo{author}{Zieli{\'{n}}ski\xfnm[ M.]}, \bibinfo{author}{Zo{\'{n}}\xfnm[
  N.]}, \bibinfo{author}{Bia{\l}as\xfnm[ P.]}, \bibinfo{author}{Gajos\xfnm[
  A.]}, \bibinfo{author}{Kochanowski\xfnm[ A.]}, \bibinfo{author}{Korcyl\xfnm[
  G.]}, \bibinfo{author}{Kowal\xfnm[ J.]}, \bibinfo{author}{Kowalski\xfnm[
  P.]}, \bibinfo{author}{Kozik\xfnm[ T.]},
  \bibinfo{author}{Krzemie{\'{n}}\xfnm[ W.]}, \bibinfo{author}{Molenda\xfnm[
  M.]}, \bibinfo{author}{Pa{\l}ka\xfnm[ M.]},
  \bibinfo{author}{Raczy{\'{n}}ski\xfnm[ L.]}, \bibinfo{author}{Rudy\xfnm[
  Z.]}, \bibinfo{author}{Salabura\xfnm[ P.]}, \bibinfo{author}{S{\l}omski\xfnm[
  A.]}, \bibinfo{author}{Smyrski\xfnm[ J.]}, \bibinfo{author}{Strzelecki\xfnm[
  A.]}, \bibinfo{author}{Wieczorek\xfnm[ A.]},
  \bibinfo{author}{Wi{\'{s}}licki\xfnm[ W.]}.
\newblock \bibinfo{title}{{Test of a single module of the J-PET scanner based
  on plastic scintillators}}.
\newblock \bibinfo{journal}{Nucl Instrum Methods Phys Res A}
  \bibinfo{year}{2014};\bibinfo{volume}{764}:\bibinfo{pages}{317--321}.
\newblock \URLprefix
  \url{https://linkinghub.elsevier.com/retrieve/pii/S0168900214009103}.
  \DOIprefix\doi{10.1016/j.nima.2014.07.052}.
\bibitem[{Moskal et~al.(2016)Moskal, Rundel, Alfs, Bednarski, Bia{\l}as,
  Czerwi{\'{n}}ski, Gajos, Giergiel, Gorgol, Jasi{\'{n}}ska, Kami{\'{n}}ska,
  Kap{\l}on, Korcyl, Kowalski, Kozik, Krzemie{\'{n}}, Kubicz,
  Nied{\'{z}}wiecki, Pa{\l}ka, Raczy{\'{n}}ski, Rudy, Sharma, S{\l}omski,
  Silarski, Strzelecki, Wieczorek, Wi{\'{s}}licki, Witkowski, Zieli{\'{n}}ski
  and Zo{\'{n}}}]{Moskal2016}
\bibinfo{author}{Moskal\xfnm[ P.]}, \bibinfo{author}{Rundel\xfnm[ O.]},
  \bibinfo{author}{Alfs\xfnm[ D.]}, \bibinfo{author}{Bednarski\xfnm[ T.]},
  \bibinfo{author}{Bia{\l}as\xfnm[ P.]},
  \bibinfo{author}{Czerwi{\'{n}}ski\xfnm[ E.]}, \bibinfo{author}{Gajos\xfnm[
  A.]}, \bibinfo{author}{Giergiel\xfnm[ K.]}, \bibinfo{author}{Gorgol\xfnm[
  M.]}, \bibinfo{author}{Jasi{\'{n}}ska\xfnm[ B.]},
  \bibinfo{author}{Kami{\'{n}}ska\xfnm[ D.]}, \bibinfo{author}{Kap{\l}on\xfnm[
  {\L}.]}, \bibinfo{author}{Korcyl\xfnm[ G.]}, \bibinfo{author}{Kowalski\xfnm[
  P.]}, \bibinfo{author}{Kozik\xfnm[ T.]},
  \bibinfo{author}{Krzemie{\'{n}}\xfnm[ W.]}, \bibinfo{author}{Kubicz\xfnm[
  E.]}, \bibinfo{author}{Nied{\'{z}}wiecki\xfnm[ S.]},
  \bibinfo{author}{Pa{\l}ka\xfnm[ M.]}, \bibinfo{author}{Raczy{\'{n}}ski\xfnm[
  L.]}, \bibinfo{author}{Rudy\xfnm[ Z.]}, \bibinfo{author}{Sharma\xfnm[ N.G.]},
  \bibinfo{author}{S{\l}omski\xfnm[ A.]}, \bibinfo{author}{Silarski\xfnm[ M.]},
  \bibinfo{author}{Strzelecki\xfnm[ A.]}, \bibinfo{author}{Wieczorek\xfnm[
  A.]}, \bibinfo{author}{Wi{\'{s}}licki\xfnm[ W.]},
  \bibinfo{author}{Witkowski\xfnm[ P.]}, \bibinfo{author}{Zieli{\'{n}}ski\xfnm[
  M.]}, \bibinfo{author}{Zo{\'{n}}\xfnm[ N.]}.
\newblock \bibinfo{title}{{Time resolution of the plastic scintillator strips
  with matrix photomultiplier readout for J-PET tomograph}}.
\newblock \bibinfo{journal}{Phys Med Biol}
  \bibinfo{year}{2016};\bibinfo{volume}{61}(\bibinfo{number}{5}):\bibinfo{pages}{2025--2047}.
\newblock \URLprefix
  \url{https://iopscience.iop.org/article/10.1088/0031-9155/61/5/2025}.
  \DOIprefix\doi{10.1088/0031-9155/61/5/2025}.
\bibitem[{Moskal and Smyrski(2013)}]{MoskalPatent2013}
\bibinfo{author}{Moskal\xfnm[ P.]}, \bibinfo{author}{Smyrski\xfnm[ J.]}.
\newblock \bibinfo{title}{{Urz{\c{a}}dzenie detekcyjne do wyznaczania miejsca
  reakcji kwant{\'{o}}w gamma oraz spos{\'{o}}b wyznaczania reakcji
  kwant{\'{o}}w gamma w emisyjnej tomografii pozytonowej, Polish Patent}}.
\newblock \bibinfo{year}{2013}.
\bibitem[{Moskal and St{\c{e}}pie{\'{n}}(2020)}]{Moskal2020}
\bibinfo{author}{Moskal\xfnm[ P.]}, \bibinfo{author}{St{\c{e}}pie{\'{n}}\xfnm[
  E.{\L}.]}.
\newblock \bibinfo{title}{{Prospects and Clinical Perspectives of Total-Body
  PET Imaging Using Plastic Scintillators}}.
\newblock \bibinfo{journal}{PET Clin}
  \bibinfo{year}{2020};\bibinfo{volume}{15}(\bibinfo{number}{4}):\bibinfo{pages}{439--452}.
\newblock \URLprefix
  \url{https://linkinghub.elsevier.com/retrieve/pii/S155685982030047X}.
  \DOIprefix\doi{10.1016/j.cpet.2020.06.009}.
\bibitem[{Moskal et~al.(2015)Moskal, Zo{\'{n}}, Bednarski, Bia{\l}as,
  Czerwi{\'{n}}ski, Gajos, Kami{\'{n}}ska, Kap{\l}on, Kochanowski, Korcyl,
  Kowal, Kowalski, Kozik, Krzemie{\'{n}}, Kubicz, Nied{\'{z}}wiecki, Pa{\l}ka,
  Raczy{\'{n}}ski, Rudy, Rundel, Salabura, Sharma, Silarski, S{\l}omski,
  Smyrski, Strzelecki, Wieczorek, Wi{\'{s}}licki and
  Zieli{\'{n}}ski}]{Moskal2015}
\bibinfo{author}{Moskal\xfnm[ P.]}, \bibinfo{author}{Zo{\'{n}}\xfnm[ N.]},
  \bibinfo{author}{Bednarski\xfnm[ T.]}, \bibinfo{author}{Bia{\l}as\xfnm[ P.]},
  \bibinfo{author}{Czerwi{\'{n}}ski\xfnm[ E.]}, \bibinfo{author}{Gajos\xfnm[
  A.]}, \bibinfo{author}{Kami{\'{n}}ska\xfnm[ D.]},
  \bibinfo{author}{Kap{\l}on\xfnm[ {\L}.]}, \bibinfo{author}{Kochanowski\xfnm[
  A.]}, \bibinfo{author}{Korcyl\xfnm[ G.]}, \bibinfo{author}{Kowal\xfnm[ J.]},
  \bibinfo{author}{Kowalski\xfnm[ P.]}, \bibinfo{author}{Kozik\xfnm[ T.]},
  \bibinfo{author}{Krzemie{\'{n}}\xfnm[ W.]}, \bibinfo{author}{Kubicz\xfnm[
  E.]}, \bibinfo{author}{Nied{\'{z}}wiecki\xfnm[ S.]},
  \bibinfo{author}{Pa{\l}ka\xfnm[ M.]}, \bibinfo{author}{Raczy{\'{n}}ski\xfnm[
  L.]}, \bibinfo{author}{Rudy\xfnm[ Z.]}, \bibinfo{author}{Rundel\xfnm[ O.]},
  \bibinfo{author}{Salabura\xfnm[ P.]}, \bibinfo{author}{Sharma\xfnm[ N.G.]},
  \bibinfo{author}{Silarski\xfnm[ M.]}, \bibinfo{author}{S{\l}omski\xfnm[ A.]},
  \bibinfo{author}{Smyrski\xfnm[ J.]}, \bibinfo{author}{Strzelecki\xfnm[ A.]},
  \bibinfo{author}{Wieczorek\xfnm[ A.]}, \bibinfo{author}{Wi{\'{s}}licki\xfnm[
  W.]}, \bibinfo{author}{Zieli{\'{n}}ski\xfnm[ M.]}.
\newblock \bibinfo{title}{{A novel method for the line-of-response and
  time-of-flight reconstruction in TOF-PET detectors based on a library of
  synchronized model signals}}.
\newblock \bibinfo{journal}{Nucl Instrum Methods Phys Res A}
  \bibinfo{year}{2015};\bibinfo{volume}{775}:\bibinfo{pages}{54--62}.
\newblock \URLprefix
  \url{https://linkinghub.elsevier.com/retrieve/pii/S0168900214014466}.
  \DOIprefix\doi{10.1016/j.nima.2014.12.005}.
\bibitem[{Mullani et~al.(1980)Mullani, Markham and Ter-Pogossian}]{Mullani1980}
\bibinfo{author}{Mullani\xfnm[ N.A.]}, \bibinfo{author}{Markham\xfnm[ J.]},
  \bibinfo{author}{Ter-Pogossian\xfnm[ M.M.]}.
\newblock \bibinfo{title}{{Feasibility of time-of-flight reconstruction in
  positron emission tomography}}.
\newblock \bibinfo{journal}{J Nucl Med}
  \bibinfo{year}{1980};\bibinfo{volume}{21}(\bibinfo{number}{11}):\bibinfo{pages}{1095--1097}.
\newblock \URLprefix \url{http://www.ncbi.nlm.nih.gov/pubmed/6968822}.
\bibitem[{Nelder and Mead(1965)}]{Nelder1965}
\bibinfo{author}{Nelder\xfnm[ J.A.]}, \bibinfo{author}{Mead\xfnm[ R.]}.
\newblock \bibinfo{title}{{A Simplex Method for Function Minimization}}.
\newblock \bibinfo{journal}{The Computer Journal}
  \bibinfo{year}{1965};\bibinfo{volume}{7}(\bibinfo{number}{4}):\bibinfo{pages}{308--313}.
\newblock \DOIprefix\doi{10.1093/comjnl/7.4.308}.
\bibitem[{NEMA(2013)}]{NEMA2012}
\bibinfo{author}{NEMA\xfnm[]}.
\newblock \bibinfo{title}{Performance measurements of positron emission
  tomographs, nema nu 2-2012 standard}.
\newblock \bibinfo{year}{2013}.
\bibitem[{Nemallapudi et~al.(2015)Nemallapudi, Gundacker, Lecoq, Auffray,
  Ferri, Gola and Piemonte}]{Nemallapudi2015}
\bibinfo{author}{Nemallapudi\xfnm[ M.V.]}, \bibinfo{author}{Gundacker\xfnm[
  S.]}, \bibinfo{author}{Lecoq\xfnm[ P.]}, \bibinfo{author}{Auffray\xfnm[ E.]},
  \bibinfo{author}{Ferri\xfnm[ A.]}, \bibinfo{author}{Gola\xfnm[ A.]},
  \bibinfo{author}{Piemonte\xfnm[ C.]}.
\newblock \bibinfo{title}{{Sub-100 ps coincidence time resolution for positron
  emission tomography with LSO:Ce codoped with Ca.}}
\newblock \bibinfo{journal}{Phys Med Biol}
  \bibinfo{year}{2015};\bibinfo{volume}{60}(\bibinfo{number}{12}):\bibinfo{pages}{4635--4649}.
\newblock \URLprefix \url{http://www.ncbi.nlm.nih.gov/pubmed/26020610}.
  \DOIprefix\doi{10.1088/0031-9155/60/12/4635}.
\bibitem[{Nied{\'{z}}wiecki et~al.(2017)Nied{\'{z}}wiecki, Bia{\l}as, Curceanu,
  Czerwi{\'{n}}ski, Dulski, Gajos, G{\l}owacz, Gorgol, Hiesmayr,
  Jasi{\'{n}}ska, Kap{\l}on, Kisielewska-Kami{\'{n}}ska, Korcyl, Kowalski,
  Kozik, Krawczyk, Krzemie{\'{n}}, Kubicz, Mohammed, Pawlik-Nied{\'{z}}wiecka,
  Pa{\l}ka, Raczy{\'{n}}ski, Rudy, Sharma, Sharma, Shopa, Silarski, Skurzok,
  Wieczorek, Wi{\'{s}}licki, Zgardzi{\'{n}}ska, Zieli{\'{n}}ski and
  Moskal}]{Niedzwiecki2017}
\bibinfo{author}{Nied{\'{z}}wiecki\xfnm[ S.]}, \bibinfo{author}{Bia{\l}as\xfnm[
  P.]}, \bibinfo{author}{Curceanu\xfnm[ C.]},
  \bibinfo{author}{Czerwi{\'{n}}ski\xfnm[ E.]}, \bibinfo{author}{Dulski\xfnm[
  K.]}, \bibinfo{author}{Gajos\xfnm[ A.]}, \bibinfo{author}{G{\l}owacz\xfnm[
  B.]}, \bibinfo{author}{Gorgol\xfnm[ M.]}, \bibinfo{author}{Hiesmayr\xfnm[
  B.C.]}, \bibinfo{author}{Jasi{\'{n}}ska\xfnm[ B.]},
  \bibinfo{author}{Kap{\l}on\xfnm[ {\L}.]},
  \bibinfo{author}{Kisielewska-Kami{\'{n}}ska\xfnm[ D.]},
  \bibinfo{author}{Korcyl\xfnm[ G.]}, \bibinfo{author}{Kowalski\xfnm[ P.]},
  \bibinfo{author}{Kozik\xfnm[ T.]}, \bibinfo{author}{Krawczyk\xfnm[ N.]},
  \bibinfo{author}{Krzemie{\'{n}}\xfnm[ W.]}, \bibinfo{author}{Kubicz\xfnm[
  E.]}, \bibinfo{author}{Mohammed\xfnm[ M.]},
  \bibinfo{author}{Pawlik-Nied{\'{z}}wiecka\xfnm[ M.]},
  \bibinfo{author}{Pa{\l}ka\xfnm[ M.]}, \bibinfo{author}{Raczy{\'{n}}ski\xfnm[
  L.]}, \bibinfo{author}{Rudy\xfnm[ Z.]}, \bibinfo{author}{Sharma\xfnm[ N.G.]},
  \bibinfo{author}{Sharma\xfnm[ S.]}, \bibinfo{author}{Shopa\xfnm[ R.Y.]},
  \bibinfo{author}{Silarski\xfnm[ M.]}, \bibinfo{author}{Skurzok\xfnm[ M.]},
  \bibinfo{author}{Wieczorek\xfnm[ A.]}, \bibinfo{author}{Wi{\'{s}}licki\xfnm[
  W.]}, \bibinfo{author}{Zgardzi{\'{n}}ska\xfnm[ B.]},
  \bibinfo{author}{Zieli{\'{n}}ski\xfnm[ M.]}, \bibinfo{author}{Moskal\xfnm[
  P.]}.
\newblock \bibinfo{title}{{J-PET: A New Technology for the Whole-body PET
  Imaging}}.
\newblock \bibinfo{journal}{APPB}
  \bibinfo{year}{2017};\bibinfo{volume}{48}(\bibinfo{number}{10}):\bibinfo{pages}{1567}.
\newblock \URLprefix
  \url{http://www.actaphys.uj.edu.pl/findarticle?series=Reg\&vol=48\&page=1567}.
  \DOIprefix\doi{10.5506/APhysPolB.48.1567}.
\bibitem[{Nuyts et~al.(2011)Nuyts, Stute, {Van Slambrouck}, van Velden,
  Boellaard and Comtat}]{Nuyts2011}
\bibinfo{author}{Nuyts\xfnm[ J.]}, \bibinfo{author}{Stute\xfnm[ S.]},
  \bibinfo{author}{{Van Slambrouck}\xfnm[ K.]}, \bibinfo{author}{van
  Velden\xfnm[ F.]}, \bibinfo{author}{Boellaard\xfnm[ R.]},
  \bibinfo{author}{Comtat\xfnm[ C.]}.
\newblock \bibinfo{title}{{Maximum-likelihood reconstruction based on a
  modified Poisson distribution to reduce bias in PET}}.
\newblock In: \bibinfo{booktitle}{2011 IEEE Nucl. Sci. Symp. Conf. Rec.}
  \bibinfo{publisher}{IEEE}; \bibinfo{year}{2011}. p.
  \bibinfo{pages}{4337--4341}.
\newblock \URLprefix \url{http://ieeexplore.ieee.org/document/6153835/}.
  \DOIprefix\doi{10.1109/NSSMIC.2011.6153835}.
\bibitem[{Pa{\l}ka et~al.(2017)Pa{\l}ka, Strzempek, Korcyl, Bednarski,
  Nied{\'{z}}wiecki, Bia{\l}as, Czerwi{\'{n}}ski, Dulski, Gajos, G{\l}owacz,
  Gorgol, Jasi{\'{n}}ska, Kami{\'{n}}ska, Kajetanowicz, Kowalski, Kozik,
  Krzemie{\'{n}}, Kubicz, Mohhamed, Raczy{\'{n}}ski, Rudy, Rundel, Salabura,
  Sharma, Silarski, Smyrski, Strzelecki, Wieczorek, Wi{\'{s}}licki,
  Zieli{\'{n}}ski, Zgardzi{\'{n}}ska and Moskal}]{Palka2017}
\bibinfo{author}{Pa{\l}ka\xfnm[ M.]}, \bibinfo{author}{Strzempek\xfnm[ P.]},
  \bibinfo{author}{Korcyl\xfnm[ G.]}, \bibinfo{author}{Bednarski\xfnm[ T.]},
  \bibinfo{author}{Nied{\'{z}}wiecki\xfnm[ S.]},
  \bibinfo{author}{Bia{\l}as\xfnm[ P.]},
  \bibinfo{author}{Czerwi{\'{n}}ski\xfnm[ E.]}, \bibinfo{author}{Dulski\xfnm[
  K.]}, \bibinfo{author}{Gajos\xfnm[ A.]}, \bibinfo{author}{G{\l}owacz\xfnm[
  B.]}, \bibinfo{author}{Gorgol\xfnm[ M.]},
  \bibinfo{author}{Jasi{\'{n}}ska\xfnm[ B.]},
  \bibinfo{author}{Kami{\'{n}}ska\xfnm[ D.]},
  \bibinfo{author}{Kajetanowicz\xfnm[ M.]}, \bibinfo{author}{Kowalski\xfnm[
  P.]}, \bibinfo{author}{Kozik\xfnm[ T.]},
  \bibinfo{author}{Krzemie{\'{n}}\xfnm[ W.]}, \bibinfo{author}{Kubicz\xfnm[
  E.]}, \bibinfo{author}{Mohhamed\xfnm[ M.]},
  \bibinfo{author}{Raczy{\'{n}}ski\xfnm[ L.]}, \bibinfo{author}{Rudy\xfnm[
  Z.]}, \bibinfo{author}{Rundel\xfnm[ O.]}, \bibinfo{author}{Salabura\xfnm[
  P.]}, \bibinfo{author}{Sharma\xfnm[ N.G.]}, \bibinfo{author}{Silarski\xfnm[
  M.]}, \bibinfo{author}{Smyrski\xfnm[ J.]}, \bibinfo{author}{Strzelecki\xfnm[
  A.]}, \bibinfo{author}{Wieczorek\xfnm[ A.]},
  \bibinfo{author}{Wi{\'{s}}licki\xfnm[ W.]},
  \bibinfo{author}{Zieli{\'{n}}ski\xfnm[ M.]},
  \bibinfo{author}{Zgardzi{\'{n}}ska\xfnm[ B.]}, \bibinfo{author}{Moskal\xfnm[
  P.]}.
\newblock \bibinfo{title}{{Multichannel FPGA based MVT system for high
  precision time (20 ps RMS) and charge measurement}}.
\newblock \bibinfo{journal}{JINST}
  \bibinfo{year}{2017};\bibinfo{volume}{12}(\bibinfo{number}{08}):\bibinfo{pages}{P08001--P08001}.
\newblock \URLprefix
  \url{https://iopscience.iop.org/article/10.1088/1748-0221/12/08/P08001}.
  \DOIprefix\doi{10.1088/1748-0221/12/08/P08001}.
\bibitem[{Pawlik-Nied{\'{z}}wiecka et~al.(2017)Pawlik-Nied{\'{z}}wiecka,
  Nied{\'{z}}wiecki, Alfs, Bia{\l}as, Curceanu, Czerwi{\'{n}}ski, Dulski,
  Gajos, G{\l}owacz, Gorgol, Hiesmayr, Jasi{\'{n}}ska, Kisielewska, Korcyl,
  Kowalski, Kozik, Krawczyk, Krzemie{\'{n}}, Kubicz, Mohammed, Pa{\l}ka,
  Raczy{\'{n}}ski, Raj, Rudy, Shivani, Silarski, Skurzok, Sharma, Sharma,
  Shopa, Strzelecki, Wieczorek, Wi{\'{s}}licki, Zgardzi{\'{n}}ska,
  Zieli{\'{n}}ski and Moskal}]{Pawlik-Niedzwiecka2017}
\bibinfo{author}{Pawlik-Nied{\'{z}}wiecka\xfnm[ M.]},
  \bibinfo{author}{Nied{\'{z}}wiecki\xfnm[ S.]}, \bibinfo{author}{Alfs\xfnm[
  D.]}, \bibinfo{author}{Bia{\l}as\xfnm[ P.]}, \bibinfo{author}{Curceanu\xfnm[
  C.]}, \bibinfo{author}{Czerwi{\'{n}}ski\xfnm[ E.]},
  \bibinfo{author}{Dulski\xfnm[ K.]}, \bibinfo{author}{Gajos\xfnm[ A.]},
  \bibinfo{author}{G{\l}owacz\xfnm[ B.]}, \bibinfo{author}{Gorgol\xfnm[ M.]},
  \bibinfo{author}{Hiesmayr\xfnm[ B.C.]}, \bibinfo{author}{Jasi{\'{n}}ska\xfnm[
  B.]}, \bibinfo{author}{Kisielewska\xfnm[ D.]}, \bibinfo{author}{Korcyl\xfnm[
  G.]}, \bibinfo{author}{Kowalski\xfnm[ P.]}, \bibinfo{author}{Kozik\xfnm[
  T.]}, \bibinfo{author}{Krawczyk\xfnm[ N.]},
  \bibinfo{author}{Krzemie{\'{n}}\xfnm[ W.]}, \bibinfo{author}{Kubicz\xfnm[
  E.]}, \bibinfo{author}{Mohammed\xfnm[ M.]}, \bibinfo{author}{Pa{\l}ka\xfnm[
  M.]}, \bibinfo{author}{Raczy{\'{n}}ski\xfnm[ L.]}, \bibinfo{author}{Raj\xfnm[
  J.]}, \bibinfo{author}{Rudy\xfnm[ Z.]}, \bibinfo{author}{Shivani\xfnm[]},
  \bibinfo{author}{Silarski\xfnm[ M.]}, \bibinfo{author}{Skurzok\xfnm[ M.]},
  \bibinfo{author}{Sharma\xfnm[ N.G.]}, \bibinfo{author}{Sharma\xfnm[ S.]},
  \bibinfo{author}{Shopa\xfnm[ R.Y.]}, \bibinfo{author}{Strzelecki\xfnm[ A.]},
  \bibinfo{author}{Wieczorek\xfnm[ A.]}, \bibinfo{author}{Wi{\'{s}}licki\xfnm[
  W.]}, \bibinfo{author}{Zgardzi{\'{n}}ska\xfnm[ B.]},
  \bibinfo{author}{Zieli{\'{n}}ski\xfnm[ M.]}, \bibinfo{author}{Moskal\xfnm[
  P.]}.
\newblock \bibinfo{title}{{Preliminary Studies of J-PET Detector Spatial
  Resolution}}.
\newblock \bibinfo{journal}{APPA}
  \bibinfo{year}{2017};\bibinfo{volume}{132}(\bibinfo{number}{5}):\bibinfo{pages}{1645--1649}.
\newblock \URLprefix
  \url{http://przyrbwn.icm.edu.pl/APP/PDF/132/app132z5p47.pdf}.
  \DOIprefix\doi{10.12693/APhysPolA.132.1645}.
\bibitem[{{R Core Team}(2020)}]{RCoreTeam2020}
\bibinfo{author}{{R Core Team}\xfnm[]}.
\newblock \bibinfo{title}{{R: A Language and Environment for Statistical
  Computing}}.
\newblock \bibinfo{year}{2020}.
\newblock \URLprefix \url{https://www.r-project.org/}.
\bibitem[{Raczy{\'{n}}ski et~al.(2015)Raczy{\'{n}}ski, Moskal, Kowalski,
  Wi{\'{s}}licki, Bednarski, Bia{\l}as, Czerwi{\'{n}}ski, Gajos, Kap{\l}on,
  Kochanowski, Korcyl, Kowal, Kozik, Krzemie{\'{n}}, Kubicz, Nied{\'{z}}wiecki,
  Pa{\l}ka, Rudy, Rundel, Salabura, Sharma, Silarski, S{\l}omski, Smyrski,
  Strzelecki, Wieczorek, Zieli{\'{n}}ski and Zo{\'{n}}}]{Raczynski2015}
\bibinfo{author}{Raczy{\'{n}}ski\xfnm[ L.]}, \bibinfo{author}{Moskal\xfnm[
  P.]}, \bibinfo{author}{Kowalski\xfnm[ P.]},
  \bibinfo{author}{Wi{\'{s}}licki\xfnm[ W.]}, \bibinfo{author}{Bednarski\xfnm[
  T.]}, \bibinfo{author}{Bia{\l}as\xfnm[ P.]},
  \bibinfo{author}{Czerwi{\'{n}}ski\xfnm[ E.]}, \bibinfo{author}{Gajos\xfnm[
  A.]}, \bibinfo{author}{Kap{\l}on\xfnm[ {\L}.]},
  \bibinfo{author}{Kochanowski\xfnm[ A.]}, \bibinfo{author}{Korcyl\xfnm[ G.]},
  \bibinfo{author}{Kowal\xfnm[ J.]}, \bibinfo{author}{Kozik\xfnm[ T.]},
  \bibinfo{author}{Krzemie{\'{n}}\xfnm[ W.]}, \bibinfo{author}{Kubicz\xfnm[
  E.]}, \bibinfo{author}{Nied{\'{z}}wiecki\xfnm[ S.]},
  \bibinfo{author}{Pa{\l}ka\xfnm[ M.]}, \bibinfo{author}{Rudy\xfnm[ Z.]},
  \bibinfo{author}{Rundel\xfnm[ O.]}, \bibinfo{author}{Salabura\xfnm[ P.]},
  \bibinfo{author}{Sharma\xfnm[ N.G.]}, \bibinfo{author}{Silarski\xfnm[ M.]},
  \bibinfo{author}{S{\l}omski\xfnm[ A.]}, \bibinfo{author}{Smyrski\xfnm[ J.]},
  \bibinfo{author}{Strzelecki\xfnm[ A.]}, \bibinfo{author}{Wieczorek\xfnm[
  A.]}, \bibinfo{author}{Zieli{\'{n}}ski\xfnm[ M.]},
  \bibinfo{author}{Zo{\'{n}}\xfnm[ N.]}.
\newblock \bibinfo{title}{{Compressive sensing of signals generated in plastic
  scintillators in a novel J-PET instrument}}.
\newblock \bibinfo{journal}{Nucl Instrum Methods Phys Res A}
  \bibinfo{year}{2015};\bibinfo{volume}{786}:\bibinfo{pages}{105--112}.
\newblock \URLprefix
  \url{https://linkinghub.elsevier.com/retrieve/pii/S0168900215003484}.
  \DOIprefix\doi{10.1016/j.nima.2015.03.032}.
\bibitem[{Raczy{\'{n}}ski et~al.(2014)Raczy{\'{n}}ski, Moskal, Kowalski,
  Wi{\'{s}}licki, Bednarski, Bia{\l}as, Czerwi{\'{n}}ski, Kap{\l}on,
  Kochanowski, Korcyl, Kowal, Kozik, Krzemie{\'{n}}, Kubicz, Molenda, Moskal,
  Nied{\'{z}}wiecki, Pa{\l}ka, Pawlik-Nied{\'{z}}wiecka, Rudy, Salabura,
  Sharma, Silarski, S{\l}omski, Smyrski, Strzelecki, Wieczorek, Zieli{\'{n}}ski
  and Zo{\'{n}}}]{Raczynski2014}
\bibinfo{author}{Raczy{\'{n}}ski\xfnm[ L.]}, \bibinfo{author}{Moskal\xfnm[
  P.]}, \bibinfo{author}{Kowalski\xfnm[ P.]},
  \bibinfo{author}{Wi{\'{s}}licki\xfnm[ W.]}, \bibinfo{author}{Bednarski\xfnm[
  T.]}, \bibinfo{author}{Bia{\l}as\xfnm[ P.]},
  \bibinfo{author}{Czerwi{\'{n}}ski\xfnm[ E.]},
  \bibinfo{author}{Kap{\l}on\xfnm[ {\L}.]}, \bibinfo{author}{Kochanowski\xfnm[
  A.]}, \bibinfo{author}{Korcyl\xfnm[ G.]}, \bibinfo{author}{Kowal\xfnm[ J.]},
  \bibinfo{author}{Kozik\xfnm[ T.]}, \bibinfo{author}{Krzemie{\'{n}}\xfnm[
  W.]}, \bibinfo{author}{Kubicz\xfnm[ E.]}, \bibinfo{author}{Molenda\xfnm[
  M.]}, \bibinfo{author}{Moskal\xfnm[ I.]},
  \bibinfo{author}{Nied{\'{z}}wiecki\xfnm[ S.]},
  \bibinfo{author}{Pa{\l}ka\xfnm[ M.]},
  \bibinfo{author}{Pawlik-Nied{\'{z}}wiecka\xfnm[ M.]},
  \bibinfo{author}{Rudy\xfnm[ Z.]}, \bibinfo{author}{Salabura\xfnm[ P.]},
  \bibinfo{author}{Sharma\xfnm[ N.G.]}, \bibinfo{author}{Silarski\xfnm[ M.]},
  \bibinfo{author}{S{\l}omski\xfnm[ A.]}, \bibinfo{author}{Smyrski\xfnm[ J.]},
  \bibinfo{author}{Strzelecki\xfnm[ A.]}, \bibinfo{author}{Wieczorek\xfnm[
  A.]}, \bibinfo{author}{Zieli{\'{n}}ski\xfnm[ M.]},
  \bibinfo{author}{Zo{\'{n}}\xfnm[ N.]}.
\newblock \bibinfo{title}{{Novel method for hit-position reconstruction using
  voltage signals in plastic scintillators and its application to Positron
  Emission Tomography}}.
\newblock \bibinfo{journal}{Nucl Instrum Methods Phys Res A}
  \bibinfo{year}{2014};\bibinfo{volume}{764}:\bibinfo{pages}{186--192}.
\newblock \URLprefix
  \url{https://linkinghub.elsevier.com/retrieve/pii/S0168900214008754}.
  \DOIprefix\doi{10.1016/j.nima.2014.07.032}.
\bibitem[{Raczy{\'{n}}ski et~al.(2020)Raczy{\'{n}}ski, Wi{\'{s}}licki,
  Klimaszewski, Krzemie{\'{n}}, Kopka, Kowalski, Shopa, Ba{\l}a, Chhokar,
  Curceanu, Czerwi{\'{n}}ski, Dulski, Gajewski, Gajos, Gorgol, {Del Grande},
  Hiesmayr, Jasi{\'{n}}ska, Kacprzak, Kap{\l}on, Kisielewska, Korcyl, Kozik,
  Krawczyk, Kubicz, Mohammed, Nied{\'{z}}wiecki, Pa{\l}ka,
  Pawlik-Nied{\'{z}}wiecka, Raj, Rakoczy, Ruci{\'{n}}ski, Sharma, Shivani,
  Silarski, Skurzok, Stepie{\'{n}}, Zgardzi{\'{n}}ska and
  Moskal}]{Raczynski2020}
\bibinfo{author}{Raczy{\'{n}}ski\xfnm[ L.]},
  \bibinfo{author}{Wi{\'{s}}licki\xfnm[ W.]},
  \bibinfo{author}{Klimaszewski\xfnm[ K.]},
  \bibinfo{author}{Krzemie{\'{n}}\xfnm[ W.]}, \bibinfo{author}{Kopka\xfnm[
  P.]}, \bibinfo{author}{Kowalski\xfnm[ P.]}, \bibinfo{author}{Shopa\xfnm[
  R.]}, \bibinfo{author}{Ba{\l}a\xfnm[ M.]}, \bibinfo{author}{Chhokar\xfnm[
  J.]}, \bibinfo{author}{Curceanu\xfnm[ C.]},
  \bibinfo{author}{Czerwi{\'{n}}ski\xfnm[ E.]}, \bibinfo{author}{Dulski\xfnm[
  K.]}, \bibinfo{author}{Gajewski\xfnm[ J.]}, \bibinfo{author}{Gajos\xfnm[
  A.]}, \bibinfo{author}{Gorgol\xfnm[ M.]}, \bibinfo{author}{{Del Grande}\xfnm[
  R.]}, \bibinfo{author}{Hiesmayr\xfnm[ B.]},
  \bibinfo{author}{Jasi{\'{n}}ska\xfnm[ B.]}, \bibinfo{author}{Kacprzak\xfnm[
  K.]}, \bibinfo{author}{Kap{\l}on\xfnm[ L.]},
  \bibinfo{author}{Kisielewska\xfnm[ D.]}, \bibinfo{author}{Korcyl\xfnm[ G.]},
  \bibinfo{author}{Kozik\xfnm[ T.]}, \bibinfo{author}{Krawczyk\xfnm[ N.]},
  \bibinfo{author}{Kubicz\xfnm[ E.]}, \bibinfo{author}{Mohammed\xfnm[ M.]},
  \bibinfo{author}{Nied{\'{z}}wiecki\xfnm[ S.]},
  \bibinfo{author}{Pa{\l}ka\xfnm[ M.]},
  \bibinfo{author}{Pawlik-Nied{\'{z}}wiecka\xfnm[ M.]},
  \bibinfo{author}{Raj\xfnm[ J.]}, \bibinfo{author}{Rakoczy\xfnm[ K.]},
  \bibinfo{author}{Ruci{\'{n}}ski\xfnm[ A.]}, \bibinfo{author}{Sharma\xfnm[
  S.]}, \bibinfo{author}{Shivani\xfnm[ S.]}, \bibinfo{author}{Silarski\xfnm[
  M.]}, \bibinfo{author}{Skurzok\xfnm[ M.]},
  \bibinfo{author}{Stepie{\'{n}}\xfnm[ E.]},
  \bibinfo{author}{Zgardzi{\'{n}}ska\xfnm[ B.]}, \bibinfo{author}{Moskal\xfnm[
  P.]}.
\newblock \bibinfo{title}{{3D TOF-PET image reconstruction using total
  variation regularization}}.
\newblock \bibinfo{journal}{Physica Medica}
  \bibinfo{year}{2020};\bibinfo{volume}{80}:\bibinfo{pages}{230--242}.
\newblock \URLprefix
  \url{https://linkinghub.elsevier.com/retrieve/pii/S1120179720302544}.
  \DOIprefix\doi{10.1016/j.ejmp.2020.10.011}.
\bibitem[{Raczy{\'{n}}ski et~al.(2017)Raczy{\'{n}}ski, Wi{\'{s}}licki,
  Krzemie{\'{n}}, Kowalski, Alfs, Bednarski, Bia{\l}as, Curceanu,
  Czerwi{\'{n}}ski, Dulski, Gajos, G{\l}owacz, Gorgol, Hiesmayr,
  Jasi{\'{n}}ska, Kami{\'{n}}ska, Korcyl, Kozik, Krawczyk, Kubicz, Mohammed,
  Pawlik-Nied{\'{z}}wiecka, Nied{\'{z}}wiecki, Pa{\l}ka, Rudy, Rundel, Sharma,
  Silarski, Smyrski, Strzelecki, Wieczorek, Zgardzi{\'{n}}ska, Zieli{\'{n}}ski
  and Moskal}]{Raczynski2017}
\bibinfo{author}{Raczy{\'{n}}ski\xfnm[ L.]},
  \bibinfo{author}{Wi{\'{s}}licki\xfnm[ W.]},
  \bibinfo{author}{Krzemie{\'{n}}\xfnm[ W.]}, \bibinfo{author}{Kowalski\xfnm[
  P.]}, \bibinfo{author}{Alfs\xfnm[ D.]}, \bibinfo{author}{Bednarski\xfnm[
  T.]}, \bibinfo{author}{Bia{\l}as\xfnm[ P.]}, \bibinfo{author}{Curceanu\xfnm[
  C.]}, \bibinfo{author}{Czerwi{\'{n}}ski\xfnm[ E.]},
  \bibinfo{author}{Dulski\xfnm[ K.]}, \bibinfo{author}{Gajos\xfnm[ A.]},
  \bibinfo{author}{G{\l}owacz\xfnm[ B.]}, \bibinfo{author}{Gorgol\xfnm[ M.]},
  \bibinfo{author}{Hiesmayr\xfnm[ B.]}, \bibinfo{author}{Jasi{\'{n}}ska\xfnm[
  B.]}, \bibinfo{author}{Kami{\'{n}}ska\xfnm[ D.]},
  \bibinfo{author}{Korcyl\xfnm[ G.]}, \bibinfo{author}{Kozik\xfnm[ T.]},
  \bibinfo{author}{Krawczyk\xfnm[ N.]}, \bibinfo{author}{Kubicz\xfnm[ E.]},
  \bibinfo{author}{Mohammed\xfnm[ M.]},
  \bibinfo{author}{Pawlik-Nied{\'{z}}wiecka\xfnm[ M.]},
  \bibinfo{author}{Nied{\'{z}}wiecki\xfnm[ S.]},
  \bibinfo{author}{Pa{\l}ka\xfnm[ M.]}, \bibinfo{author}{Rudy\xfnm[ Z.]},
  \bibinfo{author}{Rundel\xfnm[ O.]}, \bibinfo{author}{Sharma\xfnm[ N.G.]},
  \bibinfo{author}{Silarski\xfnm[ M.]}, \bibinfo{author}{Smyrski\xfnm[ J.]},
  \bibinfo{author}{Strzelecki\xfnm[ A.]}, \bibinfo{author}{Wieczorek\xfnm[
  A.]}, \bibinfo{author}{Zgardzi{\'{n}}ska\xfnm[ B.]},
  \bibinfo{author}{Zieli{\'{n}}ski\xfnm[ M.]}, \bibinfo{author}{Moskal\xfnm[
  P.]}.
\newblock \bibinfo{title}{{Calculation of the time resolution of the J-PET
  tomograph using kernel density estimation}}.
\newblock \bibinfo{journal}{Phys Med Biol}
  \bibinfo{year}{2017};\bibinfo{volume}{62}(\bibinfo{number}{12}):\bibinfo{pages}{5076--5097}.
\newblock \URLprefix
  \url{https://iopscience.iop.org/article/10.1088/1361-6560/aa7005}.
  \DOIprefix\doi{10.1088/1361-6560/aa7005}.
\bibitem[{Schaart et~al.(2010)Schaart, Seifert, Vinke, van Dam, Dendooven,
  L{\"{o}}hner and Beekman}]{Schaart2010}
\bibinfo{author}{Schaart\xfnm[ D.R.]}, \bibinfo{author}{Seifert\xfnm[ S.]},
  \bibinfo{author}{Vinke\xfnm[ R.]}, \bibinfo{author}{van Dam\xfnm[ H.T.]},
  \bibinfo{author}{Dendooven\xfnm[ P.]}, \bibinfo{author}{L{\"{o}}hner\xfnm[
  H.]}, \bibinfo{author}{Beekman\xfnm[ F.J.]}.
\newblock \bibinfo{title}{{LaBr3:Ce and SiPMs for time-of-flight PET: achieving
  100 ps coincidence resolving time}}.
\newblock \bibinfo{journal}{Physics in Medicine {\&} Biology}
  \bibinfo{year}{2010};\bibinfo{volume}{55}(\bibinfo{number}{7}):\bibinfo{pages}{N179--N189}.
\newblock \URLprefix
  \url{https://iopscience.iop.org/article/10.1088/0031-9155/55/7/N02}.
  \DOIprefix\doi{10.1088/0031-9155/55/7/N02}.
\bibitem[{Scott(1992)}]{Scott1992}
\bibinfo{author}{Scott\xfnm[ D.W.]}.
\newblock \bibinfo{title}{{Multivariate Density Estimation: Theory, Practice,
  and Visualization}}.
\newblock \bibinfo{address}{New York}: \bibinfo{publisher}{Wiley},
  \bibinfo{year}{1992}.
\bibitem[{Segars et~al.(2008)Segars, Mahesh, Beck, Frey and Tsui}]{Segars2008}
\bibinfo{author}{Segars\xfnm[ W.P.]}, \bibinfo{author}{Mahesh\xfnm[ M.]},
  \bibinfo{author}{Beck\xfnm[ T.J.]}, \bibinfo{author}{Frey\xfnm[ E.C.]},
  \bibinfo{author}{Tsui\xfnm[ B.M.W.]}.
\newblock \bibinfo{title}{{Realistic CT simulation using the 4D XCAT phantom}}.
\newblock \bibinfo{journal}{Medical Physics}
  \bibinfo{year}{2008};\bibinfo{volume}{35}(\bibinfo{number}{8}):\bibinfo{pages}{3800--3808}.
\newblock \URLprefix \url{http://doi.wiley.com/10.1118/1.2955743}.
  \DOIprefix\doi{10.1118/1.2955743}.
\bibitem[{Segars et~al.(2018)Segars, Tsui, Cai, Yin, Fung and
  Samei}]{Segars2018}
\bibinfo{author}{Segars\xfnm[ W.P.]}, \bibinfo{author}{Tsui\xfnm[ B.M.W.]},
  \bibinfo{author}{Cai\xfnm[ J.]}, \bibinfo{author}{Yin\xfnm[ F.F.]},
  \bibinfo{author}{Fung\xfnm[ G.S.K.]}, \bibinfo{author}{Samei\xfnm[ E.]}.
\newblock \bibinfo{title}{{Application of the 4-D XCAT Phantoms in Biomedical
  Imaging and Beyond}}.
\newblock \bibinfo{journal}{IEEE Transactions on Medical Imaging}
  \bibinfo{year}{2018};\bibinfo{volume}{37}(\bibinfo{number}{3}):\bibinfo{pages}{680--692}.
\newblock \URLprefix \url{http://ieeexplore.ieee.org/document/8007279/}.
  \DOIprefix\doi{10.1109/TMI.2017.2738448}.
\bibitem[{Sharma et~al.(2020)Sharma, Chhokar, Curceanu, Czerwi{\'{n}}ski,
  Dadgar, Dulski, Gajewski, Gajos, Gorgol, Gupta-Sharma, {Del Grande},
  Hiesmayr, Jasi{\'{n}}ska, Kacprzak, Kap{\l}on, Karimi, Kisielewska,
  Klimaszewski, Korcyl, Kowalski, Kozik, Krawczyk, Krzemie{\'{n}}, Kubicz,
  Mohammed, Niedzwiecki, Pa{\l}ka, Pawlik-Nied{\'{z}}wiecka, Raczy{\'{n}}ski,
  Raj, Ruci{\'{n}}ski, Shivani, Shopa, Silarski, Skurzok, St{\c{e}}pie{\'{n}},
  Wi{\'{s}}licki, Zgardzi{\'{n}}ska and Moskal}]{Sharma2020}
\bibinfo{author}{Sharma\xfnm[ S.]}, \bibinfo{author}{Chhokar\xfnm[ J.]},
  \bibinfo{author}{Curceanu\xfnm[ C.]}, \bibinfo{author}{Czerwi{\'{n}}ski\xfnm[
  E.]}, \bibinfo{author}{Dadgar\xfnm[ M.]}, \bibinfo{author}{Dulski\xfnm[ K.]},
  \bibinfo{author}{Gajewski\xfnm[ J.]}, \bibinfo{author}{Gajos\xfnm[ A.]},
  \bibinfo{author}{Gorgol\xfnm[ M.]}, \bibinfo{author}{Gupta-Sharma\xfnm[ N.]},
  \bibinfo{author}{{Del Grande}\xfnm[ R.]}, \bibinfo{author}{Hiesmayr\xfnm[
  B.C.]}, \bibinfo{author}{Jasi{\'{n}}ska\xfnm[ B.]},
  \bibinfo{author}{Kacprzak\xfnm[ K.]}, \bibinfo{author}{Kap{\l}on\xfnm[
  {\L}.]}, \bibinfo{author}{Karimi\xfnm[ H.]},
  \bibinfo{author}{Kisielewska\xfnm[ D.]}, \bibinfo{author}{Klimaszewski\xfnm[
  K.]}, \bibinfo{author}{Korcyl\xfnm[ G.]}, \bibinfo{author}{Kowalski\xfnm[
  P.]}, \bibinfo{author}{Kozik\xfnm[ T.]}, \bibinfo{author}{Krawczyk\xfnm[
  N.]}, \bibinfo{author}{Krzemie{\'{n}}\xfnm[ W.]},
  \bibinfo{author}{Kubicz\xfnm[ E.]}, \bibinfo{author}{Mohammed\xfnm[ M.]},
  \bibinfo{author}{Niedzwiecki\xfnm[ S.]}, \bibinfo{author}{Pa{\l}ka\xfnm[
  M.]}, \bibinfo{author}{Pawlik-Nied{\'{z}}wiecka\xfnm[ M.]},
  \bibinfo{author}{Raczy{\'{n}}ski\xfnm[ L.]}, \bibinfo{author}{Raj\xfnm[ J.]},
  \bibinfo{author}{Ruci{\'{n}}ski\xfnm[ A.]}, \bibinfo{author}{Shivani\xfnm[
  S.]}, \bibinfo{author}{Shopa\xfnm[ R.Y.]}, \bibinfo{author}{Silarski\xfnm[
  M.]}, \bibinfo{author}{Skurzok\xfnm[ M.]},
  \bibinfo{author}{St{\c{e}}pie{\'{n}}\xfnm[ E.]},
  \bibinfo{author}{Wi{\'{s}}licki\xfnm[ W.]},
  \bibinfo{author}{Zgardzi{\'{n}}ska\xfnm[ B.]}, \bibinfo{author}{Moskal\xfnm[
  P.]}.
\newblock \bibinfo{title}{{Estimating relationship between the time over
  threshold and energy loss by photons in plastic scintillators used in the
  J-PET scanner}}.
\newblock \bibinfo{journal}{EJNMMI Phys}
  \bibinfo{year}{2020};\bibinfo{volume}{7}(\bibinfo{number}{1}):\bibinfo{pages}{39}.
\newblock \URLprefix
  \url{https://ejnmmiphys.springeropen.com/articles/10.1186/s40658-020-00306-x}.
  \DOIprefix\doi{10.1186/s40658-020-00306-x}.
\bibitem[{Shopa(2020)}]{Shopa2020}
\bibinfo{author}{Shopa\xfnm[ R.Y.]}.
\newblock \bibinfo{title}{{Estimation of Spatial Resolution for 3-layer J-PET
  Scanner Using TOF FBP Based on Event-by-Event Approach}}.
\newblock \bibinfo{journal}{APPB}
  \bibinfo{year}{2020};\bibinfo{volume}{51}(\bibinfo{number}{1}):\bibinfo{pages}{181--190}.
\newblock \URLprefix
  \url{http://www.actaphys.uj.edu.pl/findarticle?series=Reg\&vol=51\&page=181}.
  \DOIprefix\doi{10.5506/APhysPolB.51.181}.
\bibitem[{Shopa et~al.(2017)Shopa, Klimaszewski, Kowalski, Krzemie{\'{n}},
  Raczy{\'{n}}ski, Wi{\'{s}}licki, Bia{\l}as, Curceanu, Czerwi{\'{n}}ski,
  Dulski, Gajos, G{\l}owacz, Gorgol, Hiesmayr, Jasi{\'{n}}ska,
  Kisielewska-Kami{\'{n}}ska, Korcyl, Kozik, Krawczyk, Kubicz, Mohammed,
  Pawlik-Nied{\'{z}}wiecka, Nied{\'{z}}wiecki, Pa{\l}ka, Rudy, Sharma, Sharma,
  Silarski, Skurzok, Wieczorek, Zgardzi{\'{n}}ska, Zieli{\'{n}}ski and
  Moskal}]{Shopa2017}
\bibinfo{author}{Shopa\xfnm[ R.Y.]}, \bibinfo{author}{Klimaszewski\xfnm[ K.]},
  \bibinfo{author}{Kowalski\xfnm[ P.]}, \bibinfo{author}{Krzemie{\'{n}}\xfnm[
  W.]}, \bibinfo{author}{Raczy{\'{n}}ski\xfnm[ L.]},
  \bibinfo{author}{Wi{\'{s}}licki\xfnm[ W.]}, \bibinfo{author}{Bia{\l}as\xfnm[
  P.]}, \bibinfo{author}{Curceanu\xfnm[ C.]},
  \bibinfo{author}{Czerwi{\'{n}}ski\xfnm[ E.]}, \bibinfo{author}{Dulski\xfnm[
  K.]}, \bibinfo{author}{Gajos\xfnm[ A.]}, \bibinfo{author}{G{\l}owacz\xfnm[
  B.]}, \bibinfo{author}{Gorgol\xfnm[ M.]}, \bibinfo{author}{Hiesmayr\xfnm[
  B.]}, \bibinfo{author}{Jasi{\'{n}}ska\xfnm[ B.]},
  \bibinfo{author}{Kisielewska-Kami{\'{n}}ska\xfnm[ D.]},
  \bibinfo{author}{Korcyl\xfnm[ G.]}, \bibinfo{author}{Kozik\xfnm[ T.]},
  \bibinfo{author}{Krawczyk\xfnm[ N.]}, \bibinfo{author}{Kubicz\xfnm[ E.]},
  \bibinfo{author}{Mohammed\xfnm[ M.]},
  \bibinfo{author}{Pawlik-Nied{\'{z}}wiecka\xfnm[ M.]},
  \bibinfo{author}{Nied{\'{z}}wiecki\xfnm[ S.]},
  \bibinfo{author}{Pa{\l}ka\xfnm[ M.]}, \bibinfo{author}{Rudy\xfnm[ Z.]},
  \bibinfo{author}{Sharma\xfnm[ N.G.]}, \bibinfo{author}{Sharma\xfnm[ S.]},
  \bibinfo{author}{Silarski\xfnm[ M.]}, \bibinfo{author}{Skurzok\xfnm[ M.]},
  \bibinfo{author}{Wieczorek\xfnm[ A.]},
  \bibinfo{author}{Zgardzi{\'{n}}ska\xfnm[ B.]},
  \bibinfo{author}{Zieli{\'{n}}ski\xfnm[ M.]}, \bibinfo{author}{Moskal\xfnm[
  P.]}.
\newblock \bibinfo{title}{{Three-dimensional Image Reconstruction in J-PET
  Using Filtered Back-projection Method}}.
\newblock \bibinfo{journal}{APPB}
  \bibinfo{year}{2017};\bibinfo{volume}{48}(\bibinfo{number}{10}):\bibinfo{pages}{1757}.
\newblock \URLprefix
  \url{http://www.actaphys.uj.edu.pl/findarticle?series=Reg\&vol=48\&page=1757}.
  \DOIprefix\doi{10.5506/APhysPolB.48.1757}.
\bibitem[{Silverman(1986)}]{Silverman1986}
\bibinfo{author}{Silverman\xfnm[ B.]}.
\newblock \bibinfo{title}{{Density Estimation for Statistics and Data
  Analysis}}.
\newblock \bibinfo{address}{London, New York}: \bibinfo{publisher}{Chapman and
  Hall}, \bibinfo{year}{1986}.
\bibitem[{Slomka et~al.(2016)Slomka, Pan and Germano}]{Slomka2016}
\bibinfo{author}{Slomka\xfnm[ P.J.]}, \bibinfo{author}{Pan\xfnm[ T.]},
  \bibinfo{author}{Germano\xfnm[ G.]}.
\newblock \bibinfo{title}{{Recent Advances and Future Progress in PET
  Instrumentation}}.
\newblock \bibinfo{journal}{Semin Nucl Med}
  \bibinfo{year}{2016};\bibinfo{volume}{46}(\bibinfo{number}{1}):\bibinfo{pages}{5--19}.
\newblock \URLprefix \url{http://www.ncbi.nlm.nih.gov/pubmed/26687853}.
  \DOIprefix\doi{10.1053/j.semnuclmed.2015.09.006}.
\bibitem[{van Sluis et~al.(2019)van Sluis, de~Jong, Schaar, Noordzij, van
  Snick, Dierckx, Borra, Willemsen and Boellaard}]{VanSluis2019}
\bibinfo{author}{van Sluis\xfnm[ J.]}, \bibinfo{author}{de~Jong\xfnm[ J.]},
  \bibinfo{author}{Schaar\xfnm[ J.]}, \bibinfo{author}{Noordzij\xfnm[ W.]},
  \bibinfo{author}{van Snick\xfnm[ P.]}, \bibinfo{author}{Dierckx\xfnm[ R.]},
  \bibinfo{author}{Borra\xfnm[ R.]}, \bibinfo{author}{Willemsen\xfnm[ A.]},
  \bibinfo{author}{Boellaard\xfnm[ R.]}.
\newblock \bibinfo{title}{{Performance Characteristics of the Digital Biograph
  Vision PET/CT System}}.
\newblock \bibinfo{journal}{J Nucl Med}
  \bibinfo{year}{2019};\bibinfo{volume}{60}(\bibinfo{number}{7}):\bibinfo{pages}{1031--1036}.
\newblock \URLprefix
  \url{http://jnm.snmjournals.org/lookup/doi/10.2967/jnumed.118.215418}.
  \DOIprefix\doi{10.2967/jnumed.118.215418}.
\bibitem[{Smyrski et~al.(2017)Smyrski, Alfs, Bednarski, Bia{\l}as,
  Czerwi{\'{n}}ski, Dulski, Gajos, G{\l}owacz, Gupta-Sharma, Gorgol,
  Jasi{\'{n}}ska, Kajetanowicz, Kami{\'{n}}ska, Korcyl, Kowalski,
  Krzemie{\'{n}}, Krawczyk, Kubicz, Mohammed, Nied{\'{z}}wiecki,
  Pawlik-Nied{\'{z}}wiecka, Raczy{\'{n}}ski, Rudy, Salabura, Silarski,
  Strzelecki, Wieczorek, Wi{\'{s}}licki, Wojnarska, Zgardzi{\'{n}}ska,
  Zieli{\'{n}}ski and Moskal}]{Smyrski2017}
\bibinfo{author}{Smyrski\xfnm[ J.]}, \bibinfo{author}{Alfs\xfnm[ D.]},
  \bibinfo{author}{Bednarski\xfnm[ T.]}, \bibinfo{author}{Bia{\l}as\xfnm[ P.]},
  \bibinfo{author}{Czerwi{\'{n}}ski\xfnm[ E.]}, \bibinfo{author}{Dulski\xfnm[
  K.]}, \bibinfo{author}{Gajos\xfnm[ A.]}, \bibinfo{author}{G{\l}owacz\xfnm[
  B.]}, \bibinfo{author}{Gupta-Sharma\xfnm[ N.]}, \bibinfo{author}{Gorgol\xfnm[
  M.]}, \bibinfo{author}{Jasi{\'{n}}ska\xfnm[ B.]},
  \bibinfo{author}{Kajetanowicz\xfnm[ M.]},
  \bibinfo{author}{Kami{\'{n}}ska\xfnm[ D.]}, \bibinfo{author}{Korcyl\xfnm[
  G.]}, \bibinfo{author}{Kowalski\xfnm[ P.]},
  \bibinfo{author}{Krzemie{\'{n}}\xfnm[ W.]}, \bibinfo{author}{Krawczyk\xfnm[
  N.]}, \bibinfo{author}{Kubicz\xfnm[ E.]}, \bibinfo{author}{Mohammed\xfnm[
  M.]}, \bibinfo{author}{Nied{\'{z}}wiecki\xfnm[ S.]},
  \bibinfo{author}{Pawlik-Nied{\'{z}}wiecka\xfnm[ M.]},
  \bibinfo{author}{Raczy{\'{n}}ski\xfnm[ L.]}, \bibinfo{author}{Rudy\xfnm[
  Z.]}, \bibinfo{author}{Salabura\xfnm[ P.]}, \bibinfo{author}{Silarski\xfnm[
  M.]}, \bibinfo{author}{Strzelecki\xfnm[ A.]},
  \bibinfo{author}{Wieczorek\xfnm[ A.]}, \bibinfo{author}{Wi{\'{s}}licki\xfnm[
  W.]}, \bibinfo{author}{Wojnarska\xfnm[ J.]},
  \bibinfo{author}{Zgardzi{\'{n}}ska\xfnm[ B.]},
  \bibinfo{author}{Zieli{\'{n}}ski\xfnm[ M.]}, \bibinfo{author}{Moskal\xfnm[
  P.]}.
\newblock \bibinfo{title}{{Measurement of gamma quantum interaction point in
  plastic scintillator with WLS strips}}.
\newblock \bibinfo{journal}{Nucl Instrum Methods Phys Res A}
  \bibinfo{year}{2017};\bibinfo{volume}{851}:\bibinfo{pages}{39--42}.
\newblock \URLprefix
  \url{https://linkinghub.elsevier.com/retrieve/pii/S0168900217301079}.
  \DOIprefix\doi{10.1016/j.nima.2017.01.045}.
\bibitem[{Snyder and Politte(1983)}]{Snyder1983}
\bibinfo{author}{Snyder\xfnm[ D.L.]}, \bibinfo{author}{Politte\xfnm[ D.G.]}.
\newblock \bibinfo{title}{{Image Reconstruction from List-Mode Data in an
  Emission Tomography System Having Time-of-Flight Measurements}}.
\newblock \bibinfo{journal}{IEEE Trans Nucl Sci}
  \bibinfo{year}{1983};\bibinfo{volume}{30}(\bibinfo{number}{3}):\bibinfo{pages}{1843--1849}.
\newblock \URLprefix \url{http://ieeexplore.ieee.org/document/4332660/}.
  \DOIprefix\doi{10.1109/TNS.1983.4332660}.
\bibitem[{Strzelecki(2016)}]{Strzelecki2016}
\bibinfo{author}{Strzelecki\xfnm[ A.]}.
\newblock \bibinfo{title}{{Image reconstruction and simulation of strip
  positron emission tomography scanner using computational accelerators}}.
\newblock \bibinfo{type}{Ph.d. thesis}; Polish Academy of Sciences,
  Krak{\'{o}}w; \bibinfo{year}{2016}.
\newblock \URLprefix
  \url{https://oldwww.ippt.pan.pl/_download/doktoraty/2016strzelecki_a_doktorat.pdf}.
\bibitem[{Surti and Karp(2008)}]{Surti2008}
\bibinfo{author}{Surti\xfnm[ S.]}, \bibinfo{author}{Karp\xfnm[ J.S.]}.
\newblock \bibinfo{title}{{Design considerations for a limited angle, dedicated
  breast, TOF PET scanner}}.
\newblock \bibinfo{journal}{Phys Med Biol}
  \bibinfo{year}{2008};\bibinfo{volume}{53}(\bibinfo{number}{11}):\bibinfo{pages}{2911--2921}.
\newblock \URLprefix \url{http://www.ncbi.nlm.nih.gov/pubmed/18460745}.
  \DOIprefix\doi{10.1088/0031-9155/53/11/010}.
\bibitem[{Surti et~al.(2007)Surti, Kuhn, Werner, Perkins, Kolthammer and
  Karp}]{Surti2007}
\bibinfo{author}{Surti\xfnm[ S.]}, \bibinfo{author}{Kuhn\xfnm[ A.]},
  \bibinfo{author}{Werner\xfnm[ M.E.]}, \bibinfo{author}{Perkins\xfnm[ A.E.]},
  \bibinfo{author}{Kolthammer\xfnm[ J.]}, \bibinfo{author}{Karp\xfnm[ J.S.]}.
\newblock \bibinfo{title}{{Performance of Philips Gemini TF PET/CT scanner with
  special consideration for its time-of-flight imaging capabilities.}}
\newblock \bibinfo{journal}{J Nucl Med}
  \bibinfo{year}{2007};\bibinfo{volume}{48}(\bibinfo{number}{3}):\bibinfo{pages}{471--480}.
\newblock \URLprefix \url{http://www.ncbi.nlm.nih.gov/pubmed/17332626}.
\bibitem[{Tadashi et~al.(2020)Tadashi, Sadahiro, Fumihiko, Takashi, Hidetaka,
  Daisuke, Hiroki, Mitsuaki, Eku, Yoshikatsu and Jun}]{Watabe2020}
\bibinfo{author}{Tadashi\xfnm[ W.]}, \bibinfo{author}{Sadahiro\xfnm[ N.]},
  \bibinfo{author}{Fumihiko\xfnm[ S.]}, \bibinfo{author}{Takashi\xfnm[ K.]},
  \bibinfo{author}{Hidetaka\xfnm[ S.]}, \bibinfo{author}{Daisuke\xfnm[ K.]},
  \bibinfo{author}{Hiroki\xfnm[ K.]}, \bibinfo{author}{Mitsuaki\xfnm[ T.]},
  \bibinfo{author}{Eku\xfnm[ S.]}, \bibinfo{author}{Yoshikatsu\xfnm[ K.]},
  \bibinfo{author}{Jun\xfnm[ H.]}.
\newblock \bibinfo{title}{{First in human dosimetry of 18F-NKO-035: a new PET
  probe targeting L-type amino acid transporter 1 (LAT1)}}.
\newblock \bibinfo{journal}{Journal of nuclear medicine : official publication,
  Society of Nuclear Medicine}
  \bibinfo{year}{2020};\bibinfo{volume}{61}(\bibinfo{number}{supplement
  1}):\bibinfo{pages}{627}.
\newblock \URLprefix
  \url{https://jnm.snmjournals.org/content/61/supplement_1/627}.
\bibitem[{Thielemans et~al.(2006)Thielemans, Mustafovic and
  Tsoumpas}]{STIR2006}
\bibinfo{author}{Thielemans\xfnm[ K.]}, \bibinfo{author}{Mustafovic\xfnm[ S.]},
  \bibinfo{author}{Tsoumpas\xfnm[ C.]}.
\newblock \bibinfo{title}{{STIR: Software for Tomographic Image Reconstruction
  Release 2}}.
\newblock In: \bibinfo{booktitle}{2006 IEEE Nucl. Sci. Symp. Conf. Rec.}
  \bibinfo{publisher}{IEEE}; \bibinfo{year}{2006}. p.
  \bibinfo{pages}{2174--2176}.
\newblock \URLprefix \url{http://ieeexplore.ieee.org/document/4179459/}.
  \DOIprefix\doi{10.1109/NSSMIC.2006.354345}.
\bibitem[{Thielemans et~al.(2012)Thielemans, Tsoumpas, Mustafovic, Beisel,
  Aguiar, Dikaios and Jacobson}]{STIR2012}
\bibinfo{author}{Thielemans\xfnm[ K.]}, \bibinfo{author}{Tsoumpas\xfnm[ C.]},
  \bibinfo{author}{Mustafovic\xfnm[ S.]}, \bibinfo{author}{Beisel\xfnm[ T.]},
  \bibinfo{author}{Aguiar\xfnm[ P.]}, \bibinfo{author}{Dikaios\xfnm[ N.]},
  \bibinfo{author}{Jacobson\xfnm[ M.W.]}.
\newblock \bibinfo{title}{{STIR: software for tomographic image reconstruction
  release 2}}.
\newblock \bibinfo{journal}{Phys Med Biol}
  \bibinfo{year}{2012};\bibinfo{volume}{57}(\bibinfo{number}{4}):\bibinfo{pages}{867--883}.
\newblock \URLprefix \url{http://www.ncbi.nlm.nih.gov/pubmed/22290410}.
  \DOIprefix\doi{10.1088/0031-9155/57/4/867}.
\bibitem[{Tomitani(1981)}]{Tomitani1981}
\bibinfo{author}{Tomitani\xfnm[ T.]}.
\newblock \bibinfo{title}{{Image Reconstruction and Noise Evaluation in Photon
  Time-of-Flight Assisted Positron Emission Tomography}}.
\newblock \bibinfo{journal}{IEEE Trans Nucl Sci}
  \bibinfo{year}{1981};\bibinfo{volume}{28}(\bibinfo{number}{6}):\bibinfo{pages}{4581--4589}.
\newblock \URLprefix \url{http://ieeexplore.ieee.org/document/4335769/}.
  \DOIprefix\doi{10.1109/TNS.1981.4335769}.
\bibitem[{Vandenberghe and Karp(2006)}]{Vandenberghe2006}
\bibinfo{author}{Vandenberghe\xfnm[ S.]}, \bibinfo{author}{Karp\xfnm[ J.]}.
\newblock \bibinfo{title}{{Rebinning and reconstruction techniques for 3D
  TOF-PET}}.
\newblock \bibinfo{journal}{Nucl Instrum Methods Phys Res A}
  \bibinfo{year}{2006};\bibinfo{volume}{569}(\bibinfo{number}{2}):\bibinfo{pages}{421--424}.
\newblock \URLprefix
  \url{https://linkinghub.elsevier.com/retrieve/pii/S016890020601494X}.
  \DOIprefix\doi{10.1016/j.nima.2006.08.065}.
\bibitem[{Vandenberghe et~al.(2016)Vandenberghe, Mikhaylova, D'Hoe, Mollet and
  Karp}]{Vandenberghe2016}
\bibinfo{author}{Vandenberghe\xfnm[ S.]}, \bibinfo{author}{Mikhaylova\xfnm[
  E.]}, \bibinfo{author}{D'Hoe\xfnm[ E.]}, \bibinfo{author}{Mollet\xfnm[ P.]},
  \bibinfo{author}{Karp\xfnm[ J.S.]}.
\newblock \bibinfo{title}{{Recent developments in time-of-flight PET}}.
\newblock \bibinfo{journal}{EJNMMI Phys}
  \bibinfo{year}{2016};\bibinfo{volume}{3}(\bibinfo{number}{3}):\bibinfo{pages}{1}.
\newblock \URLprefix \url{http://www.ejnmmiphys.com/content/3/1/3}.
  \DOIprefix\doi{10.1186/s40658-016-0138-3}.
\bibitem[{Vandenberghe et~al.(2020)Vandenberghe, Moskal and
  Karp}]{Vandenberghe2020}
\bibinfo{author}{Vandenberghe\xfnm[ S.]}, \bibinfo{author}{Moskal\xfnm[ P.]},
  \bibinfo{author}{Karp\xfnm[ J.S.]}.
\newblock \bibinfo{title}{{State of the art in total body PET}}.
\newblock \bibinfo{journal}{EJNMMI Phys}
  \bibinfo{year}{2020};\bibinfo{volume}{7}(\bibinfo{number}{35}):\bibinfo{pages}{1}.
\newblock \URLprefix
  \url{https://ejnmmiphys.springeropen.com/articles/10.1186/s40658-020-00290-2}.
  \DOIprefix\doi{10.1186/s40658-020-00290-2}.
\bibitem[{Walker et~al.(2011)Walker, Asselin, Julyan, Feldmann, Talbot, Jones
  and Matthews}]{Walker2011}
\bibinfo{author}{Walker\xfnm[ M.D.]}, \bibinfo{author}{Asselin\xfnm[ M.C.]},
  \bibinfo{author}{Julyan\xfnm[ P.J.]}, \bibinfo{author}{Feldmann\xfnm[ M.]},
  \bibinfo{author}{Talbot\xfnm[ P.S.]}, \bibinfo{author}{Jones\xfnm[ T.]},
  \bibinfo{author}{Matthews\xfnm[ J.C.]}.
\newblock \bibinfo{title}{{Bias in iterative reconstruction of low-statistics
  PET data: benefits of a resolution model.}}
\newblock \bibinfo{journal}{Phys Med Biol}
  \bibinfo{year}{2011};\bibinfo{volume}{56}(\bibinfo{number}{4}):\bibinfo{pages}{931--949}.
\newblock \URLprefix \url{http://www.ncbi.nlm.nih.gov/pubmed/21248391}.
  \DOIprefix\doi{10.1088/0031-9155/56/4/004}.
\bibitem[{Wand and Jones(1994)}]{Wand1994}
\bibinfo{author}{Wand\xfnm[ M.P.]}, \bibinfo{author}{Jones\xfnm[ C.]}.
\newblock \bibinfo{title}{{Multivariate plug-in bandwidth selection}}.
\newblock \bibinfo{journal}{Comp Stat}
  \bibinfo{year}{1994};\bibinfo{volume}{9}(\bibinfo{number}{2}):\bibinfo{pages}{97--116}.
\newblock \URLprefix \url{http://oro.open.ac.uk/28244/}.
\bibitem[{Westerwoudt et~al.(2014)Westerwoudt, Conti and
  Eriksson}]{Westerwoudt2014}
\bibinfo{author}{Westerwoudt\xfnm[ V.]}, \bibinfo{author}{Conti\xfnm[ M.]},
  \bibinfo{author}{Eriksson\xfnm[ L.]}.
\newblock \bibinfo{title}{{Advantages of Improved Time Resolution for TOF PET
  at Very Low Statistics}}.
\newblock \bibinfo{journal}{IEEE Trans Nucl Sci}
  \bibinfo{year}{2014};\bibinfo{volume}{61}(\bibinfo{number}{1}):\bibinfo{pages}{126--133}.
\newblock \URLprefix \url{http://ieeexplore.ieee.org/document/6704829/}.
  \DOIprefix\doi{10.1109/TNS.2013.2287175}.
\bibitem[{Whiteley et~al.(2021)Whiteley, Panin, Zhou, Cabello, Bharkhada and
  Gregor}]{Whiteley2021}
\bibinfo{author}{Whiteley\xfnm[ W.]}, \bibinfo{author}{Panin\xfnm[ V.]},
  \bibinfo{author}{Zhou\xfnm[ C.]}, \bibinfo{author}{Cabello\xfnm[ J.]},
  \bibinfo{author}{Bharkhada\xfnm[ D.]}, \bibinfo{author}{Gregor\xfnm[ J.]}.
\newblock \bibinfo{title}{{FastPET: Near Real-Time Reconstruction of PET
  Histo-Image Data Using a Neural Network}}.
\newblock \bibinfo{journal}{IEEE Transactions on Radiation and Plasma Medical
  Sciences}
  \bibinfo{year}{2021};\bibinfo{volume}{5}(\bibinfo{number}{1}):\bibinfo{pages}{65--77}.
\newblock \URLprefix \url{https://ieeexplore.ieee.org/document/9211797/}.
  \DOIprefix\doi{10.1109/TRPMS.2020.3028364}.
\bibitem[{Wong et~al.(1983)Wong, Mullani, Philippe, Hartz and Gould}]{Wong1983}
\bibinfo{author}{Wong\xfnm[ W.H.]}, \bibinfo{author}{Mullani\xfnm[ N.A.]},
  \bibinfo{author}{Philippe\xfnm[ E.A.]}, \bibinfo{author}{Hartz\xfnm[ R.]},
  \bibinfo{author}{Gould\xfnm[ K.L.]}.
\newblock \bibinfo{title}{{Image improvement and design optimization of the
  time-of-flight PET}}.
\newblock \bibinfo{journal}{J Nucl Med}
  \bibinfo{year}{1983};\bibinfo{volume}{24}(\bibinfo{number}{1}):\bibinfo{pages}{52--60}.
\newblock \URLprefix \url{http://www.ncbi.nlm.nih.gov/pubmed/6600276}.
\bibitem[{Yamamoto et~al.(1982)Yamamoto, Ficke and
  Ter-Pogossian}]{Yamamoto1982}
\bibinfo{author}{Yamamoto\xfnm[ M.]}, \bibinfo{author}{Ficke\xfnm[ D.C.]},
  \bibinfo{author}{Ter-Pogossian\xfnm[ M.M.]}.
\newblock \bibinfo{title}{{Experimental Assessment of the Gain Achieved by the
  Utilization of Time-of-Flight Information in a Positron Emission Tomograph
  (Super PETT I)}}.
\newblock \bibinfo{journal}{IEEE Trans Med Imaging}
  \bibinfo{year}{1982};\bibinfo{volume}{1}(\bibinfo{number}{3}):\bibinfo{pages}{187--192}.
\newblock \URLprefix \url{http://www.ncbi.nlm.nih.gov/pubmed/18238274}.
  \DOIprefix\doi{10.1109/TMI.1982.4307571}.
\bibitem[{Zeng et~al.(2019)Zeng, Li and Huang}]{Zeng2019}
\bibinfo{author}{Zeng\xfnm[ G.L.]}, \bibinfo{author}{Li\xfnm[ Y.]},
  \bibinfo{author}{Huang\xfnm[ Q.]}.
\newblock \bibinfo{title}{{Analytic time-of-flight positron emission tomography
  reconstruction: two-dimensional case}}.
\newblock \bibinfo{journal}{Vis Comput Ind Biomed Art}
  \bibinfo{year}{2019};\bibinfo{volume}{2}(\bibinfo{number}{1}):\bibinfo{pages}{22}.
\newblock \URLprefix
  \url{https://vciba.springeropen.com/articles/10.1186/s42492-019-0035-4}.
  \DOIprefix\doi{10.1186/s42492-019-0035-4}.

\end{thebibliography}

\end{document}